# GLOBAL GENE EXPRESSION ANALYSIS

# USING MACHINE LEARNING METHODS

XU MIN

(B.Eng. Beihang Univ., China)

A THESIS SUBMITTED

FOR THE DEGREE OF MASTER OF SCIENCE

DEPARTMENT OF INFORMATION SYSTEMS

NATIONAL UNIVERSITY OF SINGAPORE

2003



## Acknowledgements

I am very grateful to my supervisor Dr. Rudy Setiono for his insightful suggestions in both the content and presentation of this thesis. It was his encouragement, support and patience that saw me through and I am ever grateful to him. I am full of gratitude to my boss, Dr. Peng Jinrong, for his understanding and support on allowing me to take the part-time Master of Science research work.

I would like to thank Ms. Jane Lo, Mr. Guan Bin, and other members of Lab of Functional Genomics of Institute of Molecular and Cell Biology, Singapore, for their numerous help throughout my research work.

For the study of the hybrid of Likelihood method and Recursive Feature Elimination method, I would like to thank Dr. Isabelle Guyon for providing the supplementary data. I also thank Dr. Cui Lirong, Mr. Wang Yang, Dr. Oilian Kon, Dr. Wolfgang Hartmann, and Mr. Li Guozheng for their numerous helpful consultations.

I would also like to thank my parents for their love, encouragement, guidance and patience throughout my studies.



# Table of Contents













# List of Figures





# List of Tables





# Summary


Microarray is a technology to quantitatively monitor the expression of large number of genes in parallel. It has become one of the main tools for global gene expression analysis in molecular biology research in recent years. The large amount of expression data generated by this technology makes the study of certain complex biological problems possible and machine learning methods are playing a crucial role in the analysis process. At present, many machine learning methods have been or have the potential to be applied to major areas of gene expression analysis. These areas include clustering, classification, dynamic modeling and reverse engineering.

In this thesis, we focus our work on using machine learning methods to solve the classification problems arising from microarray data. We first identify the major types of the classification problems; then apply several machine learning methods to solve the problems and perform systematic tests on real and artificial datasets. We propose improvement to existing methods. Specifically, we develop a multivariate and a hybrid feature selection method to obtain high classification performance for high dimension classification problems. Using the hybrid feature selection method, we are able to identify small sets of features that give predictive accuracy that is as good as that from other methods which require many more features.




# 1 Introduction

## 1.1 Background

### 1.1.1 Functional genomics

With the completion of Human Genome Project, biology research is entering the post genome era. Although biologists have collected a vast amount of DNA sequence data, the details of how these sequences function still remain largely unknown. Genomes of even the simplest organisms are very complex. Nowadays, biologists are still trying to find answers to the following questions (Brazma and Vilo, 2000):

- What are the functional roles of different genes and in what cellular process do they participate?

- How are the genes regulated? How do the genes and gene products interact? What are the interaction networks?

- How does the gene expression level differ in various cell types and states? How is the gene expression changed by the various diseases of compound treatments?

Biology used to be data-poor science. With more advanced techniques developed in recent years, biologists are now able to transform vast amount of biological information into useful data. This makes it possible to study gene function globally, and a new field, functional genomics emerges. Specifically, *functional genomics* refers to the development and application of global (genome-wide or system-wide) experimental approaches to assess gene function by making use of the information and reagents provided by structural genomics. It is characterized by high throughput or large scale experimental methodologies combined with statistical and computational analysis of the results (Hieter and Boguski, 1997).



## 1.1.2 Microarray Technology

Several methods have been developed to understand the behavior of genes. Microarray technology is an important one among them. It is used to monitor large amount of genes' expression level in parallel. Here *gene expression* refers to the process to transcribe a gene's DNA sequence into the RNA that serves as a template for protein production, and gene expression level indicates how active a gene is in certain tissue, at certain time, or under certain experimental condition. The monitored gene expression level provides an overall picture of the genes being studied. It also reflects the activities of the corresponding protein under certain conditions.

Several steps are involved in this technology. First, complementary DNA (cDNA) molecules or oligos are printed onto slides as spots. Then, two kinds of dye labeled samples, i.e. *sample* and *control* , are hybridized. Finally, the hybridization is scanned and stored as images (see example in Figure 1.1, a sample from Zebra fish). Using a suitable image-processing algorithm, these images are quantified into a set of expression values representing the intensity of spots. Usually, the dye intensity may be biased by factors like its physical property, experimental variability in probe coupling and processing procedures, and scanner settings. To minimize the undesirable effects caused by this biased dye intensity, normalization is done to balance dye intensities and make expression value comparable across experiments (Yang et al., 2001). Here the term comparable means that the difference of any measured expression value of a gene between two experiments should reflect the difference of its true expression levels.



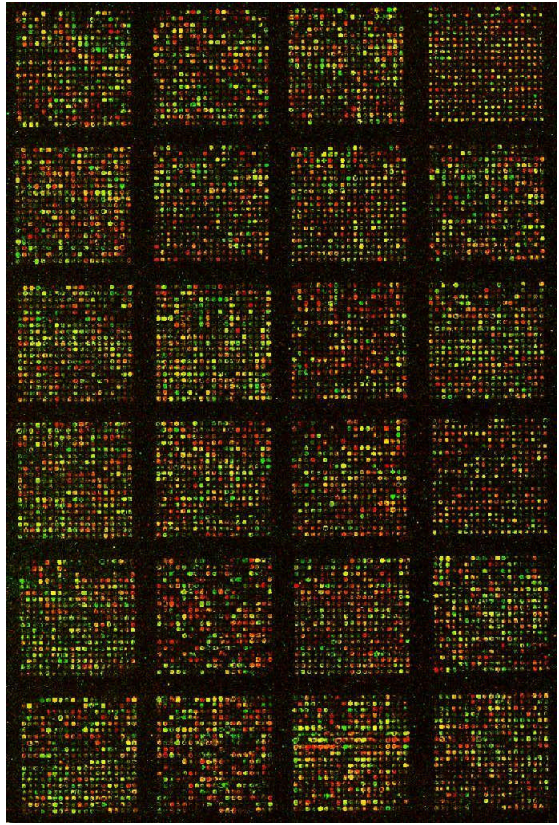

Figure 1.1: A scanned microarray image

### 1.1.3  Machine learning methods for global analysis

Molecular biology also used to be a data poor field, and most of gene expression analysis work was done manually with very limited information derived from experiment. The focus of a molecular biologist was on a few genes or proteins. With the application of large-scale biological information quantification methods like microarray and DNA sequencing, the behavior of genes can be studied globally. Currently, there is an increasing demand for automatic analysis of the overall relationship hidden behind large amount of genes from their expression.



Machine learning is the study of algorithms that could learn from experience and then predict. The theoretical aspects of machine learning are rooted in statistics and informatics, but computational considerations are also indispensable. Due to the complex nature of biological information, machine learning could play an important role in the analysis process.

## 1.2  Research Objectives

Microarray technology based gene expression profiling is one of hottest research topics in biology at present. The experimental part of this technology is already mature. Compared with this, the exploration of automatic analysis methods is still at its early stage. In this thesis, we study several machine learning approaches to solving several typical gene expression analysis problems.

The main objectives of this research are:

- To identify typical gene expression analysis problems from machine learning point of view.

- To apply suitable machine learning methods to the problems from public datasets, and to improve these methods when necessary.

- To find new approaches to the problems.

- To study the experimental results.

- To apply the methods to new datasets and validate the result.

## 1.3  Organization of chapters

This thesis is organized as follows. Chapter 2 provides a brief review of the current methods that can be applied to microarray data analysis. Chapter 3 gives detailed illustrations of



several important machine learning methods that can be applied to classification using gene expression data. In particular, we improve a neural network feature selector method, developed multivariate likelihood feature selection method, and propose a hybrid framework of univariate and multivariate feature selection method. Chapter 4 describes the experimental results of these methods on two different kinds of gene expression analysis problems, and discusses the experiment results. Specifically, we perform systematic tests of the hybrid of Likehood method and Recursive Feature Elimination method because we have obtained very good feature selection performance on several microarray datasets; we also apply Support Vector Machine on a recently obtained Zebra fish dataset to perform gene function prediction. Finally, Chapter 5 concludes the thesis and future works are illustrated.



# 2  Literature review

Various automatic methods have been applied or developed for the gene expression analysis. They are basically from fields such as machine learning, statistics, signal processing, and informatics. The following are the relevant works categorized according to the analysis tasks.

## 2.1  Finding gene groups

Some methods could be used to find useful information or pattern from biological data, which indicates relationship among the genes. These methods are unsupervised (Haykin, 1999), i.e., the learning models are optimized using pre-specified task-independent measures, which reflect the difference or similarity of the training samples. Once the model has become tuned to the statistical regularities of the input gene expression data, it develops the ability to form internal representations for encoding features of the input and thereby to create new classes automatically (Becker, 1991).

Principle Component Analysis (PCA) is an exploratory multivariate statistical technique for simplifying complex data sets (Raychaudhuri et al., 2000). Given an expression matrix with a number of features, a set of new features is generated by PCA. These new features account for most of the information in the original features, but the number of dimensions is smaller than that of the original data. There are several neural network algorithms that support PCA, these algorithms are mainly Hebbian-based algorithms (Haykin, 1999) which are self-organizing and adaptive. Singular Value Decomposition (SVD) can also be used to perform PCA. SVD is a linear transformation that decomposes the gene expression matrix into a product of three matrices that represent the underlying characteristics of the original matrix. Alter et al. (2000) applied SVD for gene expression analysis. They first obtained the principal components from the decomposed matrices by applying SVD to expression data of yeast



genes; then rejected the genes that contribute little information to the principal components. In their work, the information contribution was measured by Shannon entropy of the expression values of the genes, where Shannon entropy characterizes the complexity of the expression values. Finally, the remaining genes were sorted, and the results reflect the strong relationship between the groups of these genes and their functional categories.

Clustering is a typical way to group genes together according to their features. Certain distance measure, which reflects the similarity of genes' expression, is needed for clustering process. Most clustering methods that have been studied in the gene expression analysis literature use either Euclidean distance or Pearson correlation between expression profiles as a distance measure (D'haeseleer, 2000). Other measures include Euclidean distance between expression profiles and slopes (Wen et al., 1998), squared Pearson correlation (D'haeseleer et al., 1997), Euclidean distance between pairwise correlations to all other genes (Ewing et al., 1999), Spearman rank correlation (D'haeseleer et al., 1997), and mutual information represented by pairwise entropy (D'haeseleer et al., 1997; Michaels et al., 1998; and Butte and Kohane, 2000).

There are two kinds of clustering methods: hierarchical and non-hierarchical ones. A hierarchical clustering method starts from individual genes, merging them into bigger clusters until there is only one cluster left, in an agglomerative way. The method can also divisively start from all genes, splitting them until no two of them are together. The output of the method is a hierarchy of clusters, where the higher-level clusters are the sum of the lower-level ones. On the other hand, a non-hierarchical clustering method first divides genes into a certain number of clusters, and then iteratively refines them until certain optimization criterion is met.



Clustering methods that have been applied for gene expression analysis were reviewed in (D'haeseleer, 2000) and (Tibshirani et al., 1999). Because different algorithms may be applicable to different datasets, in (Yeung et al., 2001) a data-driven method to evaluate these algorithms was proposed.

## 2.2  Finding relationships between genes and function

Some methods can be used to find the relationship between gene expression and other information, i.e. properties of genes and samples. These properties can be type of hybridization sample, experimental condition, or the biological process that they are involved in. These are basically supervised methods for classification and regression. The methods try to construct learning models that could represent the relationship when given gene expression data as input and other information as output.

Machine learning classification methods have been applied to gene expression analysis in recent years. These methods usually employ class labels to represent different groups of expression data. An important application is cancer tissue classification, i.e. to construct learning model to predict whether a tissue is cancerous or to predict the type of cancer using gene expression. Cancer tissue classification is crucial for the diagnosis of patients. It used to be based on morphological appearances, which is often hard to measure and differentiate, and the classification result is very subjective. With the emergence of microarray technology, the classification is improved greatly by going to the molecular level. The various machine learning methods applied for classification are briefly described in the following paragraphs.



Neural networks are learning models that are based on the structure and behavior of neurons in the human brain and can be trained to recognize and categorize complex patterns (Bishop, 1995). Khan et al. (2001) used neural networks to classify cancer tissues using gene expression data as input. In their experiments, PCA was used to select a set of candidate genes. A number of neural networks were then trained on the training dataset. The prediction on test samples was achieved by summarizing all the outputs of the trained neural networks.

Support Vector Machine (SVM), which is rooted in statistical learning theory, is another method that can also be used to perform classification. It can achieve good generalization performance by minimizing both the training error and a generalization criterion that depends on *Vapnik-Chervonenkis (VC) dimension* (Vapnik, 1998). In (Brown et al., 2000), SVM was applied to classify yeast genes according to the biological process they involve in as represented by their expression data. In (Furey et al., 2000), it was also used to classify cancer tissues.

Decision tree generates a tree structure consisting of leaves and decision nodes. Each leaf indicates a class, and each decision node specifies some test to be carried out on a single attribute value, with one branch and subtree for each possible outcome of the test (Quinlan, 1993). C4.5 is a well-known decision tree induction algorithm, which uses information gain measure. Cai et al. (2000) used it to classify cancer tissue samples. A comparison was done with SVM, which was found to outperform C4.5.

Naïve Bayes Classification is a statistical discrimination method based on Bayes rule. In (Keller et al., 2000), this method was used to classify cancer tissues. The algorithm is simple



and can be easily extended from two-class to multi-class classification. Gaussian distribution of the data and class independence are assumed by the method.

A related technique is Bayesian Networks, which is also based on Bayesian rule. It is a probabilistic graphical model that represents the unique joint probability distribution of random variables efficiently. Nodes of a Bayesian network could correspond to genes and class labels, and represent the probability of the class label given some gene expression levels. Hwang et al. (2001) used Bayesian Networks to classify acute leukemia samples. A simple Bayesian Network with four gene nodes and one class label node was constructed from gene expression data. The high prediction performance indicated that the constructed network model can correctly represent the causal relationships of certain genes that are relevant to the classification.

Radial Basis Function (RBF) networks is a type of neural networks whose hidden neurons contain RBFs, a statistical transformation based on a Gaussian distribution, and whose output neuron computes a linear combination of its inputs. Hwang et al. (2001) also used an RBF network to classify acute leukemia samples. The network was larger than the constructed Bayesian network, but test results showed the prediction accuracy of RBF networks was higher.

Besides classification, feature selection i.e. the process of selecting genes that are most relevant to the class labels is also an important task for gene expression analysis. In (Slonim et al., 2000), a statistical method involving mean and variance was used to reflect the relevance between individual genes and class labels. In their work, the acute leukemia samples were divided into two groups according to their class labels. Those genes whose



expression values had small variance in both groups and big mean difference between the two groups were selected. In (Keller et al., 2000), a likelihood gene selection method was proposed based on likelihood. It outperformed the Baseline method in (Slonim et al., 2000) on the same cancer dataset by choosing less number of genes while achieving similar classification performance. In (Li, 2002), the linear relationship between the logarithm of measurement of classification ability of genes and the logarithm of rank of classification ability of genes was found to obey Zipf's law (Zipf, 1965). Plots of this relationship provided a useful tool in estimating the number of genes that is necessary for classification. Gouyon et al. (2002) proposed a Recursive Feature Elimination method based on Support Vector Machine. It made use of the magnitude of weights of trained SVMs as indicators of the discrimination ability of the genes. The algorithm keeps eliminating the genes that have relatively small contribution to the classification. In the test of the method on a leukemia dataset, small sets of genes with high discrimination contribution is obtained.

Neural trees represent multilayer feed-forward neural networks as tree structures. They have heterogeneous neuron types in a single network, and their connectivity is irregular and sparse (Zhang et al., 1997). Compared with the conventional neural networks, neural trees are more flexible. They can represent more complex relationship and permit structural learning and feature selection. Evolutionary algorithms can be used to construct neural trees. Hwang et al. (2001) constructed neural trees using gene expression, and selected relevant genes according to the connections in the trees. Neural trees were found to have better classification performance than the other two methods in the paper, i.e. Redial Basis networks and Bayesian networks. Genes with significant contribution to the classification could also be found from the constructed neural trees.



## 2.3  Dynamic modeling and time series analysis

It is also important to the study significant patterns and infer the dynamic model of gene expression from hybridization samples collected at different experimental time points. The dynamics can provide clues to the role of genes in the biological processes. Some of these methods have been successfully applied to discrete signal analysis.

Filkov et al. (2001) proposed a set of analysis methods that are suitable for short-term discrete time series data. These methods include: period detection, phase detection, correlation significance of short sequences of different length, and edge detection of group of regularity genes. The prediction analysis on yeast microarray data (Spellman et al., 1998) showed that the amount of data is not sufficient for large regularity pathway inference.

Singular Value Decomposition (SVD) constructs characteristic models of gene expression. These characteristic models can also be used to construct dynamic models of gene expression by deducing time translational matrices (Alter et al., 2000; Dewey and Bhan 2001; Holter et al., 2001). In (Dewey and Bhan, 2001), the change of expression level in a gene was modeled as first order Markov process. The time translational coefficient matrix was computed using least squares method based on a combination of SVD and linear response theory. The network model inferred from the matrix provided a way to cluster genes using their function. The clusters derived by applying the method to yeast time series expression data were in agreement with previously reported experimental work.

The dynamics of gene expression could also be modeled as differential equations. In (Chen et al., 1999) a linear transcription model is proposed, and two methods, Minimum Weight



Solutions for Linear Equations and Fourier Transform for Stable Systems, are proposed for constructing the model.

Shannon entropy can be used as a measure of the information content or complexity of a measurement series. It indicates the amount of information contained in expression of a gene pattern over time, or across anatomical regions, and therefore reveals the amount of information carried by the gene during a disease process or during normal phenotypic change. Shannon entropy is used by Fuhrman et al. (2000) to identify the most likely drug target candidate genes from temporal gene expression patterns.

## 2.4  Reverse engineering and gene network inference

Gene network inference attempts to construct and study coarse-scale network models of regulatory interactions between genes. It shows relationship between individual genes, and then provides a richer structure than clustering, which only reveals relationship between groups of genes. Gene network inference requires inference of the causal relationships among genes, i.e. reverse engineering the network architecture from its activity profiles. Reverse engineering is generally an unsupervised system identification process, which involves the following issues: choosing hybridization sample or expression data, choosing network model, and choosing method to construct the model, studying the structure and dynamics of the model. The study of network dynamics often involves time series analysis techniques mentioned in Section 2.3.

The simplest gene network model is Boolean network proposed by Kauffman (1969). In a Boolean network model, each node is in one of two possible states: express or not-express. The actual state depends on the states of other nodes that are linked to it.  A variety of



Boolean network construction algorithms have been developed. Somogyi et al. (1996) employed a phylogenetic tree construction algorithm (Fitch and Margoliash, 1967) to create and visualize the network. In (Liang et al., 1998), a more systematic and general algorithm was developed using mutual information to identify a minimal set of inputs that uniquely defines the output for each gene at next time step. Akutsu et al. (1999) improved Liang's algorithm to accept noisy expression data. Ideker et al. (2000) developed an alternative algorithm, which introduced perturbations in the expression data to iteratively and interactively refine the sensitivity and specificity of the constructed networks. At each of the iterations, a set of networks was inferred according to the expression data from different experimental perturbations. They were then discriminated using entropy based approach. The discriminations provide guide in further experimental perturbation design. Samsonova and Serov (1999) proposed an interactive Java applet tools for visualization and analysis of the Boolean network constructed. Maki et al. (2001) proposed a system that uses a top-down approach for the inference of Boolean network. The inferred networks on the simulated expression data matched the original ones well even when one of the genes was disrupted.

The advantage of Boolean network is its low construction cost. But it has the disadvantage being too coarse to represent the true regulation relationship between genes. Linear modeling tries to overcome this disadvantage by using weighted sum to represent the influence of other genes on one particular gene. In the model, the overall relationship is then represented as a matrix. Someren et al. (2000) used linear algebra methods to construct the model. Partial Least Squares method is a statistical method that is particularly useful for modeling large number of variables each with few observations (Stone and Brooks, 1990). Datta (2001) applied it to Sacccharomyces cerevisiae yeast microarray data to get linear regression model



and then predicted expression level of a gene according to that of other genes. The result appeared to be consistent with the known biological knowledge.

The dependency of the expression of one gene on the expression of other genes can also modeled using nonlinear functions. The nonlinear approach provides the model more ability to reveal the biological reality. However, it often introduces more difficulty in solving the model at the same time. Maki et al. (2001) also modeled the gene interaction as an S-system (Savageau, 1976). S-system is one of the best formalisms to estimate the complex gene interaction mechanisms. The disadvantage of the S-system network is the number of parameters to be estimated is vary large compared with that of Boolean network. To analysis large scale network, S-system approach was combined with their Boolean approach.

Bayesian networks can also be used for gene network inference. The inference process estimates the statistical confidence of dependencies between the activities of genes. Friedman et al. (2000) used it to analyze Sacccharomyces cerevisiae yeast microarray data from (Spellman et al., 1998). Their order relation and Markov relation analysis showed that the constructed Bayesian network had strong link to cell cycle regulated genes. Pe'er et al. (2001) extended this framework by the following steps: adding new kinds of factors such as mediator, activator, and inhibitor; enabling construction of subnets of strong confidence; enabling handling of mutation; and employing better discretization on the data for preprocessing. Their experiment on yeast microarray data showed that the constructed significant subnets could reveal biological pathways.

Several works have been conducted on the study of the dynamics of constructed network model. Huang (1999) used Boolean network to interpret gene activity profiles as entities



related to the dynamics of both the regularity network and functional cellular states. In this approach, the dynamics were mapped into state space and the system property of the network like stability, trajectories and attractors were studied.

This dynamics could be modeled more precisely as a set of differential equations. Neural network is one of the methods that could effectively solve these equations. Both Vohradský (2001) and D'haeseleer (2000) modeled gene network as recurrent neural networks. Vohradský (2001) used recurrent back-propagation (Pineda 1987), and simulated annealing to construct the network. However, D'haeseleer (2000) tried back-propagation through time (Werbos, 1990) in the training process, with techniques such as weight decay and weight elimination (Weigend et al., 1991) applied to simplify the model. Compared with simulated annealing, back-propagation is a more effective training method, but its scalability is worse because it attempts to unfold the temporal operation of the network into a layered feed-forward network. When doing their experiments, both Vohradský and D'haeseleer did not have microarray dataset that was large enough for the network construction. Instead, they used artificial data for experiment. The trained networks appeared to match original ones.

Szallasi, (1999) illustrated some basic natures of gene networks that could affect modeling. They include the stochastic nature, effective size, compartmentalization, and information content of expression matrix. In (Wessels et al., 2001; Someren et al., 2001), different network models were categorized and compared under criteria like inferential power, predictive power, robustness, consistency, stability, and computational cost.



## *2.5  Overview of the field*

Clustering based on gene expression reflects the correlation of genes. Classification links the expression of genes to functions. To study the change of expression of genes through time, dynamic modeling and time series analysis method have been used. In order to obtain the causal relationship or regulation of genes globally, the gene networks are needed to be inferred from the expression data. The inference work is a reverse engineering process (D'haeseleer, 2000). Reverse engineering is one of the major focuses of systems biology at present. The field of *Systems biology* studies biology at system level by examining the structure and dynamics of cellular and organismal function (Kitano, 2002). When the field of systems biology advances to the stage of trying to unify the biological knowledge across different levels of living organisms, we expect the understanding of the inherent complexity of living organisms will become a central issue (Michigan, 1999).



# 3   Machine learning methods used

Our work focuses on classification and feature selection methods for global gene expression analysis. In Section 3.1, the classification problems are described. Section 3.2 is about feature integration and univariate feature selection methods. Section 3.3 describes multivariate feature selection methods and classification methods.

## *3.1   Description of problem*

In this section, we present the gene expression data and class information in a mathematical form, and illustrate two types of classification problems that are commonly encountered in microarray data analysis.

### 3.1.1  Structure of microarray data

An expression matrix can be generated when quantified expression values of different hybridizations are available. Suppose there are $m$ genes and $n$ hybridizations. The expression matrix $\mathbf{A}$ is an $m \times n$ matrix

$$\mathbf{A} = \begin{bmatrix} a_{11} & a_{12} & \cdots & a_{1n} \\ a_{21} & a_{22} & \cdots & a_{2n} \\ \vdots & \vdots & \ddots & \vdots \\ a_{m1} & a_{m2} & \cdots & a_{mn} \end{bmatrix} = \begin{bmatrix} \mathbf{a}_1 \\ \mathbf{a}_2 \\ \vdots \\ \mathbf{a}_m \end{bmatrix}, \qquad \text{Eq. 3.1.1.1}$$

where $a_{ij}$ represents the expression value of the $i$ th gene in the $j$ th hybridization.

Certain property of gene or hybridization sample needs to be defined, i.e. labeling is required, in order to find the relationship between the genes and their expression matrix. The gene's property is represented by an $m \times 1$ vector



$$\mathbf{x} = \begin{bmatrix} x_1 \\ x_2 \\ \vdots \\ x_m \end{bmatrix},$$

Eq. 3.1.1.2

where each element represents one possible value of this property. For example, $x_i = +1$ means the $i$ th gene belongs to some biological process, while $x_i = -1$ means the $i$ th gene does not belong to this process. Similarly, the property of hybridization sample is defined as an $n \times 1$ vector

$$\mathbf{y} = \begin{bmatrix} y_1 \\ y_2 \\ \vdots \\ y_n \end{bmatrix},$$

Eq. 3.1.1.3

where each element represents one possible value of this property. For example, $y_i = +1$ means the $i$ th hybridization sample is cancerous, while $y_i = -1$ means the $i$ th hybridization sample is non-cancerous.

## 3.1.2  Two types of classification problems

From a micoarray experiment, we can only obtain a very limited number of hybridizations which involve a large number of genes. That is to say, $n$ is usually no more than a hundred, but $m$ can be a few thousands. So there are basically two types of classification problems:

- First type: large number of samples with low dimension. When the relationship between genes' expression and their property (function) is studied, the classification problem consists of $\mathbf{A}$ as input of learning model and $\mathbf{x}$ as output. There are $n$ features, each of them corresponds to one hybridization, and $m$ samples, each of them corresponds to one gene.



- Second type: small number of samples with large number of features and high dimension. When the relationship between expression of all genes under consideration and a certain property of hybridization sample is studied, the classification problem can be expressed as $\mathbf{B} = \mathbf{A}^{\mathrm{T}} = \begin{bmatrix} \mathbf{b}_1 \\ \mathbf{b}_2 \\ \vdots \\ \mathbf{b}_n \end{bmatrix}$, which is transpose of $\mathbf{A}$, as input of learning model and $\mathbf{y}$ as output. There are $m$ features, each corresponds to one gene, and $n$ samples, each corresponds to one hybridization.

In this thesis, our work is focused mainly on solving the classification problems of the second type, because this type of problems has distinct nature from the ordinary classification problems, and many cancer tissue classification problems based on gene expression is of this kind. We also have obtained a newly released Zebra fish developmental microarray dataset, which can be used to form classification problems of the first type. Because we are able to validate our prediction results using more precise biological experiments with the help of researchers who generated this dataset, we have applied Support Vector Machine classification method to this dataset as well.

## 3.2 Feature integration and univariate feature selection methods

Preprocessing is needed for the second type of classification problem (large number of features). The main goal of preprocessing is to reduce the number of inputs for a learning model without much loss, or even with some improvement of classification accuracy. Two kinds of methods can be used for the reduction: feature selection and feature integration (Liu and Motoda, 1998). *Feature selection* selects a subset of features as classifier input, while



*feature integration* generates a new feature set from original features as input. The feature integration method used in this thesis is Principle Component Analysis (PCA) (Muirhead, 1982); the two univariate feature selection methods used in our research are Information Gain (Quinlan, 1993) and Likelihood method (Keller et al., 2000). Due to space limitation, only Information gain and Likelihood method will be described in this section. Here *univariate* means the selection method only takes the contribution of individual features to the classification into consideration. The multivariate feature selection method will be described in Section 3.3, because most of them are based on classification methods. The term *multivariate* means the selection method accounts for the combinatorial effect of the features on the classification.

### 3.2.1 Information gain

Information gain method can be used to rank the individual discrimination ability of the features. It comes from information theory. In this method, information amount is measured by entropy. Let $S$ denote the set of all samples, $|S| = n$. Let $k$ denote the number of classes, and let $C_i, i = 1, \ldots, k$ denote the set of samples that belong to a class, $\bigcup_{i=1}^{k} C_i = S$, $\forall 1 \leq i < j \leq k : C_i \cap C_j = \Phi$. Suppose we select one sample from $S$ and label it as a member of class $C_i$. This message has probability $\dfrac{|C_i|}{|S|}$ of being correct. So the information it conveys is $-\log_2\left(\dfrac{|C_i \cap S|}{|S|}\right)$. The expected information of such a message is

$$\text{info}(S) = -\sum_{i=1}^{k} \frac{|C_i \cap S|}{|S|} \times \log_2\left(\frac{|C_i \cap S|}{|S|}\right). \qquad \text{Eq. 3.2.1.1}$$



Similarly, the expected information amount of any subset of $S$ can also be determined. If $S$ is partitioned into $l$ subsets $T_i, i = 1, \ldots, l$, $\bigcup_{i=1}^{l} T_i = S$, $\forall 1 \leq i < j \leq l : T_i \cap T_j = \Phi$. Then the expected information requirement is

$$\text{info}_X(S) = \sum_{i=1}^{l} \frac{|T_i|}{|S|} \times \text{info}(T_i).$$

Eq. 3.2.1.2

The difference

$$\text{gain} = \text{info}(S) - \text{info}_X(S)$$

Eq. 3.2.1.3

represents the amount of information gained from this partitioning.

Information gain tends to be greater when $l$ becomes larger, which may not truly reflect the quality of the partition. So it needs to be normalized by taking split information into consideration. Split information is defined as

$$\text{split info} = -\sum_{i=1}^{l} \frac{|T_i|}{|S|} \times \log_2\left(\frac{|T_i|}{|S|}\right).$$

Eq. 3.2.1.4

The normalized gain is thereby defined as

$$\text{gain ratio} = \frac{\text{gain}}{\text{split info}}.$$

Eq. 3.2.1.5

Given sample expression values and their class labels, each feature's information gain ratio could be calculated this way: sort samples according to expression values of this feature, partition them and calculate gain ratio of every possible split, and choose the maximum one as this feature's information gain ratio. This ratio provides a measure to evaluate the classification ability of one feature. Feature selection is done by choosing features that have the highest gain ratios.



## 3.2.2 Likelihood method (LIK)

Keller et al. (2000) proposed the Maximum Likelihood gene selection (LIK) method. Denote the event that a sample belongs to class $a$ or class $b$ by $M_a$ and $M_b$, respectively. The difference in the log likelihood is used to rank the usefulness of gene $g$ for distinguishing the samples of one class from the other. The LIK score is computed as follows:

$$LIK_{a\to b}^g = \log(P(M_a \mid x_{a,1}^g, \ldots, x_{a,n_a}^g)) - \log(P(M_b \mid x_{a,1}^g, \ldots, x_{a,n_a}^g)) \qquad \text{Eq. 3.2.2.1}$$

and

$$LIK_{b\to a}^g = \log(P(M_b \mid x_{b,1}^g, \ldots, x_{b,n_b}^g)) - \log(P(M_a \mid x_{b,1}^g, \ldots, x_{b,n_b}^g)) \qquad \text{Eq. 3.2.2.2}$$

where $P(M_i \mid x_{j,1}^g, \ldots, x_{j,n_j}^g)$ is *a posteriori* probability that $M_i$ is true given the expression values of the $g$ th gene of all the training samples that belong to class $j$, where $n_j$ is the number of training samples that belong to class $j$. Bayes rule

$$P(M \mid X)P(X) = P(X \mid M)P(M) \qquad \text{Eq. 3.2.2.3}$$

is used, with three assumptions required by the method. First is the assumption of equal prior probabilities of the classes

$$P(M_a) = P(M_b), \qquad \text{Eq. 3.2.2.4}$$

and second is the assumption that the conditional probability of $X$ falling within a small non-zero interval centered at $x$ given $M$ can be modeled by a normal distribution

$$P(x \mid M) = \frac{1}{d\sqrt{2\pi}} e^{\frac{-(x-m)^2}{2d^2}} \qquad \text{Eq. 3.2.2.5}$$

where $m$ and $d$ are the mean and standard deviation of $X$ respectively. The values $m$ and $d$ can be estimated from the training data. With the third assumption that the distributions of the expression values of the genes are independent, we obtain the LIK ranking of class $a$ over class $b$ for the $g$ th gene as follows:



$$LIK_{a \to b}^g = \log(P(x_{a,1}^g, \ldots, x_{a,n_a}^g \mid M_a)) - \log(P(x_{a,1}^g, \ldots, x_{a,n_a}^g \mid M_b))$$

$$= \log(\prod_{i=1}^{n_a} P(x_{a,i}^g \mid M_a)) - \log(\prod_{i=1}^{n_a} P(x_{a,i}^g \mid M_b)) \qquad \text{Eq. 3.2.2.6}$$

$$= \sum_{i=1}^{n_a} \left( -\log(\boldsymbol{d}_a^g) - \frac{(x_{a,i}^g - \boldsymbol{m}_a^g)^2}{2(\boldsymbol{d}_a^g)^2} + \log(\boldsymbol{d}_b^g) + \frac{(x_{a,i}^g - \boldsymbol{m}_b^g)^2}{2(\boldsymbol{d}_b^g)^2} \right),$$

and similarly, the LIK ranking of class $b$ over class $a$ for this gene is

$$LIK_{b \to a}^g = \log(P(x_{b,1}^g, \ldots, x_{b,n_b}^g \mid M_b)) - \log(P(x_{b,1}^g, \ldots, x_{b,n_b}^g \mid M_a))$$

$$= \sum_{i=1}^{n_b} \left( -\log(\boldsymbol{d}_b^g) - \frac{(x_{b,i}^g - \boldsymbol{m}_b^g)^2}{2(\boldsymbol{d}_b^g)^2} + \log(\boldsymbol{d}_a^g) + \frac{(x_{b,i}^g - \boldsymbol{m}_a^g)^2}{2(\boldsymbol{d}_a^g)^2} \right). \qquad \text{Eq. 3.2.2.7}$$

Genes that have higher likelihood scores are expected to have better ability to distinguish one class from the other.

### 3.3  Classification and multivariate feature selection methods

Classification is also called *pattern recognition*. It is a process to assign one of the prescribed number of classes (categories) given an input pattern (Haykin, 1999). Training of a classifier is usually needed to establish the learning model that could reflect the relationship between input patterns and class labels. This thesis employs four classification methods. They are decision tree (Quinlan, 1993), neural networks (Haykin, 1999), support vector machines (Vapnik, 1998) and Bayesian classification (Keller et al., 2000). Due to the space limitation, only Neural Network, Support Vector Machines and Bayesian classification will be described. Boosting technique, which combines output of multiple classifiers to form more accurate hypothesis, will also be illustrated in this section. Three multivariate feature selection methods: neural network feature selector, recursive feature elimination method and multivariate likelihood method are also described in this section.



### 3.3.1 Neural Networks

A neural network is a massively parallel distributed processor made up of simple processing units, which has a natural property for storing experimental knowledge and making it available for future use. Similar to human brain, it acquires knowledge from environment through a learning process, and its interneuron connection strengths store knowledge (Haykin, 1999; and Aleksander and Morton 1990). Neural networks as an analysis method has advantages such as nonlinearity, input-output mapping, adaptivity, evindential response, contextual information, and fault tolerance.

Three-layer feed-forward neural network is a typical neural network architecture. It has a simple hierarchical structure with high synaptic connections. A three-layer neural network model consists of an input layer, a hidden layer and an output layer. Input neurons in the input layer receive input signals, and send them to neurons in the hidden layer. Hidden neurons are computational units. They combine their inputs as their local fields, and send their output to neurons at the output layer. Output neurons then perform a similar combination and their output is the output of the whole model. Each of the synaptic links in the model has a weight associated with it, which is used to amplify/reduce the signal when the signal is passing by.

Computationally, this model can also be described as follows: Suppose there are $k_0$ input neurons, $k_1$ hidden neurons and $k_2$ output neurons. The synaptic links between the input and hidden layer can be represented as a $k_1 \times (k_0 + 1)$ matrix $\mathbf{w}^{(1)}$, including biases, and the links between hidden layer and output layer can similarly be represented as a $k_2 \times (k_1 + 1)$ matrix



$\mathbf{w}^{(2)}$, also including biases. Let a vector $\mathbf{x} = [+1, x_1 \cdots x_{k_0}]^T$ be the input of the network. The neurons in the hidden layer first sum up their input together with the bias associated with the first element of $\mathbf{x}$

$$\mathbf{v}^{(1)} = \mathbf{w}^{(1)}\mathbf{x},$$ 

Eq. 3.3.1.1

and perform a certain transformation using activation $\boldsymbol{j}^{(1)}$ function and get their output

$$\mathbf{y}^{(1)} = \begin{bmatrix} +1 \\ \boldsymbol{j}^{(1)}(\mathbf{v}^{(1)}) \end{bmatrix} = [1, y_1^{(1)} \cdots y_{k_1}^{(1)}]^T .$$

Eq. 3.3.1.2

Here we also add an additional element $+1$ as the first element of $\mathbf{y}$ in order to handle bias of the output neurons. Similarly, the neurons in the output layer perform the summation

$$\mathbf{v}^{(2)} = \mathbf{w}^{(2)}\mathbf{y}^{(1)}$$

Eq. 3.3.1.3

and transformation using activation function $\boldsymbol{j}^{(2)}$ to get the output

$$\mathbf{y}^{(2)} = \boldsymbol{j}^{(2)}(\mathbf{v}^{(2)}) .$$

Eq. 3.3.1.4

Here the activation functions $\boldsymbol{j}$ usually take following forms:

- *Threshold function*

  $$y = \boldsymbol{j}(v) = \begin{cases} 1 & \text{if } v \geq 0 \\ 0 & \text{if } v < 0 \end{cases}$$

  Eq. 3.3.1.5

- *Picewise-Linear function*

  $$y = \boldsymbol{j}(v) = \begin{cases} 1 & \text{if } v \geq a \\ v & \text{if } a \geq v > -a \\ 0 & \text{if } v < -a \end{cases} \quad \text{, with } a > 0$$

  Eq. 3.3.1.6

- *Sigmoid function (logistic function)*

  $$y = \boldsymbol{j}(v) = \frac{1}{1 + e^{-av}} \quad \text{with } a > 0$$

  Eq. 3.3.1.7

- *Hyperbolic tangent function*



$$y = \boldsymbol{j}(v) = a\tanh(bv) = a\frac{e^{bv} - e^{-bv}}{e^{bv} + e^{-bv}} \quad \text{with } a > 0, b > 0 \qquad \text{Eq. 3.3.1.8}$$

- *Hyperbolic tangent sigmoid function*

$$y = \boldsymbol{j}(v) = \frac{2}{1 + e^{-2v}} - 1 \qquad \text{Eq. 3.3.1.9}$$

Here $v$ and $y$ are scalars corresponding to the elements of the vectors.

A neural network provides a mapping from its input to its output, and this mapping is determined by the network's structure and synaptic weights. With a proper mapping established, given certain inputs, the network can produce the desired outputs. This mechanism could be used for classification. Training is needed to obtain a neural network that can give correct predictions. The training is done using an algorithm which usually consists of two steps: generate initial weights, and then iteratively refine these weights until certain stopping criterion is met. In essence, these algorithms are optimization algorithms, which attempt to meet certain criteria when refining the weights. These criteria are measured in terms of error functions because they reflect the difference between neural network outputs and the desired outputs.

Among many training algorithms, back-propagation is most popular (Hertz et al., 1991). There are two types of training in back-propagation: sequential mode and batch mode. In sequential mode, the algorithm calculates training error and updates weights each time it receives a training sample. On the other hand, in batch mode, the algorithm calculates overall error of all the training samples, and then updates the network's weights. The sequential mode is computationally slower than the batch model. But the order of training samples that are presented to the training algorithm can be randomly assigned, and the stochastic nature of



samples is able to be modeled. Below is the description of back-propagation algorithm in sequential mode. Its batch mode version could be easily adapted from the sequential mode.

Suppose the errors at output neurons are defined as

$$\mathbf{e} = \mathbf{d} - \mathbf{y}^{(2)} = [e_1 \cdots e_{k_2}]^{\mathrm{T}}. \qquad \text{Eq. 3.3.1.10}$$

Back-propagation uses

$$E = \mathbf{e}^{\mathrm{T}} \mathbf{e} \qquad \text{Eq. 3.3.1.11}$$

as the instantaneous error function of the network. The objective of back-propagation training is to minimize the average instantaneous error function to a certain extent, given all training samples.

Given a sample vector $\mathbf{x}$ with its class labels $\mathbf{d}$, $E$ can be computed using Eq. 3.3.1.1 to Eq. 3.3.1.4 and Eq. 3.3.1.10 to Eq. 3.3.1.11. To meet the objective, the correction of weights $\Delta\mathbf{w}^{(1)}$ and $\Delta\mathbf{w}^{(2)}$ should be proportional to the partial derivatives $\dfrac{\partial \mathbf{E}}{\partial \mathbf{w}^{(1)}}$ and $\dfrac{\partial \mathbf{E}}{\partial \mathbf{w}^{(2)}}$ respectively. According to the chain rule, we have:

$$\Delta\mathbf{w}^{(2)} = -\boldsymbol{h}\,\frac{\partial \mathbf{E}}{\partial \mathbf{w}^{(2)}}$$

$$= -\boldsymbol{h}\,\frac{\partial \mathbf{E}}{\partial \mathbf{e}}\,\frac{\partial \mathbf{e}}{\partial \mathbf{y}^{(2)}}\,\frac{\partial \mathbf{y}^{(2)}}{\partial \mathbf{v}^{(2)}}\,\frac{\partial \mathbf{v}^{(2)}}{\partial \mathbf{w}^{(2)}}$$

$$= -\boldsymbol{h}\times\mathbf{e}\times(-1)\otimes\boldsymbol{j}^{(2)'}(\mathbf{v}^{(2)})\times\mathbf{y}^{(1)\mathrm{T}} \qquad \text{Eq. 3.3.1.12}$$

$$= \boldsymbol{h}\times(\mathbf{e}\otimes\boldsymbol{j}^{(2)'}(\mathbf{v}^{(2)}))\times\mathbf{y}^{(1)\mathrm{T}}$$

$$= \boldsymbol{h}\times\boldsymbol{d}^{(2)}\times\mathbf{y}^{(1)\mathrm{T}},$$

here we let $\boldsymbol{d}^{(2)} = \mathbf{e}\otimes\boldsymbol{j}^{(2)'}(\mathbf{v}^{(2)})$; and similarly,



$$\Delta \mathbf{w}^{(1)} = -h \frac{\partial \mathbf{E}}{\partial \mathbf{w}^{(1)}}$$

$$= -h \frac{\partial \mathbf{E}}{\partial \mathbf{y}^{(1)}} \frac{\partial \mathbf{y}^{(1)}}{\partial \mathbf{w}^{(1)}}$$

$$= h \times ((\mathbf{w}^{(2)\mathrm{T}} \times (\mathbf{e} \otimes \boldsymbol{j}^{(2)'}(\mathbf{v}^{(2)}))) \otimes \boldsymbol{j}^{(1)'}(\mathbf{v}^{(1)})) \times \mathbf{x}^{\mathrm{T}} \qquad \text{Eq. 3.3.1.13}$$

$$= h \times ((\mathbf{w}^{(2)\mathrm{T}} \times \boldsymbol{d}^{(2)}) \otimes \boldsymbol{j}^{(1)'}(\mathbf{v}^{(1)})) \times \mathbf{x}^{\mathrm{T}}$$

$$= h \times \boldsymbol{d}^{(1)} \times \mathbf{x}^{\mathrm{T}} ,$$

here we let $\boldsymbol{d}^{(1)} = (\mathbf{w}^{(2)\mathrm{T}} \times \boldsymbol{d}^{(2)}) \otimes \boldsymbol{j}^{(1)'}(\mathbf{v}^{(1)})$ .

In Eq. 3.3.1.12 and Eq. 3.3.1.13, $h$ is a positive learning rate parameter which controls the amount of weight adjustment. The operator $\otimes$ is an element-by-element multiplication. Activation function $\boldsymbol{j}^{(1)}$ and $\boldsymbol{j}^{(2)}$ must be differentiable for back-propagation training. The bias elements need to be included or excluded in certain steps of matrix multiplication in order to maintain consistency.

The back-propagation algorithm could be easily extended to training multi-layer feed-forward neural networks. Adjustment of weights involves only the neuron signals of the successive layers they connected, so the algorithm is a *local* method. This also makes it computationally efficient.

Feed-forward neural networks can be used for classification. For two class problems, a neural network with a single output neuron is needed. Samples are labeled $+1$ or $-1$. The label value will be the desired values at the output side of the network when training the network. For multi-class problems, a neural network that has the number of output neurons identical to



the number of classes is usually required. The desired output value in this case is a vector. The elements of the vector are $-1$, except the one that corresponds to the class of the sample, which is $+1$. If the training samples are biased, for example, the number of positive samples is much less than that of the negative ones, label values other than $\pm 1$ can be used to adjust the feed-back signal to obtain better performance. Many gene expression classification problems are two class problems.

### 3.3.2  Ensemble of neural networks (boosting)

One of the main disadvantages of neural network for classification is that the training result also depends on initial weights, which are generated randomly. Boosting can be used to enhance the robustness of the neural network. The term *Boosting* refers to a machine learning framework that combines a set of simple decision rules, which is generated by a set of learners with different learning abilities, into a complex one that has higher accuracy and lower variance. It is especially useful in handling real world problems that have the following properties (Freund and Schapire, 1996): the samples have various degrees of hardness to learn and the learner is sensitive to the change of training samples. The complexity of learning hardness often occurs when applying machine learning method to tackle biological problems. There are three boosting approaches (Haykin, 1999):

- Boosting by filtering: If the number of training samples is large, the samples are either discarded or kept during training.
- Boosting by subsampling: With a fixed training sample set size, the probability to include samples into training sample set for learning algorithms is adjusted.



- Boosting by reweighting: This approach assumes that the training samples could be weighted by the learning algorithms. The training errors are calculated by making use of these weights.

AdaBoost (Freund and Schapire, 1996) is a simple and effective boosting algorithm through subsampling. During learning process, the algorithm tries to make the learners focus on different portions of the training samples by refining the sampling distribution. Figure 3.1 shows one of two versions of AdaBoost training algorithm.

The input of the algorithm are $n$ training samples, $\{(\mathbf{b}_1, y_1), \ldots, (\mathbf{b}_n, y_n)\}$ , where $\mathbf{b}_i$ $(i = 1 \ldots n)$ are the sample vectors, and $y_i \in Y$ $(i = 1 \ldots n)$ are their associated class labels, which are also the desired outputs of the learning models. $Y$ is the set of class labels. The algorithm starts by setting the initial sampling distribution $d_{1,1 \ldots n}$ to a uniform one. It then enters an iterative process. At the $t$ th iteration, the algorithm calls the function train_learner() with training samples $\mathbf{b}_{1..n}$ and sampling distribution $d_{t,1 \ldots n}$, to get the trained learning model $M_t$. By calling function get_hypothesis() with $M_t$ and $\mathbf{b}_{1..n}$, the algorithm generates the hypothesized classes of training samples. The algorithm then calculates total error $\boldsymbol{e}_t$ by adding up all the wrongly predicted samples' distribution. If the error is bigger than $\frac{1}{2}$, the algorithm stops. Otherwise, it proceeds to calculate a factor $\boldsymbol{b}_t$. This factor is used to reduce the portion of the correctly predicted training samples, in the sampling distribution $d_{t+1,1 \ldots n}$ of the next iteration. When the algorithm terminates, $T$ learning models $M_{1..T}$ are trained with factors $\boldsymbol{b}_{1..T}$ indicating the contribution of the learning models to the combined hypothesis. The combined hypothesis for a test sample $\mathbf{b}$ can then be calculated as



$$h(\mathbf{b}) = \arg\max_{y \in Y} \left( \sum_{t:\text{get\_hypothesis}(M_t, \mathbf{b}) = y} \log\left(\frac{1}{\boldsymbol{b}_t}\right) \right).$$

Eq. 3.3.2.1

When number of training samples is small, $\boldsymbol{b}_t$ could be zero. We set a lower threshold to $\boldsymbol{b}_t$ when implementing the algorithm so that $h(\mathbf{b})$ can be calculated.

---

function AdaBoost($\{(\mathbf{b}_1, y_1), \ldots, (\mathbf{b}_n, y_n)\}$)     // Input: training samples

$d_{1,i} = \frac{1}{n}$   $(i = 1 \ldots n)$                // Initial sampling distribution

for $t = 1$ to $T$ do

$M_t = \text{train\_learner}(\mathbf{b}_{1..n}, d_{t,1..n})$

$h_{t,1..n} = \text{get\_hypothesis}(M_t, \mathbf{b}_{1..n})$

$\boldsymbol{e}_t = \sum_{i:h_{t,i} \neq y_i} d_{t,i}$

if $\boldsymbol{e}_t > \frac{1}{2}$ then

$T = t - 1$

terminate loop          // Terminate the algorithm

end

// Update sampling distribution to focus on the wrongly classified samples

$\boldsymbol{b}_t = \frac{\boldsymbol{e}_t}{1 - \boldsymbol{e}_t}$

$d'_{t,i} = \begin{cases} \boldsymbol{b}_t d_{t,i} & \text{if } h_{t,i} = y_i \\ d_{t,i} & \text{if } h_{t,i} \neq y_i \end{cases}$   $(i = 1 \ldots n)$

$d_{t+1,i} = \frac{d'_{t,i}}{\sum_{i=1}^{n} d'_{t,i}}$   $(i = 1 \ldots n)$

end

return $(M_{1..T}, \boldsymbol{b}_{1..T})$

end

---

Figure 3.1: AdaBoost algorithm



Theoretical study shows that if the hypothesis obtained by individual classifiers constantly has error that is slightly better than random guess, the number of prediction errors of the final hypothesis $h$ drops to zero exponentially fast when $T$ increases (Freund and Schapire, 1996).

Three layer neural networks can be the learning model integrated with AdaBoost. In this case, train_learner() consists of initializing, training and optimizing neural network weights, and get_hypothesis() corresponds to neural network decision making. The running of the algorithm will construct $T$ neural networks in total for making combined decision.

### 3.3.3  Neural network feature selector

This section first analyses the limitation of applying information gain measure for ranking of features having continuous value, then describes the neural network feature selector method.

### 3.3.3.1 Disadvantage of information gain measure

Information gain can be applied in measuring both discrete and continuous feature values. But in continuous case, it only takes the order of the values into consideration, which may not be sufficient. For example, suppose the expression matrix is



$$\mathbf{B} = \begin{bmatrix} 6 & 1 & 7 \\ 7 & 2 & 7.5 \\ 8 & 3 & 8 \\ 9 & 4 & 8.5 \\ 10 & 5 & 9 \\ 11 & 16 & 12 \\ 12 & 17 & 12.5 \\ 13 & 18 & 13 \\ 14 & 19 & 13.5 \\ 15 & 20 & 14 \end{bmatrix}$$

and the corresponding class labels are $\mathbf{y} = \begin{bmatrix} -1,-1,-1,-1,-1,+1,+1,+1,+1,+1 \end{bmatrix}^{\mathrm{T}}$. The three features' values are plotted in Figure 3.2 at line $y=1$, $y=2$ and $y=3$. Intuitively we can see that the discrimination abilities of the feature at line $y=2$ and $y=3$ are better than the one at line $y=1$. Compared with feature values at line $y=1$, the distance between feature values of different classes at line $y=2$ are longer, and the density of feature values within different classes at line $y=3$ are higher. But the information gain ratios of these three features are the same because it only takes the order of the feature values into account, hence achieving the same maximum gain ratio for all three features when split in the middle.



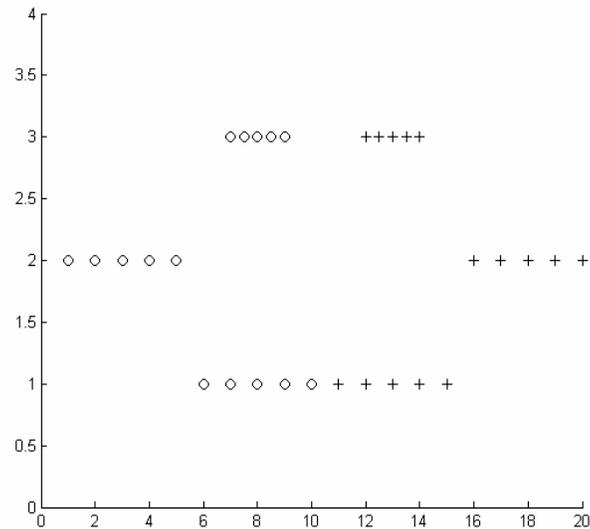

Figure 3.2: Plot of feature values with class labels, 'o' for '-1' and '+' for '+1'

### 3.3.3.2 Neural network feature selector

Since feature selection is a preprocessing process for classification, classification accuracy can also be a selection criterion. The advantage of integrating a classifier into the feature selection process is that the feature set is optimized by the classification accuracy. Moreover, the training of the classifier and the selection of the features use the same bias. The consistency improves the classification performance. However, the computational cost of the integration may be high. In (Liu and Motoda, 1998) this approach is called *wrapper model*. This framework is shown in Figure 3.3.

Procedure *wrapper* accepts full feature set (F), training samples (D) and testing samples (T) as input. It first generates a subset of features (S), and then performs cross validation using S and D to get classification accuracy (A). These two steps are repeated until A is sufficiently



high under certain criterion. A classification model (M) is then obtained from D with selected S. Finally M and S are used to perform the test to measure the performance. In the algorithm, cross validation can be done by dividing D into a training set and a validating set; or by the leave-one-out method (see more details below).

```
procedure wrapper (F, D, T)
        do
                S=feature_set_gen(F)
                A=cross_validation(S,D)
        until sufficient(A)
        M=train(S,D)
        test(M,S,T)
end
```

Figure 3.3: Wrapper model framework

Setiono and Liu, (1997) proposed a neural network feature selection method based on the wrapper approach. Here in this thesis, we applied it to gene expression analysis, with some modifications. The following is a detailed description of this method.

The algorithm starts with all features, and removes features that have minor contribution to classification one by one. The algorithm first trains a neural network with a given feature set, then disables each feature and estimates the classification performance of this neural network with the remaining features. If the decrease of the estimated performance is within an acceptable level, the algorithm constructs a neural network with the remaining features, and calculates the actual classification performance. If the actual performance is also acceptable, the feature is removed, and the algorithm continues searching for more features to be removed. Otherwise, it keeps this feature and continues to test other features according to



their estimated performances. It is a greedy method, and usually achieves sub-optimal solutions.

There are two ways to train and validate the neural network. Suppose there are $n$ samples. The first one is to separate these samples into two sets: training set and validating set. If $n$ is too small to produce sufficiently large training and validating sets, the leave-one-out technique can be used to perform training and validation $n$ times, and obtain the average training and validation accuracy.

Neural network feature selector method requires the neural network training to force the weights associated with an irrelevant input neuron to have small magnitude, in order to reduce the effect of the corresponding feature's removal on the classification performance. This is implemented using weight decay on the weights between the input and the hidden layer when applying the back-propagation training. After every element of the weights is updated with $\Delta \mathbf{w}^{(1)}$, according to Eq. 3.3.1.13,

$$w_{i,j}^{(1)\,'} = w_{i,j}^{(1)} + \Delta w_{i,j}^{(1)}, \qquad\qquad\qquad \text{Eq. 3.3.3.1}$$

the elements of the new weights are computed as follows

$$w_{i,j}^{(1)\,''} = \left( 1 - \frac{\boldsymbol{e}_j \boldsymbol{h}}{\left(1 + \sum_i \left(w_{i,j}^{(1)}\right)^2\right)^2} \right) w_{i,j}^{(1)\,'}, \qquad\qquad \text{Eq. 3.3.3.2}$$

where $\boldsymbol{h}$ is the learning rate parameter, and $\boldsymbol{e}_j$ is a penalty term associated with the $j$ th input neuron. Eq. 3.3.3.2 is similar to the weight decay method in (Hertz et al., 1991), but the focus is different: the later can be used to eliminate hidden neurons.



In order to improve the convergence, we also tried the cross entropy error function instead of Eq. 3.3.1.11,

$$F(\mathbf{w}^{(1)}, \mathbf{w}^{(2)}) = -\left\{ \sum_{p=1}^{n} \sum_{k}^{k_2} \left[ d_k^p \log(y_k^{(2),p}) + (1 - d_k^p) \log(1 - y_k^{(2),p}) \right] \right\}, \qquad \text{Eq. 3.3.3.3}$$

where the desired value $d_k^p$ of $p$ th sample of $k$ th class is either $1$ or $0$, which is different from the one used in Eq. 3.3.1.11. The derivation of the back propagation with error function in Eq. 3.3.3.3 is as follows:

$$\frac{\partial F}{\partial w} = \left( \sum_{p=1}^{n} \sum_{k}^{k_2} \frac{\partial F}{\partial y_k^{(2),p}} \frac{\partial y_k^{(2),p}}{\partial w} \right), \qquad \text{Eq. 3.3.3.4}$$

where

$$\frac{\partial F}{\partial y_k^{(2),p}} = -\left( \frac{d_k^p}{y_k^{(2),p}} - \frac{1 - d_k^p}{1 - y_k^{(2),p}} \right). \qquad \text{Eq. 3.3.3.5}$$

Because the three layer neural network model is

$$y_k^{(2),p} = \boldsymbol{j}^{(2)} \left[ \sum_{j=1}^{k_1} \boldsymbol{j}^{(1)} \left( \sum_{i=1}^{k_0} x_i^p w_{ji}^{(1)} \right) w_{kj}^{(2)} \right], \qquad \text{Eq. 3.3.3.6}$$

we have

$$\frac{\partial y_k^{(2),p}}{\partial w_{kj}^{(2)}} = \boldsymbol{j}^{(2)'} \left( v_k^{(2),p} \right) \cdot \boldsymbol{j}^{(1)} \left( v_j^{(1),p} \right) = \boldsymbol{j}^{(2)'} \left( v_k^{(2),p} \right) \cdot y_j^{(1),p}, \qquad \text{Eq. 3.3.3.7}$$

and

$$\frac{\partial y_k^{(2),p}}{\partial w_{ji}^{(1)}} = \boldsymbol{j}^{(2)'} \left( v_k^{(2),p} \right) \cdot \boldsymbol{j}^{(1)'} \left( v_j^{(1),p} \right) \cdot x_i^p. \qquad \text{Eq. 3.3.3.8}$$

Again according to the delta rule (Haykin, 1999) and the gradient descent rule (Hertz et al., 1991), we obtain

$$\Delta w_{kj}^{(2)} = -\boldsymbol{h} \frac{\partial F}{\partial w_{kj}^{(2)}} = \boldsymbol{h} \left\{ \sum_{p=1}^{n} \left( \frac{d_k^p}{y_k^{(2),p}} - \frac{1 - d_k^p}{1 - y_k^{(2),p}} \right) \boldsymbol{j}^{(2)'} \left( v_k^{(2),p} \right) \cdot y_j^{(1),p} \right\} \qquad \text{Eq. 3.3.3.9}$$

and



$$\Delta w_{ji}^{(1)} = -\boldsymbol{h}\frac{\partial F}{\partial w_{ji}^{(1)}} = \boldsymbol{h}\left\{\sum_{p=1}^{n}\sum_{k}^{k_2}\left(\frac{d_k^p}{y_k^{(2),p}} - \frac{1-d_k^p}{1-y_k^{(2),p}}\right)\boldsymbol{j}^{(2)'}\left(v_k^{(2),p}\right)\cdot\boldsymbol{j}^{(1)'}\left(v_j^{(1),p}\right)\cdot x_i^p\right\}. \qquad \text{Eq. 3.3.3.10}$$

Because the implementation of the algorithm is on MATLAB, we transform the computation of weight into matrix form: Let $\mathbf{F}'$ be a $n \times k_2$ matrix, the elements $f_{ij} = \left(\dfrac{d_j^i}{y_j^{(2),i}} - \dfrac{1-d_j^i}{1-y_j^{(2),i}}\right)$ which has the value opposite to that of $\dfrac{\partial F}{\partial y_k^{(2),p}}$. Let $\mathbf{X}$ be $n \times k_0$ matrix; $\mathbf{V}^{(1)}$ and $\mathbf{Y}^{(1)}$ be $n \times k_1$ matrix; $\mathbf{V}^{(2)}$ and $\mathbf{Y}^{(2)}$ be $n \times k_2$ matrix. Then we have

$$\Delta \mathbf{W}^{(2)} = \boldsymbol{h}(\mathbf{F}' \otimes \boldsymbol{j}^{(2)'}(\mathbf{V}^{(2)}))^{\mathbf{T}} \mathbf{Y}^{(1)} \qquad \text{Eq. 3.3.3.11}$$

and

$$\Delta \mathbf{W}^{(1)} = \boldsymbol{h}\left(\left((\mathbf{F}' \cdot \boldsymbol{j}^{(2)'\mathbf{T}}(\mathbf{V}^{(2)})) \otimes \mathbf{I}\right) \cdot \boldsymbol{j}^{(1)'}(\mathbf{V}^{(1)})\right)^{\mathbf{T}} \mathbf{X} \qquad \text{Eq. 3.3.3.12}$$

where $\mathbf{I}$ is a $n \times n$ identity matrix used to extract the diagonal elements out.

Figure 3.4 shows the algorithm: At line 1, the function neural_network_feature_selector() accepts six parameters: a set of $m$ feature vectors $\{\mathbf{f}_1,\ldots,\mathbf{f}_m\}$, a set of penalty parameters $\{\boldsymbol{e}_1,\ldots,\boldsymbol{e}_m\}$, a penalty parameter amplification factor $f$, penalty parameter thresholds $(\boldsymbol{e}_{\min},\boldsymbol{e}_{\max})$, a class label vector $\mathbf{y}$, an allowable maximum decrease in validation accuracy $\Delta r''$, training and validation thresholds $(r'_{\min},r''_{\min})$, and a linear combination factor $r_{\text{factor}}$ for testing and training performances.

At lines 2 to 5, the function copies feature number, feature set, penalty parameter set into three variables $m'$, $F$ and $E$, and initializes the maximum validating accuracy $r''_{\max}$ to a



small positive value $e$. It then enters an iterative process, at lines 6 to 40, to remove features one by one until no more feature in $F$ can be removed with sufficiently high performance. In the removal process, at line 7, a neural network $N$ is initialized according to $m'$, by calling function initialize(). Then at line 8, $N$ is trained and validated with features in $F$ and penalty parameters in $E$ by calling train_validate(), which returns the trained network $N$, training accuracy $r'$ and validating accuracy $r''$. The algorithm then updates $r''_{max}$. At line 10, $m'$ neural networks $N_{1,...,m'}$ are initialized with same weights as $N$.

At lines 11 to 15, $m'$ feature sets $F_{1,...,m'}$ and penalty parameter sets $E_{1,...,m'}$ are constructed, each of which has one feature omitted; and they are tested and validated with corresponding neural network by calling simulate_validate() at line 14; the corresponding estimated training accuracy $r'_{1,...,m'}$ and validating accuracy $r''_{1,...,m'}$ are returned. At line 16, $r'_{1,...,m'}$ and $r''_{1,...,m'}$ are sorted according to their linear combination in descending order. The higher the linear combination $r_{factor}r'_i + r''_i$ is, the more likely the corresponding $i$th feature to be eliminated. The factor $r_{factor}$ is usually set to be bigger than one. When the difference among $r'_{1,...,m'}$ are high, $r'_{1,...,m'}$ will be the main contributor to the ranking. However, when the difference among $r'_{1,...,m'}$ are not so high, which often occurs for small of training set, ranking is mainly affected by $r''_{1,...,m'}$.

At lines 17 to 25, according to the sorted index $s_{1,...,m'}$, the corresponding neural networks whose estimated training and validation accuracy rates, $r'_{s_i}$ and $r''_{s_i}$ are bigger than the threshold $(r'_{min}, r''_{min})$ are retrained to get actual training accuracy $r'_{s_i}$ and validating accuracy $r''_{s_i}$. If the validating accuracy is sufficiently high compared with maximum validating



accuracy, as indicated in the conditions at lines 23 and 26, then the penalty parameters is updated at lines 27 to 33. At lines 35 to 38, the selected feature $f_{s_i}$ and penalty parameter $e_{s_i}$ are removed from $F$ and $E$; feature number $m'$ is decreased; and $r''_{\max}$ is updated. At lines 29 to 33, the way to update penalty parameter is as follows: for any $r'_j$, if it is bigger than average value $\bar{r}'$, which means that the corresponding feature is likely to be removed, the corresponding $e_j$ is enlarged by the factor $f$; otherwise it is reduced by $f$. A penalty parameter's lower and upper thresholds $[e_{\min}, e_{\max}]$ are set to prevent it to be too large or too small. After the feature removal process at from lines 6 to 40, the selected feature set in $F$ is returned at line 41.

As mentioned earlier, neural network feature selector method is based on the wrapper model. But its feature generation and cross validation part are not so distinct. In general, lines 7 to 16 and lines 26 to 39 relate to feature generation; lines 17 to 25 validate the feature set; and line 40 contains the stopping criterion.

1     function
    neural_network_feature_selector($\{\mathbf{f}_1,\ldots,\mathbf{f}_m\},\{\mathbf{e}_1,\ldots,\mathbf{e}_m\},f,(\mathbf{e}_{min},\mathbf{e}_{max}),\mathbf{y},\Delta r'',(r'_{min},r''_{min}),r_{factor})$

2         $m' = m$

3         $F = \{\mathbf{f}_1,,\mathbf{f}_m\}$

4         $E = \{\mathbf{e}_1,\ldots,\mathbf{e}_m\}$

5         $r''_{max} = \mathbf{e}$

6         do

7             $N = \text{initialize}(m')$

8             $(N,r',r'') = \text{train\_validate}(N,F,E,\mathbf{y})$

9             $r''_{max} = \max(r''_{max},r'')$

10            $(N_1,\ldots,N_{m'}) = (N,\ldots,N)$

11            for $i = 1$ to $m'$

12                 $F_i = F - \{N_i\}$

13                 $E_i = E - \{\mathbf{e}_i\}$

14                 $(r'_i,r''_i) = \text{simulate\_validate}(N_i,F_i,E_i,\mathbf{y})$

15            end

16            $((r'_{s_1},\ldots,r'_{s_{m'}}),(r''_{s_1},\ldots,r''_{s_{m'}})) = \text{sort\_descending}(r_{factor} \times (r'_1,\ldots,r'_{m'}) + (r''_1,\ldots,r''_{m'}))$

17            $i = 0$

18            do

19                 $i = i + 1$

20                 if $r'_{s_i} > r'_{min}$ and $r''_{s_i} > r''_{min}$

21                     $N_{s_i} = \text{initialize}(m' - 1)$

22                     $(N_{s_i},r'_{s_i},r''_{s_i}) = \text{train\_validate}(N_{s_i},F_{s_i},E_{s_i},\mathbf{y})$

23                     $\mathbf{d} = \dfrac{r''_{max} - r''_{s_i}}{r''_{max}}$

24                 end

25            until $i \geq m'$ or $\mathbf{d} \leq \Delta r''$

26            if $\mathbf{d} \leq \Delta r''$ then

27                 $\bar{r}' = \dfrac{1}{m'}\sum_{i=1}^{m'} r'_i$

28                 for $j = 1$ to $m'$

29                     if $r'_j \geq \bar{r}'$ and $\mathbf{e}_j \in [\mathbf{e}_{min},\mathbf{e}_{max}]$ then

30                         $\mathbf{e}_j = f\mathbf{e}_j$

31                     else

32                         $\mathbf{e}_j = \dfrac{\mathbf{e}_j}{f}$

33                     end

34                 end

35                 $F = F - \{f_{s_i}\}$

36                 $E = E - \{\mathbf{e}_{s_i}\}$

37                 $m' = m' - 1$

38                 $r''_{max} = \max(r''_{max},r''_{s_i})$

39            end

40         until $i \geq m'$

41         return $F$

42     end

Figure 3.4: Neural network feature selector



The main modifications of the neural network feature selector in this thesis compared with the one proposed by Setiono and Liu (1997) are summarized below.

1) Use of back propagation training method: Setiono and Liu employed BFGS (Broyden-Fletcher-Shanno-Goldfarb) method for training of neural networks, which is a variant of quasi-Newton method that has been shown to be very effective. However, in the training process, BFGS computes a Hessian matrix with dimension equal to the square of the number of features. The size of the matrix is huge when the algorithm is applied to microarray dataset consisting of a large number of features. So we replace the BFGS algorithm with back propagation algorithm in our implementation.

2) Use of leave-one-out validation as an estimator: The neural network feature selector algorithm proposed by Setiono and Liu uses a fixed partition of training and cross-validation sample set as initial input. In a typical microarray dataset, when there are only a very limited number of samples, a fixed training and cross-validation sample set may introduce large bias in estimating the generalization performance of the trained neural networks. Instead of simply splitting the sample set into two partitions, the neural network feature selector proposed in this thesis employs leave-one-out method to reduce the bias in estimating the generalization performance.

3) Use of different of penalty function: Penalty functions are employed in both versions of neural network feature selectors to force small weights between the input layer and the hidden layer to zero, in order to reduce the effect of irrelevant features to the classification. The penalty function used in (Setiono and Liu, 1997) is



$$P(w) = e_1 (\sum_i \sum_j \frac{b\left(w_{i,j}^{(1)}\right)^2}{1 + b\left(w_{i,j}^{(1)}\right)^2}) + e_2 (\sum_i \left(w_{i,j}^{(1)}\right)^2), \qquad \text{Eq. 3.3.3.13}$$

where $e_1$, $e_2$ and $b$ are parameters for deciding the detailed penalty effect. From Eq. 3.3.1.13 it can be easily seen that the derivative of the penalty function against a weight involves only that weight itself:

$$\frac{\partial P(w)}{\partial w_{i,j}^{(1)}} = e_1 \left( \frac{2 b w_{i,j}^{(1)}}{\left(1 + b\left(w_{i,j}^{(1)}\right)^2\right)^2} \right) + 2 e_2 w_{i,j}^{(1)}. \qquad \text{Eq. 3.3.3.14}$$

As a result, the update of the weights tries to force the individual small weights to zero. By contrast, the penalty function used in this thesis forces all weights associated with an input neuron to zero, if the summed square of these weights are small. The new penalty function is implicitly implemented in Eq. 3.3.3.2. It helps reducing the effect of an input neuron on all hidden neurons when this input neuron is removed.

### 3.3.4  Support vector machines

Support vector machine is a linear learning model that can also perform classification. It was invented by Boser et al. (1992). The theoretical aspect of Support Vector Machines is based on Statistical Learning Theory (Vapnik, 1998). Section 3.3.4.1 describes the basic SVM learning model and its training; Section 3.3.4.2 illustrates the linearly non-separable case; Section 3.3.4.3 describes SVM with nonlinear mapping; Section 3.3.4.4 introduces a multivariate feature selection method based on SVM.



### 3.3.4.1 Linearly separable learning model and its training

From decision-making point of view, a linear classifier tries to obtain decision surfaces that can discriminate samples of different classes. These decision surfaces are hyperplanes. Suppose there are $n$ training samples that belong to two classes, $\{(\mathbf{x}_1, y_1), \ldots, (\mathbf{x}_n, y_n)\}$, where $\mathbf{x}_i$ $(i = 1 \ldots n)$ are sample vectors, and $y_i = \pm 1$ $(i = 1 \ldots n)$ are associated class labels. If there exists a hyperplane $\mathbf{w}^{\mathrm{T}}\mathbf{x} + b = 0$ so that for all $n$ samples $\mathbf{x}_i$,

$$\begin{cases} \mathbf{w}^{\mathrm{T}}\mathbf{x}_i + b \geq 0 & y_i = +1 \\ \mathbf{w}^{\mathrm{T}}\mathbf{x}_i + b < 0 & y_i = -1 \end{cases}$$

Eq. 3.3.4.1

hold, then these samples are *linearly separable*. The distance between the separating hyperplane and its closest sample vector is called *margin of separation*, denoted as $\mathbf{r}$. A Support Vector Machine's task is to find an optimal separating hyperplane $\mathbf{w}^{*\mathrm{T}}\mathbf{x} + b^* = 0$ that has the biggest margin of separation among all separating hyperplanes. Obviously, the distances between this hyperplane and its nearest sample vectors on both sides are equal. With $\mathbf{w}^*$ and $b^*$ properly scaled, we have

$$\begin{cases} \mathbf{w}^{*\mathrm{T}}\mathbf{x}_i + b^* \geq +1 & y_i = +1 \\ \mathbf{w}^{*\mathrm{T}}\mathbf{x}_i + b^* \leq -1 & y_i = -1 \end{cases}$$

Eq. 3.3.4.2

for all samples. The sample vectors that satisfy

$$y_i(\mathbf{w}^{*\mathrm{T}}\mathbf{x}_i + b^*) = 1$$

Eq. 3.3.4.3

are called *support vectors*. Let us denote a pair of support vectors on both sides of the separating hyperplane as $\mathbf{x}^+$ and $\mathbf{x}^-$,

$$\begin{cases} \mathbf{w}^{*\mathrm{T}}\mathbf{x}^+ + b^* = +1 \\ \mathbf{w}^{*\mathrm{T}}\mathbf{x}^- + b^* = -1 \end{cases}.$$

Eq. 3.3.4.4

We then have



$$r = \frac{1}{2}\left(\frac{\mathbf{w}^{*T}\mathbf{x}^+}{\|\mathbf{w}^*\|} - \frac{\mathbf{w}^{*T}\mathbf{x}^-}{\|\mathbf{w}^*\|}\right)$$

$$= \frac{1}{2\|\mathbf{w}^*\|}\left(\mathbf{w}^{*T}\mathbf{x}^+ - \mathbf{w}^{*T}\mathbf{x}^-\right) \qquad \text{Eq. 3.3.4.5}$$

$$= \frac{1}{\|\mathbf{w}^*\|}$$

where $\|\mathbf{w}^*\| = \mathbf{w}^{*T}\mathbf{w}^*$.

So $r$ is maximized when $\|\mathbf{w}^*\|$ is minimized. The training task can then be converted to an optimization problem

$$\begin{array}{ll} \text{minimize} & \|\mathbf{w}\| \\ \text{subject to} & y_i(\mathbf{w}^T\mathbf{x}_i + b) \geq 1 \quad i = 1,\ldots,n \end{array}. \qquad \text{Eq. 3.3.4.6}$$

This constrained optimization problem can be solved using the *method of Lagrange multipliers*. First, the Lagrange function corresponding to problem in Eq. 3.3.4.6 is constructed

$$J(\mathbf{w},b,\mathbf{a}) = \frac{1}{2}\|\mathbf{w}\| - \sum_{i=1}^{n} a_i\left[y_i(\mathbf{w}^T\mathbf{x}_i + b) - 1\right] \qquad \text{Eq. 3.3.4.7}$$

where the nonnegative variables $\mathbf{a} = [a_1,\ldots,a_n]^T$ are called *Lagrange multipliers*. The problem in Eq. 3.3.4.6 is equivalent to minimizing $J(\mathbf{w},b,\mathbf{a})$ with respect to $\mathbf{w}$ and $b$, or maximizing $J(\mathbf{w},b,\mathbf{a})$ respect to $\mathbf{a}$. The solution lies in the *saddle point* of $J(\mathbf{w},b,\mathbf{a})$

$$\begin{cases} \dfrac{\partial J(\mathbf{w},b,\mathbf{a})}{\partial \mathbf{w}} = \mathbf{0} \\ \dfrac{\partial J(\mathbf{w},b,\mathbf{a})}{\partial b} = 0 \end{cases}. \qquad \text{Eq. 3.3.4.8}$$

By solving Eq. 3.3.4.8, we have



$$\begin{cases} \mathbf{w} = \sum_{i=1}^{n} \boldsymbol{a}_i y_i \mathbf{x}_i \\ \sum_{i=1}^{n} \boldsymbol{a}_i y_i = 0 \end{cases},$$

Eq. 3.3.4.9

and by substituting Eq. 3.3.4.9 into Eq. 3.3.4.7 we get

$$J(\mathbf{w}, b, \mathbf{a}) = \sum_{i=1}^{n} \boldsymbol{a}_i - \frac{1}{2} \sum_{i=1}^{n} \sum_{j=1}^{n} y_i y_j \boldsymbol{a}_i \boldsymbol{a}_j \mathbf{x}_i^{\mathrm{T}} \mathbf{x}_j .$$

Eq. 3.3.4.10

The problem in Eq. 3.3.4.6 can then be transformed into the quadratic optimization problem

$$\begin{aligned} \text{maximize} \quad & W(\mathbf{a}) = \sum_{i=1}^{n} \boldsymbol{a}_i - \frac{1}{2} \sum_{i=1}^{n} \sum_{j=1}^{n} y_i y_j \boldsymbol{a}_i \boldsymbol{a}_j \mathbf{x}_i^{\mathrm{T}} \mathbf{x}_j \\ \text{subject to} \quad & \sum_{i=1}^{n} y_i \boldsymbol{a}_i = 0, \quad \boldsymbol{a}_i \geq 0, \quad i = 1, \ldots, n \end{aligned}.$$

Eq. 3.3.4.11

This quadratic optimization problem has a unique solution that can be expressed as a weighted combination of the training samples.

Suppose the solution of the problem is $\mathbf{a}^* = [\boldsymbol{a}_1^*, \ldots, \boldsymbol{a}_n^*]^{\mathrm{T}}$, according to Eq. 3.3.4.9, we have

$$\mathbf{w}^* = \sum_{i=1}^{n} \boldsymbol{a}_i y_i \mathbf{x}_i .$$

Eq. 3.3.4.12

and using Eq. 3.3.4.4, we get

$$b^* = 1 - \mathbf{w}^{*\mathrm{T}} \mathbf{x}^+ = -1 - \mathbf{w}^{*\mathrm{T}} \mathbf{x}^- ,$$

Eq. 3.3.4.13

where support vectors $\mathbf{x}^+$ and $\mathbf{x}^-$ could be determined using the *Karush-Kuhn-Tucker conditions* (Fletcher, 1987; and Bertsekas, 1995). According to the Karush-Kuhn-Tucker condition, following equation holds,

$$\boldsymbol{a}_i [y_i (\mathbf{w}^{\mathrm{T}} \mathbf{x}_i + b) - 1] = 0 \quad i = 1, \ldots, n .$$

Eq. 3.3.4.14

So the sample vectors with positive Lagrange multipliers are support vectors.



## 3.3.4.2 Linearly non-separable learning model and its training

In the linearly non-separable case, there is no hyperplane that can separate all samples. In this circumstance, a set of *slack variables* $\{\boldsymbol{x}_i\}$ $\boldsymbol{x}_i \geq 0, i = 1, \ldots, n$ is introduced, and the separation hyperplane is redefined as

$$y_i(\mathbf{w}^\mathrm{T}\mathbf{x}_i + b) \geq 1 - \boldsymbol{x}_i \quad i = 1, \ldots, n,$$   Eq. 3.3.4.15

and the optimization problem becomes

$$\begin{aligned} \text{minimize} &\quad \mathbf{w}^\mathrm{T}\mathbf{w} + C\sum_{i=1}^{n}\boldsymbol{x}_i \\ \text{subject to} &\quad y_i(\mathbf{w}^\mathrm{T}\mathbf{x}_i + b) \geq 1 \quad i = 1, \ldots, n \end{aligned}.$$   Eq. 3.3.4.16

where the parameter $C$ balances the generalization ability represented in the first term, and the separation ability indicated in the second term. Problem in Eq. 3.3.4.16 can be converted to its dual problem similar to that in the separable case, in which the slack variables are omitted

$$\begin{aligned} \text{maximize} &\quad W(\mathbf{a}) = \sum_{i=1}^{n}\boldsymbol{a}_i - \frac{1}{2}\sum_{i=1}^{n}\sum_{j=1}^{n}y_i y_j \boldsymbol{a}_i \boldsymbol{a}_j \mathbf{x}_i^\mathrm{T}\mathbf{x}_j \\ \text{subject to} &\quad \sum_{i=1}^{n}y_i\boldsymbol{a}_i = 0, \quad 0 \leq \boldsymbol{a}_i \leq C, \quad i = 1, \ldots, n \end{aligned},$$   Eq. 3.3.4.17

and the optimum solution becomes

$$\mathbf{w}^* = \sum_{i=1}^{n_s}\boldsymbol{a}_{s_i} y_{s_i} \mathbf{x}_{s_i},$$   Eq. 3.3.4.18

where $n_s$ is the number of support vectors, and $s_i, i = 1 \ldots n_s$ are indices corresponding to those support vectors. To identify support vectors, the Karush-Kuhn-Tucker condition is defined as

$$\begin{cases} \boldsymbol{a}_i[y_i(\mathbf{w}^\mathrm{T}\mathbf{x}_i + b) - 1 + \boldsymbol{x}_i] = 0 \\ \quad\quad \boldsymbol{x}_i(\boldsymbol{a}_i - C) = 0 \end{cases} \quad i = 1, \ldots, n.$$   Eq. 3.3.4.19



According to this condition, all sample vectors with positive Lagrange multipliers are support vectors; and the slack variable is non-zero only when its corresponding Lagrange multiplier equals to $C$. The value of $b^*$ can be determined by choosing any support vector $\mathbf{x}_i$ with Lagrange multiplier $0 < a_i < C$:

$$b^* = 1 - \mathbf{w}^{*\mathrm{T}}\mathbf{x}_i \quad \text{if } y_i = +1 \qquad\qquad \text{Eq. 3.3.4.20}$$

or

$$b^* = \mathbf{w}^{*\mathrm{T}}\mathbf{x}_i - 1 \quad \text{if } y_i = -1 . \qquad\qquad \text{Eq. 3.3.4.21}$$

When the classification task has more than two class labels, there are two ways to transform it to a binary classification problem. One way is to encode class labels using a binary representation. Suppose there are $l$ class labels in the task. $\lceil \log_2(l) \rceil$ support vector machines are needed to perform classification task together. If a sample has the $k$ th class label, a vector $\mathbf{y} = [y_1, \ldots, y_{\lceil \log_2(l) \rceil}]^{\mathrm{T}}, \quad y_1, \ldots, y_{\lceil \log_2(l) \rceil} \in \{+1, -1\}$,

where $y_i = \begin{cases} +1 & \text{if } i\text{th bit of binary form of } k \text{ is } 1 \\ -1 & \text{if } i\text{th bit of binary form of } k \text{ is } 0 \end{cases} \quad i = 1, \ldots, \lceil \log_2(l) \rceil$,

is sufficient to represent corresponding desired outputs of the support vector machines. Another way is the *one against others* method. In this method, $l$ support vector machines are needed. The corresponding desired output of a sample with $k$ th class label is $\mathbf{y} = [y_1, \ldots, y_l]^{\mathrm{T}}$, where $y_i = \begin{cases} +1 & \text{if } i = k \\ -1 & \text{otherwise} \end{cases} \quad i = 1, \ldots, l$.

### 3.3.4.3 Nonlinear Support Vector Machines

Support vector machine is basically a linear model. It can be extended to handle non-linear cases by introducing slack variables, as is shown in Section 3.3.4.2. In addition, nonlinear



mapping functions for transformation of input vectors can also be employed. This approach makes the learning more flexible. Let $\mathbf{t}(\mathbf{x}) = [t_0(\mathbf{x}), t_1(\mathbf{x}), \ldots, t_l(\mathbf{x})]^T$ be a vector of nonlinear transform functions, where $t_0(\mathbf{x}) = 1$. The optimal hyperplane is then defined as follows

$$\mathbf{w}^T \mathbf{t}(\mathbf{x}) = 0, \qquad\qquad \text{Eq. 3.3.4.22}$$

where the bias term is included implicitly in $\mathbf{w}$. By adapting Eq. 3.3.4.12 we get

$$\mathbf{w} = \sum_{i=1}^{l} \boldsymbol{a}_i y_i \mathbf{t}(\mathbf{x}_i). \qquad\qquad \text{Eq. 3.3.4.23}$$

Substituting Eq. 3.3.4.23 into Eq. 3.3.4.22, we obtain

$$\sum_{i=1}^{l} \boldsymbol{a}_i y_i \mathbf{t}^T(\mathbf{x}_i) \mathbf{t}(\mathbf{x}) = 0. \qquad\qquad \text{Eq. 3.3.4.24}$$

Let the inner product kernel $K(\mathbf{x}_i, \mathbf{x}) = \mathbf{t}^T(\mathbf{x}_i)\mathbf{t}(\mathbf{x})$ be a symmetric function, Eq. 3.3.4.24 becomes

$$\sum_{i=1}^{l} \boldsymbol{a}_i y_i K(\mathbf{x}_i, \mathbf{x}) = 0. \qquad\qquad \text{Eq. 3.3.4.25}$$

Problem Eq. 3.3.4.17 is then reformulated as

$$\begin{aligned}
\text{maximize} \quad & W(\mathbf{a}) = \sum_{i=1}^{n} \boldsymbol{a}_i - \frac{1}{2} \sum_{i=1}^{n} \sum_{j=1}^{n} y_i y_j \boldsymbol{a}_i \boldsymbol{a}_j K(\mathbf{x}_i, \mathbf{x}_j) \\
\text{subject to} \quad & \sum_{i=1}^{n} y_i \boldsymbol{a}_i = 0, \quad 0 \leq \boldsymbol{a}_i \leq C, \quad i = 1, \ldots, n
\end{aligned} \qquad \text{Eq. 3.3.4.26}$$

The optimal decision hyperplane can be found by solving problem in Eq. 3.3.4.26 and substituting the Lagrange multipliers into Eq. 3.3.4.25.

The complexity of the target function to be learned depends on the way it is represented. The kernel approach provides a means to *implicitly* map input vectors into a feature space, i.e. the kernel can be used without knowing its corresponding transforming function. The



introduction of kernel simplifies the design of a learner, and may improve generalization ability. This approach can be used not only in support vector machine but also in other learning models. Some of the examples of learning models that consist of kernel are listed below (Haykin, 1999):

- Polinomial learning machine

$$K(\mathbf{x}, \mathbf{x}_i) = (\mathbf{x}^T \mathbf{x}_i + 1)^p.$$  Eq. 3.3.4.27

- Radial-basis function network

$$K(\mathbf{x}, \mathbf{x}_i) = e^{-\frac{1}{2s^2}\|\mathbf{x}-\mathbf{x}_i\|}.$$  Eq. 3.3.4.28

- Three layer neural network

$$K(\mathbf{x}, \mathbf{x}_i) = \tanh(\boldsymbol{b}_0 \mathbf{x}^T \mathbf{x}_i + \boldsymbol{b}_1)$$  Eq. 3.3.4.29

where $p$, $\boldsymbol{s}$, $\boldsymbol{b}_0$ and $\boldsymbol{b}_1$ are pre-specified parameters.

### 3.3.4.4 SVM recursive feature elimination (RFE)

The weights of a trained SVM can indicate the importance of the corresponding features to the classification. Based on this idea, Guyon et al. (2002) proposed a recursive feature elimination method. After training a linear kernel SVM, its weight can be obtained by Eq. 3.3.4.18. The algorithm iteratively trains SVM and eliminates the feature(s) with small weights, until the feature set become empty. Figure 3.5 shows the algorithm:



---

$\text{RFE}(\, S = \{\mathbf{f}_1, \ldots, \mathbf{f}_m\}, \mathbf{y}\,)$

$\quad$ While $|S| > 1$

$\qquad \mathbf{w} = \text{svm\_training}(S, \mathbf{y})$

$\qquad f = \arg\min(w_i^{\,2}), i = 1, \ldots, m$

$\qquad S = S - \{\mathbf{f}_f\}$

$\quad$ end

---

Figure 3.5: Recursive feature elimination algorithm

### 3.3.5 Bayesian classifier

Keller et al. (2000) used a simple classification method based on naïve Bayes rule. Given an expression vector $\mathbf{x}$ of $m$ selected features, the classification of a sample is computed as follows

$$\text{class}(\mathbf{x}) = \arg\max_i(\log P(M_i \mid \mathbf{x})), \quad i = a, b,$$
Eq. 3.3.5.1

where $P(M_i \mid \mathbf{x})$ is *a posteriori* probability that $M_i$ is true given $\mathbf{x}$. Applying the Bayes rule once again, the class for vector $\mathbf{x}$ can be predicted as

$$\text{class}(\mathbf{x}) = \arg\max_i(\log P(\mathbf{x} \mid M_i))$$

$$= \arg\max_i(\sum_{g=1}^{m} \log P(x^g \mid M_i))$$
Eq. 3.3.5.2

$$= \arg\max_i(\sum_{g=1}^{m} -\log \boldsymbol{d}_i^{\,g} - \frac{(x^g - \boldsymbol{m}_i^{\,g})^2}{2(\boldsymbol{d}_i^{\,g})^2}), \quad i = a, b,$$

where $\boldsymbol{m}_i^{\,g}$ and $\boldsymbol{d}_i^{\,g}$ are the mean and standard deviation of the feature values of the training samples of class $i$. In a binary classification, we can be more confident about the



classification when the difference between $\log P(\mathbf{x}\,|\,M_a)$ and $\log P(\mathbf{x}\,|\,M_b)$ is bigger. However, in order to obtain more information regarding the confidence of the classification, we need to compute the following

$$\text{class}(\mathbf{x}) = \log P(\mathbf{x}\,|\,M_a) - \log P(\mathbf{x}\,|\,M_b)$$

$$= \sum_{g=1}^{m}\left[-\log \boldsymbol{d}_a^g - \frac{(x^g - \boldsymbol{m}_a^g)^2}{2(\boldsymbol{d}_a^g)^2}\right] - \sum_{g=1}^{m}\left[-\log \boldsymbol{d}_b^g - \frac{(x^g - \boldsymbol{m}_b^g)^2}{2(\boldsymbol{d}_b^g)^2}\right].$$

Eq. 3.3.5.3

A positive difference means that the sample is predicted to be class $a$, and a negative difference means that the sample is predicted to be class $b$. The larger the difference, the more confident we are about the classification. We also make use of this difference when computing another measure of accuracy, i.e. acceptance rate, which will be discussed in Section 4.1.6.1.

## 3.3.6  Two discriminant methods for multivariate feature selection

We found that a number of multivariate feature ranking methods can be placed in a unified framework. The methods attempt either to find a vector projection of the samples onto which maximizes or to minimize certain objective function $f(\mathbf{w})$. The function $f(\mathbf{w})$ is originally used on individual features to measure the discrimination ability or diversity of the feature. The magnitude of the elements in the vector then indicates the relative importance of the features. When $f(\mathbf{w})$ is extremal margin, the method is equivalent to RFE. When $f(\mathbf{w})$ is set to Fisher's criterion (see Section 3.3.6.1), the method becomes Fisher's Linear Discrimination method. When the function $f(\mathbf{w})$ is substituted by Eq. 3.3.3.3, the method resembles neural network feature selector in the sense that the optimized weights between input and hidden layer can indicate the relative importance of the input neuron. When $f(\mathbf{w})$ is



the standard deviation of the projection of all samples, the method becomes PCA. Section 3.3.6.1 describes Fisher's linear discriminator. In Section 3.3.6.2, we attempt to use Likelihood ranking method as the objective function to rank relative discrimination contribution of features.

### 3.3.6.1 Fisher's linear discriminant

Let $G = \{g_1, \ldots, g_m\}$ be a set of features. By performing linear transform $y_{a,i} = \sum_{g \in G} w^g x_{a,i}^g$ and

$y_{b,i} = \sum_{g \in G} w^g x_{b,i}^g$, that is projecting all samples from $m$ dimension space to a unit vector $\mathbf{w}$,

we can obtain the Fisher's criterion of the projections alone $\mathbf{w}$,

$$
\begin{aligned}
F(\mathbf{w}) &= \frac{(\mathbf{m}_a' - \mathbf{m}_b')^2}{\mathbf{d}_a'^2 + \mathbf{d}_b'^2} \\
&= \frac{(\mathbf{w}^\mathsf{T} \mathbf{u}_a - \mathbf{w}^\mathsf{T} \mathbf{u}_b)^2}{\mathbf{w}^\mathsf{T}(n_a \Sigma_a + n_b \Sigma_b)\mathbf{w}}
\end{aligned}
, \qquad \text{Eq. 3.3.6.1}
$$

where $\mathbf{m}'$ and $\mathbf{d}'$ are the mean and standard deviation of two classes of projections respectively, $\mathbf{u}_a$ and $\mathbf{u}_b$ are the mean of the original sample vectors of the two classes respectively, and $\Sigma_a$ and $\Sigma_b$ are the covariance matrix of the samples from the two classes respectively. Fisher's linear discriminant tries to find the weight $\mathbf{w}$ that maximizes $F(\mathbf{w})$, that is,

$$
\begin{aligned}
\text{maximize} \quad & F(\mathbf{w}) \\
\textbf{s.t.} \quad & \mathbf{w}^\mathsf{T}\mathbf{w} = 1
\end{aligned}
. \qquad \text{Eq. 3.3.6.2}
$$

The solution of the maximization problem is

$$
\mathbf{w}^* = (n_a \Sigma_a + n_b \Sigma_b)^{-1}(\mathbf{u}_a - \mathbf{u}_b). \qquad \text{Eq. 3.3.6.3}
$$

The value of $\mathbf{w}^*$ can be an indicator of the contribution of features to discrimination.



## 3.3.6.2 Multivariate Likelihood feature ranking

We propose a multivariate likelihood feature selection method that is based on a similar idea as that of Fisher's linear discriminant. Suppose there are $n = n_a + n_b$ samples with $m$ features. Recall Keller's Likelihood method for ranking for individual gene $g$, which are expressed in Eq. 3.2.2.6 and Eq. 3.2.2.7 in Section 3.2.2. Let $G = \{g_1, \ldots, g_m\}$ be a set of features. By performing linear transform $y_{a,i} = \sum_{g \in G} w^g x_{a,i}^g$ and $y_{b,i} = \sum_{g \in G} w^g x_{b,i}^g$, that is projecting all samples from $m$ dimension space to a unit vector $\mathbf{w}$, we can obtain the likelihood of the projections of the samples of two classes on vector $\mathbf{w}$,

$$LIK_{a \to b} = \sum_{i=1}^{n_a} \left( -\log(\boldsymbol{d}_a) - \frac{(y_{a,i} - \boldsymbol{m}_a)^2}{2(\boldsymbol{d}_a)^2} + \log(\boldsymbol{d}_b) + \frac{(y_{a,i} - \boldsymbol{m}_b)^2}{2(\boldsymbol{d}_b)^2} \right) \qquad \text{Eq. 3.3.6.4}$$

and

$$LIK_{b \to a} = \sum_{i=1}^{n_b} \left( -\log(\boldsymbol{d}_b) - \frac{(y_{b,i} - \boldsymbol{m}_b)^2}{2(\boldsymbol{d}_b)^2} + \log(\boldsymbol{d}_a) + \frac{(y_{b,i} - \boldsymbol{m}_a)^2}{2(\boldsymbol{d}_a)^2} \right). \qquad \text{Eq. 3.3.6.5}$$

Where $\boldsymbol{m}$ and $\boldsymbol{d}$ are the mean and standard deviation of two classes of projections, respectively.

The multivariate feature selection process becomes:

maximize   f($\mathbf{w}$)

**s.t.**        $\mathbf{w}^\top \mathbf{w} = 1$                                Eq. 3.3.6.6

where f($\mathbf{w}$) = $LIK_{a \to b}$, f($\mathbf{w}$) = $LIK_{b \to a}$ or f($\mathbf{w}$) = $LIK_{a \to b} + LIK_{b \to a}$. When the maximization problem in Eq. 3.3.6.6 is solved, it becomes an indicator of the contribution of features to



discrimination. We use Sequential Quadratic Programming method (Fletcher, 1987) to find a suitable $\mathbf{w}$.

### 3.3.7 Combining univariate feature ranking method and multivariate feature selection method

Information gain and likelihood method are univariate feature ranking methods in the sense that they assume genes contribute to classification independently, and rank the genes according to their individual contribution. This assumption has computational advantages. But in real world, genes are often working together for a certain function, and the combinatorial effect of these genes is not considered by univariate selection methods. On the other hand, neural network feature selector, recursive feature elimination and multivariate likelihood method consider the whole contribution of subset of features to the classification. These three approaches have the potential to select smaller subsets of features with higher classification performance. However, the selection process may be obscured when applied to microarray datasets with high dimensionality and in the presence of large number of irrelevant features. Take RFE as an example, the presence of large number of irrelevant features hides the discriminative information from relevant features. This can be seen in the formulation of the SVM dual problem where the coefficients of the quadratic terms in the dual problem are computed as the scalar products of two inputs

$$\mathbf{x}_i^{\mathrm{T}} \mathbf{x}_j = \sum_{k=1}^{m} x_{ik} x_{jk} . \qquad \text{Eq. 3.3.7.1}$$

The gene elimination process is very sensitive to change in the feature set. SVM also has the disadvantage that it is sensitive to outliers as discussed in Guyon et al. (2002). In microarray data, the outliers may be introduced by: 1) noise in the expression data, or 2) incorrectly



identified or labeled samples in the training dataset. It is therefore more beneficial to apply RFE on a dataset with a reduced number of features. A univariate feature selection algorithm can be used to first efficiently reduce the large number of features originally present in the dataset and a multivariate feature selection method such as RFE can then be applied to remove more features. To summarize, we first identify and remove genes that are expected to have low discrimination ability as indicated by LIK scores. Then, we apply RFE to remove the size of the feature set further. With this integrated approach to feature selection, we are able to achieve good classification performance with fewer genes than those reported by Guyon et al. (2002) and Keller et al. (2000).

A multivariate method is usually very time-consuming when applied to a dataset with thousands of genes. For RFE, in order to eliminate one or more genes, a new SVM has to be trained, and the overall computational cost is $\Omega(m^2 n^2)$. On the other hand, LIK ranks genes independently, which makes the computational complexity of LIK, $O(mn)$. Using LIK first to reject a large number of genes, and then using RFE to perform further selection will save significant running time compared to just using RFE alone. This is especially important when the improvement in microarray technology makes it possible to obtain gene expression values from tens or even hundreds of thousands of genes.



# 4 Experimental results and discussion

The experimental results from and discussions on applying machine learning methods to global gene expression analysis in our research will be described in this chapter. We start with the experiments on the datasets of the second type classification problem, which consist of large number of features and small number of samples. The high dimension nature of this kind of problem makes it distinct from common classification problems. We then describe the test result on a newly released dataset, which is of the first type of classification problem with large number of samples and small number of features.

## 4.1 Second type - high dimension problems

The main work in this thesis focuses on the second type of classification problems, which involves a small number of samples with a large number of features, and feature selection problems associated with this type of problems. Our strategy is to first try various analysis methods, most of which were described in Chapter 3, on a well known benchmark dataset, human acute leukemia microarray dataset (Golub et al., 1999), select one that has the best performance, then try that method on other datasets of same type, including small, round blue cell tumors (SRBCTs) (Khan et al., 2001) dataset and artificial datasets.

The human acute leukemia microarray dataset consists of 72 microarray experiments with expression values of 7129 clones from 6817 human genes. Here the term clone refers to the fragment of a gene. Each of the genes has a short description; and each clone is represented by an accession number. Each microarray is assigned with a class label, either *Acute Myeloid Leukemia* (AML) or *Acute Lymphoblastic Leukemia* (ALL), according to the organism used for the hybridization. A second type of classification problem arises from this dataset. We



used the clone id as feature id. Each of the hybridizations corresponds to a classification sample. These samples were divided into two sets by Golub et al.: The first sample set, which consists of 27 ALL samples and 11 AML samples, is for training the classifier. The second sample set, consisting of 20 ALL samples and 14 AML samples, is for testing the classifier.

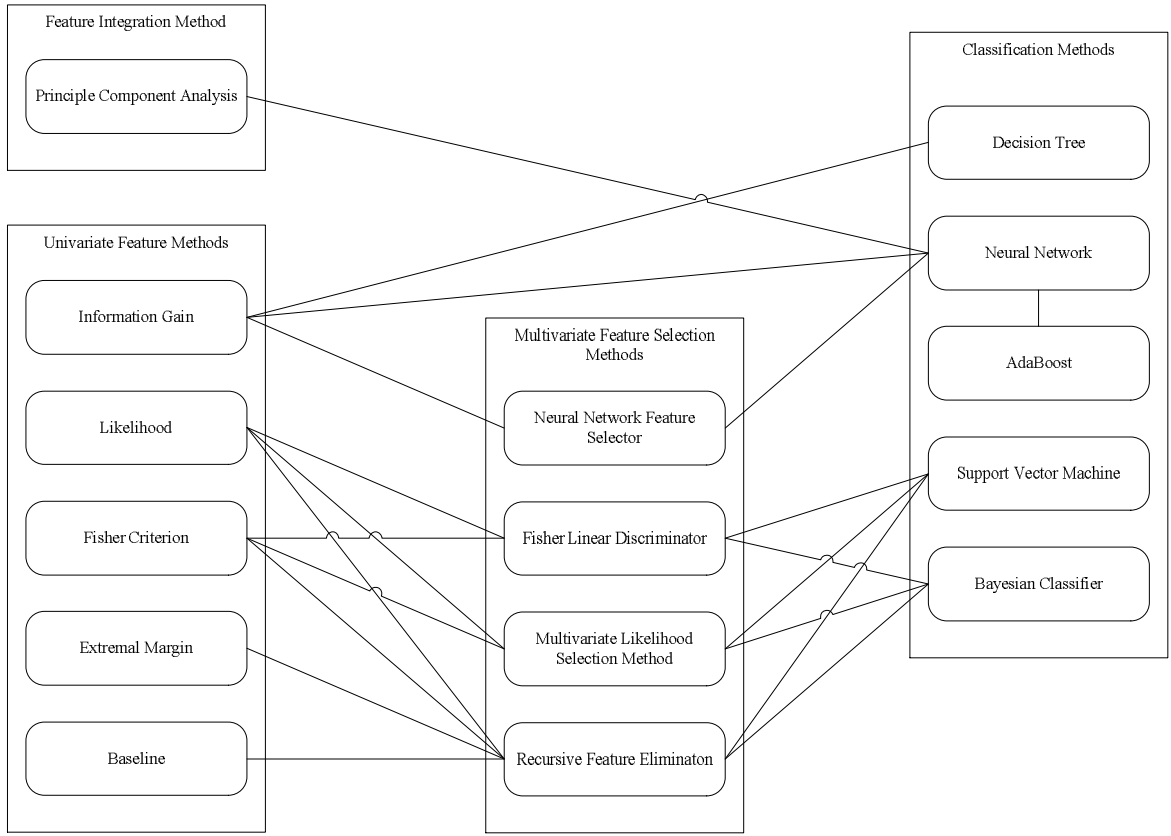

Figure 4.1: The combination of feature selection / integration methods and classification methods.

Due to the time constraint, we only tested a subset of all possible combinations of these methods. In our experiment described in the following sections, the combination of Likelihood and Recursive Feature Selection method achieved the best feature selection



performance. With the optimal feature sets selected using the above combination, Bayesian classification method achieved the best classification performance. The combinations that used are summarized in Table 4.1. Characteristic of some of the combinations are described. The term *homogenous* in the table refer to the methods that use same kind of criterion in single dimension and multi-dimension. Those combinations that are not tested correspond to blank cells in the table.

| | Neural network feature selector | Fisher Linear Discriminator | Multivariate Likelihood Selection Method | Recursive Feature Elimination |
|---|---|---|---|---|
| Information gain | Homogeneous , stopping criterion, heavy computation | | | |
| Likelihood | | Tested | Homogeneous | Best selection performance / Extensively studied |
| Fisher Criterion | | Homogeneous | Two gene sets can be obtained, each distinguish one class from the other | Tested |
| External Margin | | | | Second best combination we found |
| Baseline | | | | Tested |

Table 4.1: List of combined univariate and multivariate feature selection methods tested.

## 4.1.1  Principle Component Analysis for feature integration

Our first effort was to study the relevance of features to the class labels. The study involves both feature integration and feature selection. For feature integration, we chose principle component analysis to see how much the most informative component extracted from the features can contribute to the classification. Experiment was done using statistics toolbox and neural network toolbox in MATLAB v6.1. Computation of the principle components form large number of features consumes large amount of computer memory. Due to the limitation of our computer system, the program was unable to process all 7129 features. We generated 72 components from randomly chosen 4500 features, which is the maximum number of



features that can be processed by the program, and then used all samples with these components to perform leave-one-out training and validation using a three-layer feed-forward neural network with different number of hidden units on all samples. In the training process, batch mode was used. In each leave-one-out iteration, a test of the classification performance on the 71 samples was done after every 10 epochs. If the accuracy on training data is 100%, the training is stopped, and the remaining one sample is tested. Table 4.2 shows the performance. The result shows that although principle component extracted most informative information in terms of standard variance from the features, this information is hardly relevant to the classification.

| Hidden neuron number | Training method | Performance |
| --- | --- | --- |
| 600 | RPROP backpropagation | 51.39% |
| 600 | One Step Secant Algorithm | 51.39% |
| 50 | One Step Secant Algorithm | 43.06% |
| 50 | One Step Secant Algorithm | 48.61% |

Table 4.2: Performance of neural network using principle components as input.

### 4.1.2  C4.5 for feature selection

We applied C4.5 algorithm on all 72 samples of the leukemia dataset. The constructed decision tree was surprisingly simple, which only involves two genes. It could correctly classify 71 samples. The tree is as follows:

M84526_at > 290 : -1

M84526_at <= 290 :

|  X54489_rna1_at <= 91 : 1

|  X54489_rna1_at > 91 : -1



Using only 38 training samples, a even simpler tree involving only one feature is generated, this tree correctly classified all 38 training samples and 31 out of 34 testing samples (accuracy: 91.2%).

X95735_at <= 938 : 1

X95735_at > 938 : -1

The decision tree construction algorithm C4.5 also supports constructing trees in iterative mode. In this mode, the algorithm randomly selects an initial sample subset from training set to construct a decision tree, and then iteratively add the samples that are misclassified by the tree into the subset and reconstruct the tree until there is no misclassification among all training samples. We tested C4.5 with iterative mode for 20 trials using all 72 samples. All trials generated a two-layer tree with three nodes. These trees are listed in Table 4.3. For simplification, we just list the gene accession numbers of the first layer and the second layer to represent the tree.

| Feature of first layer | Feature of second layer | Occurrence |
|---|---|---|
| M84526_at | M83652_s_at | 4 |
| M84526_at | D86967_at | 2 |
| M84526_at | X54489_rna1_at | 2 |
| U46499_at | M98833_at | 1 |
| U46499_at | X86401_s_at | 1 |
| U46499_at | D89289_at | 1 |
| L09209_s_at | D80003_at | 1 |
| D88422_at | M31166_at | 5 |
| D88422_at | U23070_at | 2 |
| D88422_at | M83652_s_at | 1 |

Table 4.3: Decision trees constructed by iterative mode from all 72 samples.



We then tried leave-one-out test to construct decision trees from all samples. A total of 72 trees were constructed, most of which have two layers, but some had only one layer. Table 4.4 summarize these trees. When applying leave-one-out to 38 training samples, all the trees constructed had one layer. They are summarized in Table 4.5.

| Feature of first layer | Feature of second layer | Occurance |
|---|---|---|
| M23197_at | M20902_at | 1 |
| M23197_at | D80003_at | 1 |
| M23197_at | | 2 |
| M27891_at | M31166_at | 2 |
| M27891_at | M55418_at | 1 |
| M27891_at | | 2 |
| M84526_at | X06948_at | 1 |
| M84526_at | Y07604_at | 1 |
| M84526_at | M81883_at | 1 |
| M84526_at | D86967_at | 1 |
| M84526_at | X54489_rna1_at | 52 |
| U46499_at | M60527_at | 1 |
| U46499_at | M98833_at | 1 |
| U46499_at | U36922_at | 1 |
| U46499_at | X86401_s_at | 1 |
| X95735_at | HG2160-HT2230_at | 1 |
| X95735_at | | 1 |
| M83652_s_at | M31211_s_at | 1 |

Table 4.4: Decision trees constructed by leave-one-out mode from all 72 samples

| Feature of first layer | Occurance | Classification accuracy on test samples |
|---|---|---|
| X95735_at | 35 | 0.912 |
| M27891_at | 1 | 0.941 |
| M31166_at | 1 | 0.706 |
| M55150_at | 1 | 0.794 |

Table 4.5: Decision trees constructed by leave-one-out mode from 38 training samples with prediction accuracy on 34 test samples.



Table 4.3, Table 4.4 and Table 4.5 show that certain features occur very frequently in these three experiments. It appears that the selectivity of C4.5 algorithm is high for the dataset. In Table 4.3 and Table 4.4, feature M84526_at has the highest occurrence frequency, which implies that this feature is important in deciding the classes when all 72 samples are taken into account. But when only taking the 38 training samples into account, the algorithm selected a very different set of features, which is shown in Table 4.5. Only feature X95735_at appeared two times in leave-one-out mode for all 72 samples (see Table 4.4). The fact that the trees generated on all training and test samples are very simple and the trees generated on the training samples are even much simpler make us expect that some feature selection method should that can generate a very small feature set when presented with a very limited number of training samples. It may also be possible to obtain high classification accuracy on test samples by certain classifiers constructed using this small feature set.

| Rank | Feature | All samples | Training samples | Test samples |
|---|---|---|---|---|
| 1 | M84526_at | 0.652 | 0.408 | 0.689 |
| 2 | M27891_at | 0.652 | 0.685 | 0.689 |
| 3 | D88422_at | 0.651 | 0.578 | 0.584 |
| 4 | M23197_at | 0.648 | 0.581 | 0.684 |
| 5 | X95735_at | 0.647 | 0.844 | 0.522 |
| 6 | U46499_at | 0.634 | 0.565 | 0.692 |
| 7 | M31523_at | 0.590 | 0.511 | 0.851 |
| 8 | L09209_s_at | 0.589 | 0.562 | 0.577 |
| 9 | M83652_s_at | 0.550 | 0.578 | 0.420 |
| 10 | M11722_at | 0.542 | 0.332 | 0.683 |
| 22 | M31166_at | 0.405 | 0.689 | 0.182 |
| 26 | M55150_at | 0.398 | 0.671 | 0.198 |

Table 4.6: Leukemia features and their information gain of all samples, training samples and test samples, sorted by gain of all samples.



We continued to investigate the information gain computed for the features of first level. In Table 4.6, top ten ranked features are listed according to the information gain of all samples. The features that are listed in Table 4.5 whose rank are higher than ten are also listed in Table 4.6.

Information gain is a measure of relevance of a feature to classification. It appeared that the features with high gain in training samples are likely to have high gain in all samples and test samples. As mentioned before, C4.5 can construct a decision tree that consists of only one feature from training samples using information gain. The tree is very simple but it can only correctly classify 31 of 34 test samples. The generalization ability of the classifiers constructed using only information gain measure for continuous features is not high, as was discussed in Chapter 3, so we tried to use information gain measure as a feature selection method, and test neural networks based on the selected features to see whether there is any improvement in performance.

### 4.1.3 Neural networks with features selected using information gain

We tested the neural network with different configurations and different numbers of features, as is listed in the following tables, with the highest information gain obtained from training samples. The test results are listed in Table 4.7. Two kinds of training methods were used: trainoss (One Step Secant Algorithm) and traingd (Gradient descent backpropagation). The training was done in batch mode. In the training process, a test of the classification error of training samples was performed repeatedly after a certain number of epochs indicated in *test interval* column in the table until the error converge to no greater than *training error tolerance*. The reason why we choose different tolerance is to check how well the trained neural network could generalize, under a given over-fitting limit. If the number of tests



exceeded the number indicated in *number of unsuccessful trials* column, then the training was considered to be unsuccessful and the training result was rejected. Once the training was successfully terminated, we tested the classification accuracy of the trained network on test samples. For every configuration corresponding to each row in the table, we collected 100 successful trainings and then calculates the mean and standard deviation of the classification accuracy on the test samples.

| Feature number | Number of hidden units | Training method | Test interval (number of epochs) | Number of unsuccessful trials | Training error tolerance (number of samples) | Test accuracy |
|---|---|---|---|---|---|---|
| 50 | 60 | trainoss | 20 | N/A | 0 | 0.810±0.090 |
| 20 | 60 | trainoss | 20 | N/A | 0 | 0.855±0.080 |
| 10 (from all samples) | 20 | trainoss | 20 | 20 | 0 | 0.952±0.024 |
| 5 | 20 | trainoss | 20 | 20 | 0 | 0.880±0.075 |
| 100 | 40 | trainoss | 20 | 20 | 0 | 0.824±0.101 |
| 50 | 50 | trainoss | 200 | 20 | 0 | 0.843±0.089 |
| 50 | 50 | trainoss | 20 | 20 | 0 | 0.848±0.090 |
| 50 | 10 | traingd | 100 | 5 | 2 | 0.936±0.036 |
| 50 | 10 | traingd | 100 | 5 | 2 | 0.931±0.041 |
| 20 | 10 | traingd | 100 | 5 | 4 | 0.942±0.025 |
| 10 | 10 | traingd | 100 | 3 | 4 | (Repeats can hardly converge) |
| 20 | 5 | traingd | 100 | 2 | 4 | 0.947±0.021 |

Table 4.7: Prediction performance of neural network using features selected by information gain.

Note that the experiment in the third row of Table 4.7 used top ten features from information gain of all features. The aim of this test is to see how high the accuracy could be when information from test samples is used. We found that when the training error tolerance of the training sample was as large as 4, with top 20 features, the test accuracy could approximate



the test in the third row. If the training error tolerance is smaller than 4, the test performance may be affected by over-fitting.

The test in Table 4.7 gave us some indications on how to optimize the parameters to obtain better generalization ability. We continued to test whether the test errors were located on a few test samples or evenly distributed with fine tuned parameters. The method was the same in the experiments in Table 4.7. We collected 100 successful trainings and counted the number of times the test samples that were wrongly predicted as well. In Table 4.8 the configuration and prediction accuracy are listed, and the wrong prediction frequency is shown in Figure 4.2.

As can be seen from Figure 4.2, it seemed that for individual experiments of certain configurations, the wrong predication frequency was very high in certain test samples. In addition, most of trained neural networks were likely to make wrong prediction on the class of test samples 28 and 29.

| Experiment ID | Number of hidden units | Training method | Test interval (number of epochs) | Number of unsuccessful trials | Feature number | Training error tolerance (number of samples) | Mean precision | Standard deviation |
|---|---|---|---|---|---|---|---|---|
| 14 | 5 | traingd | 100 | 1 | 15 | 5 | 0.893 | 0.044 |
| 15 | 3 | traingd | 50 | 1 | 50 | 2 | 0.932 | 0.036 |
| 16 | 3 | traingd | 50 | 1 | 50 | 4 | 0.914 | 0.050 |
| 17 | 3 | traingd | 50 | 1 | 20 | 4 | 0.948 | 0.019 |
| 18 | 3 | traingd | 50 | 1 | 100 | 4 | 0.924 | 0.030 |
| 19 | 3 | traingd | 50 | 1 | 100 | 2 | 0.930 | 0.021 |

Table 4.8: Test of neural network using features selected by information gain to identify incorrectly predicted test samples.



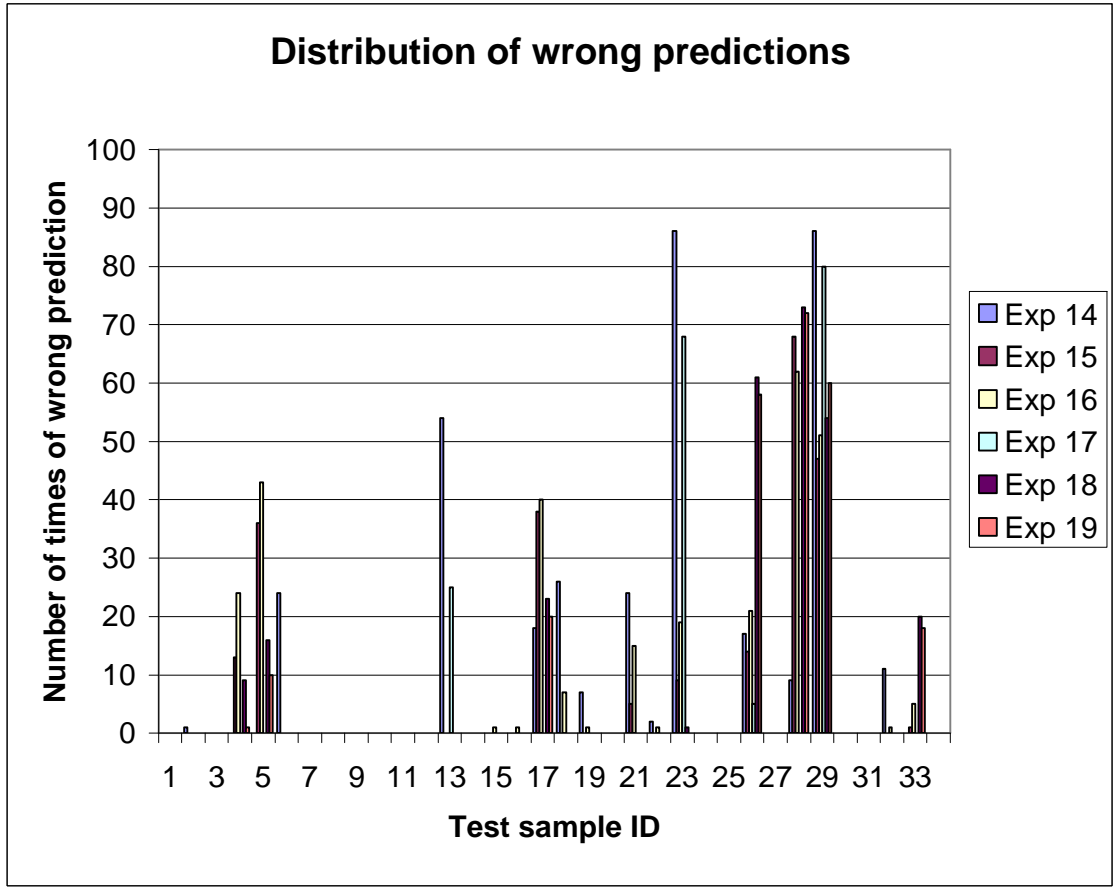

Figure 4.2: Wrong prediction times.

### 4.1.4 AdaBoost

AdaBoost has the mechanism to change the sampling distribution to focus on the training

samples that have high validation errors, and then AdaBoost may fit the training samples well

and keep good generalization ability as well. Hence we expect that AdaBoost framework

might further improve the overall classification performance. In each AdaBoost experiment, a

number of neural networks of the identical size were consecutively trained for 50 epochs, and

those neural networks with training error no higher than error tolerance were employed for

refining sampling distribution. The training process continued until 50 such neural networks



were obtained. The final hypothesis of the test samples could then be obtained by combining hypothesis of individual neural networks and factors $b_{1...T}$, according to Eq. 3.3.2.1. One of disadvantages of AdaBoost is that the training process is slow, so we only conducted the test no more than twice for different configurations. The test results are listed in Table 4.9. The configurations in tests 1 and 4 were tested once for their training time is extremely long, but the remaining configurations were tested twice.

As can be seen from Table 4.9, the test performance of tests using the top 20 to 50 features is generally better than the test results using the same number of features in Table 4.7 even when the error tolerance was as low as 2 validation errors. In AdaBoost tests, there are generally one or two errors (97.1% or 94.1% accuracy) in the test samples. The trained networks are expected to fit the training samples well, and the combined hypothesis can also achieve high prediction performance on the test samples.

| Test ID | Hidden unit number | Feature number | Training error tolerance | Test accuracy |
|---|---|---|---|---|
| 1 | 5 | 20 | 3 | 0.971 |
| 2 | 5 | 10 | 3 | 0.853 |
| 3 | 5 | 10 | 3 | 0.794 |
| 4 | 5 | 15 | 4 | 0.912 |
| 5 | 5 | 50 | 4 | 0.941 |
| 6 | 5 | 50 | 4 | 0.971 |
| 7 | 5 | 50 | 2 | 0.971 |
| 8 | 5 | 50 | 2 | 0.941 |
| 9 | 3 | 50 | 2 | 0.971 |
| 10 | 3 | 50 | 2 | 0.941 |

Table 4.9: AdaBoost test results.



### 4.1.5  Neural network feature selector

By seeing the test result in the previous section, we continued to investigate whether it is possible to further reduce the number of relevant genes without losing too much prediction accuracy. Neural network feature selector, being a wrapper feature selection approach, could exploit the relevant information between features and classes in the trained neural network models. Another advantage of the neural network feature selector is that it can decide the optimum number of selected features. Experiments were carried out using neural network feature selector. The neural network feature selector algorithm was implemented using MATLAB. In the experiments, a set of features, which had highest information gain from the training samples, was used as initial feature set whose size was to be reduced. Because the number of training samples is very small, we used leave-one-out approach to get average training accuracy and validation accuracy when calling function train_validate() and simulate_validate(). Each neural network was trained for maximum 200 epochs in the function train_validate(). The $r_{factor}$ was set to 10. After the selection process, with the selected features, 100 repetitions of training and testing were done, and the mean and standard deviation of training and testing accuracy ranks were calculated. The results are summarized in Table 4.10 and Table 4.11. The performance of both summed square and cross entropy error functions are listed in the two tables respectively. In these two tables, the column $(e_{min}, e_{max})$ contains the thresholds of penalty parameter $e$. Because we increased or decreased the penalty parameter by a factor of 1.1, the values reflect the minimum and maximum number of times that the penalty parameter may increase or decrease accumulatively.



| Experiment | Feature set size | $r'_{min}$ | $r''_{min}$ | $\Delta r''$ | $(e_{min}, e_{max})$ | Number of features selected | Accuracy on training samples | Accuracy on test samples |
|---|---|---|---|---|---|---|---|---|
| 1 | 200 | 0.9 | 0.9 | 0.05 | $1.1^{\pm30}$ | 6 | 1.00±0.00 | 0.67±0.01 |
| 2 | 50 | 0.9 | 0.9 | 0.01 | $1.1^{\pm20}$ | 3 | 1.00±0.00 | 0.79±0.00 |
| 3 | 50 | 0.95 | 0.95 | 0.01 | $1.1^{\pm20}$ | 4 | 1.00±0.00 | 0.88±0.03 |
| 4 | 50 | 0.97 | 0.97 | 0.01 | $1.1^{\pm20}$ | 4 | 1.00±0.00 | 0.88±0.03 |
| 5 | 200 | 0.95 | 0.9 | 0.03 | $1.1^{\pm20}$ | 6 | 0.98±0.01 | 0.71±0.00 |

Table 4.10: Experiment result of neural network feature selector with summed square error function.

| Experiment | Feature set size | $r'_{min}$ | $r''_{min}$ | $\Delta r''$ | $(e_{min}, e_{max})$ | Number of features selected out | Accuracy on training samples | Accuracy on test samples |
|---|---|---|---|---|---|---|---|---|
| 1 | 200 | 0.9 | 0.9 | 0.05 | $1.1^{\pm30}$ | 4 | 1.00±0.00 | 0.71±0.02 |
| 2 | 50 | 0.9 | 0.9 | 0.01 | $1.1^{\pm20}$ | 4 | 1.00±0.00 | 0.72±0.03 |
| 3 | 50 | 0.95 | 0.95 | 0.01 | $1.1^{\pm20}$ | 6 | 1.00±0.00 | 0.68±0.04 |
| 4 | 50 | 0.97 | 0.97 | 0.01 | $1.1^{\pm20}$ | 36 | 1.00±0.00 | 0.80±0.08 |
| 5 | 200 | 0.95 | 0.9 | 0.03 | $1.1^{\pm20}$ | 191 | 1.00±0.01 | 0.79±0.09 |

Table 4.11: Experiment result of neural network feature selector with cross entropy error function.

From Table 4.10, we can see that under certain settings the neural network feature selector can select a small set of four out of 50 genes and the prediction accuracy on the test samples could be as high as 88%. In comparison, the experiments in Table 4.11 show that the selection performances are generally worse in terms of the number of the features selected and the prediction accuracy. We noticed that back-propagation training of neural networks was much faster when using cross entropy error function than when using summed error function.



## 4.1.6  Hybrid Likelihood and Recursive Feature Elimination method

We tried the combination of likelihood and recursive feature elimination method. Very good feature selection performance was found using this hybrid method on Leukemia dataset, which encouraged us to continue to test the method systematically, comparing it with feature selection methods using LIK and RFE alone. We computed acceptance rate, which is a performance measure that is stricter than accuracy.

Suppose there are $n$ samples with predicted values of $o_1, \ldots, o_n$, and their corresponding class labels are $y_1, \ldots, y_n$. Each of the class labels takes value of either $+1$ or $-1$, and the values of outputs are real numbers. If the prediction output of a classifier for a sample has the same sign as that of its true class, we consider this sample to be correctly classified. The performance measure of accuracy, i.e. the number of correctly classified samples over the total number of test samples, is defined as

$$\text{accuracy} = \frac{\left| \left\{ i \mid o_i y_i > 0, i = 1, \ldots, n \right\} \right|}{n},$$

<div align="right">Eq. 4.1.6.1</div>

where $|S|$ denotes the cardinality of the set $S$. In contrast, the acceptance rate is computed as follows

$$\text{acceptance rate} = \frac{\left| \left\{ i \mid o_i y_i > -\min_{j=1,\ldots,n}(o_j y_j), i = 1, \ldots, n \right\} \right|}{n}.$$

<div align="right">Eq. 4.1.6.2</div>

The strength of correct prediction of a sample can be obtained by multiplying the output and class label of a sample together $o_i y_i$, the larger the value of the product, the better the prediction made by the classifier. When the value of the product is negative, the classifier makes a wrong prediction of the sample. To calculate the acceptance rate, we first select the



worst prediction out of all test samples. The worst prediction corresponds to the sample whose $o_j y_j$ is minimum. This minimum values is multiplied by $-1$ and is used as a threshold. All the predictions that have the $o_j y_j$ value bigger than this threshold are considered as being accepted. The acceptance rate is 1 when all the test samples are correctly predicted. Otherwise, it will not be greater than the accuracy, because the prediction of the classifier on some test samples may have small confidence, as indicated by the output-class label products that are lower than the threshold. These test samples are correctly predicted and are counted in the computation of accuracy, but they will not be counted in the computation of acceptance rate.

In the tables and figures in this section, we will denote accuracy and acceptance rate by acu and acp, respectively. Obviously, the acceptance rate cannot be higher than accuracy. Because the number of samples was small, we used the leave-one-out method for validating the classifier on training samples as well as on all samples. When there are $n$ samples, leave-one-out is a technique to iteratively choose each sample for testing, and the remaining samples for training. A total of $n$ classifiers were trained, and $n$ predictions were made. The accuracy and acceptance rates were computed from the predictions and labels of the corresponding test samples.

We ran our experiments on a Pentium 4 1.4GHz computer with 512-megabyte memory. We wrote and ran our program using MATLAB 6.1. The support vector machine was constructed with the Support Vector Machine Toolbox from http://theoval.sys.uea.ac.uk/~gcc/svm/toolbox which was developed by Gavin Cawley. For the SVM, we set $C = 100.0$ and used the linear kernel, the same as those used by Guyon et al. (2002).



### 4.1.6.1 Leukemia dataset

Figure 4.3 shows the sorted LIK scores. We chose equal numbers of genes with the highest $LIK_{ALL \to AML}$ and $LIK_{AML \to ALL}$ score as the initial gene sets for RFE. We plotted in this figure the scores of the top (2 x 80) genes. The top $i$ th gene according to $LIK_{ALL \to AML}$ always has a higher $LIK_{ALL \to AML}$ value than the corresponding top $i$ th gene's $LIK_{AML \to ALL}$ score. Genes with low LIK scores are not expected to be good discriminators. We decided to pick the top genes to check their discriminating ability. In particular, we ran experiments using the top (2 x 10), (2 x 20), (2 x 30) genes. We found the best performance was obtained when 2 x 20 top ranking genes were selected. The performance was measured by computing the prediction accuracy and acceptance rate of SVM and Bayesian classifiers built using the selected genes on the test samples.

Figure 4.4 shows the accuracy and acceptance rate using two different experimental settings: leave-one-out and train-test split. For the leave-one-out (LOO) setting, we computed the performance measures using only the 38 training samples as well as on the entire dataset consisting of 72 samples. For the train-test split, the measures shown were computed on the 34 test samples, while the measures on the 38 training samples are not reported in the table. A series of experiments were conducted to find the smallest number of genes that would give good performance measure. The experiments started with all 40 (= 2 x 20) genes selected by the LIK feature selection. One gene at a time was eliminated using RFE. RFE feature selection was conducted until there was only one gene left. For a selected subset of genes, the performance measures were computed under all experimental settings and using both SVM and Bayesian classifiers.



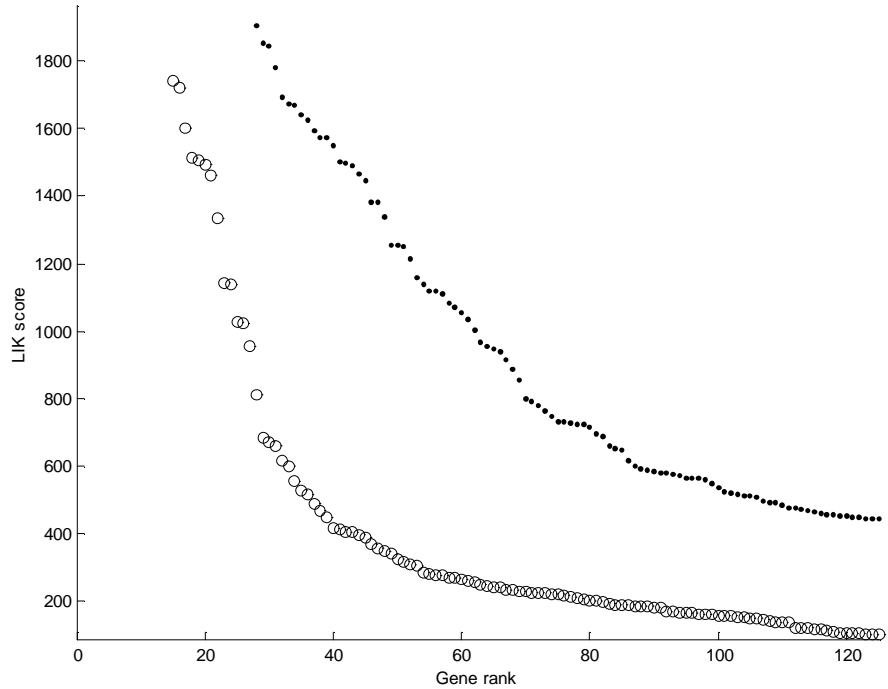

Figure 4.3: Sorted LIK score of a subset of genes in the leukemia dataset. Dots indicate $LIK_{ALL \to AML}$ scores and circles indicate $LIK_{AML \to ALL}$. The top 28 genes according to their $LIK_{ALL \to AML}$ values have scores between 92014 and 1978.3; and the top 15 genes according to their $LIK_{ALL \to AML}$ values have scores between 22852 and 2148.7; they are not shown in this figure.

As can be seen from the figure, the SVM classifier achieved almost perfect accuracy and acceptance rate when there were three to 14 genes used to find the separating hyperplane. On the other hand, when the Naïve Bayesian method was used for classification, almost perfect performance was achieved with as many as 40 genes in the model. Elimination of the genes by RFE one by one showed that the results could be maintained as long as there are at least



three genes in the model. This stability in performance indicates the robustness of the RFE feature selection method when given a pre-selected small subset of relevant genes, as identified by the LIK method. It is worth noting that the acceptance rate on the test samples was almost constant with at least three genes, both when the SVM classifier and the Naïve Bayesian classifier were used for prediction. We emphasize here that the hybrid LIK+RFE feature selection was run using the 38 training samples; the classifiers were also built using the same set of training samples without the use of any information from the data in the test set.

A set of three genes was discovered to give perfect accuracy and acceptance rate regardless of the experimental settings and the classifiers used. These genes are listed in Table 4.12. They have also been identified as relevant genes in this dataset by several researchers. Golub et al. (1999) identified U05259_rna1_at and M27891_at as relevant, while Keller et al. (2000) identified the gene X03934_at as relevant. On the other hand, Guyon et al. (2002) identified a completely different set consisting of four genes. Among these four genes, only M27891_at occurs in the previous C4.5 result.



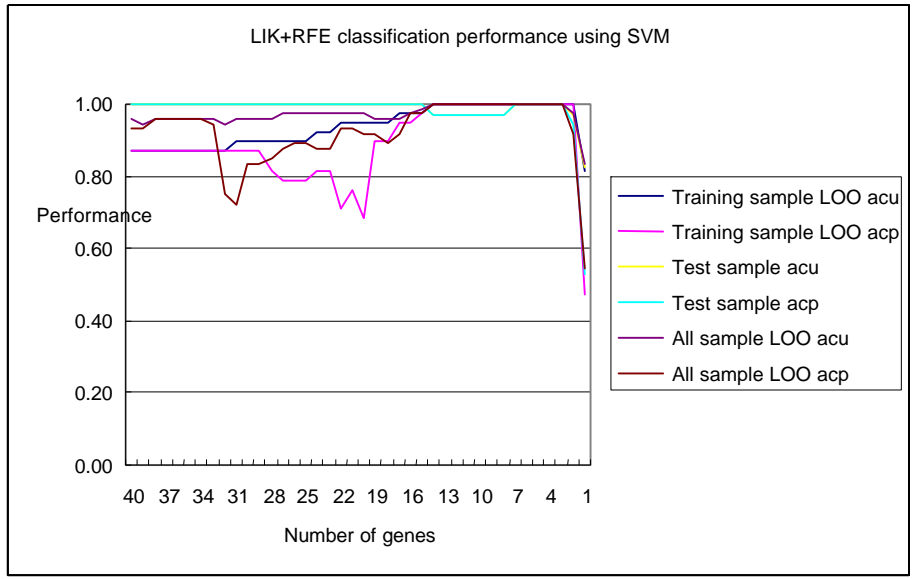

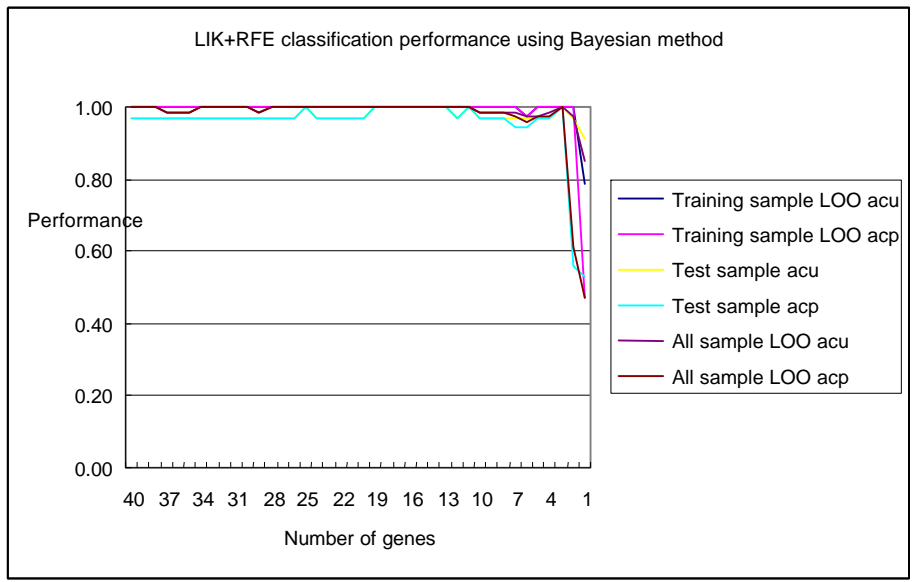

Figure 4.4:  Classification performance of genes selected using the hybrid LIK+RFE.



| Gene accession number | Description |
|---|---|
| U05259_rna1_at | MB-1 gene |
| M27891_at | CST3 Cystatin C (amyloid angiopathy and cerebral hemorrhage) |
| X03934_at | GB DEF = T-cell antigen receptor gene T3-delta |

Table 4.12: The smallest gene set found that achieves prefect classification performance.

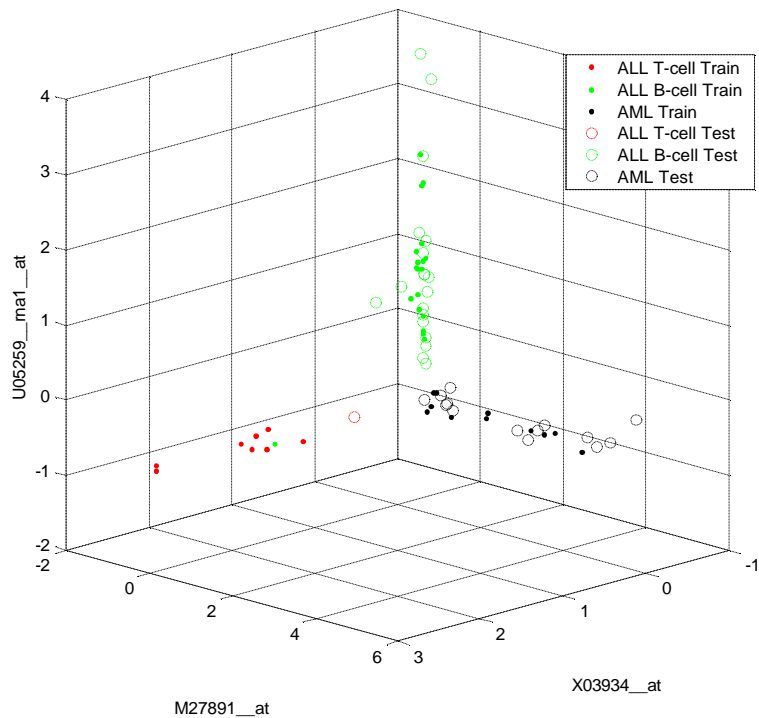

Figure 4.5: Plot of the leukemia data samples according to the expression values of the three genes selected by the hybrid LIK+RFE.

Since there were only three genes selected by the hybrid LIK+RFE method, we are able to visualize the distribution of both the training and test samples in a three-dimensional space. Figure 4.5 shows the plot of the samples. In this figure, we differentiate between acute



myeloid leukemia (AML) and acute lymphoblastic leukemia (ALL) samples. There were actually two different types of ALL samples. These were B-cells or T-cells as determined by whether they arose from a B or a T cell lineage (Keller et al., 2000). From the figure, we can see that all except one B-cell sample had almost constant expression values for two genes, namely M27891_at and X02934_at. Training sample number 17 was the one ALL B-cell that was an outlier. On the other hand, all T-cell samples had almost constant expression values for genes U05259_mal_at and M27891_at, while all AML samples had similar expression values for U05259_mal_at and X03934_at. The plot shows that the three selected genes were also useful in differentiating ALL B-cell and T-cell samples.

For comparison purposes, the classification performance of SVM and Naïve Bayesian classifiers built using genes selected according to their LIK scores only is shown in Figure 4.6. For the results shown in this figure, we started with the same set of 40 genes and removed one gene at a time according to their LIK scores. As can be seen from the figure, the results were not as good as those shown in Figure 4.4. In particular, using SVM classifiers, the accuracy and the acceptance rate were more than 80 percent when there were still more than 20 genes in the model. The acceptance rate drops drastically when there are fewer genes. Naïve Bayesian classifiers performed well when there were more than 21 genes. Further removal of more genes according to their LIK scores caused the acceptance rate to drop considerably. When there were fewer than five genes, the accuracy and the acceptance rate of the classifiers were low.



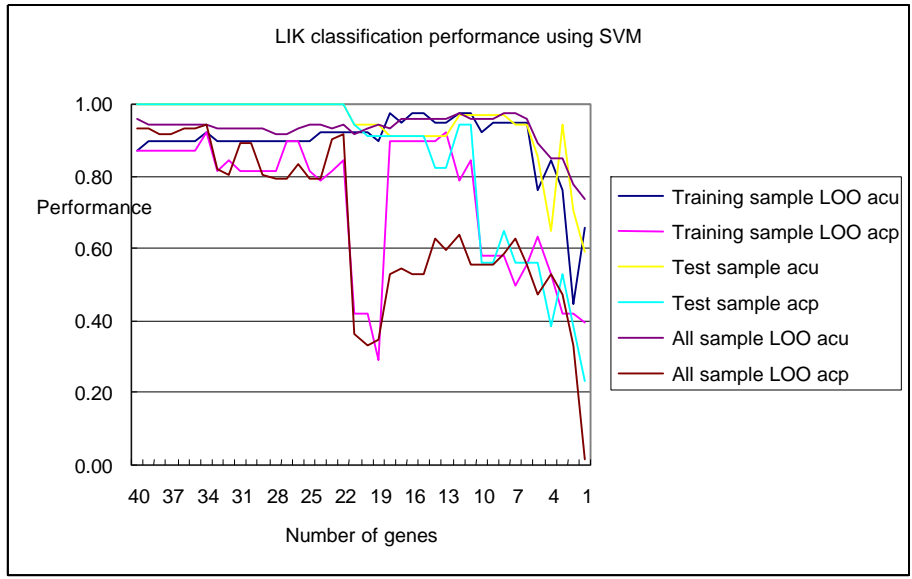

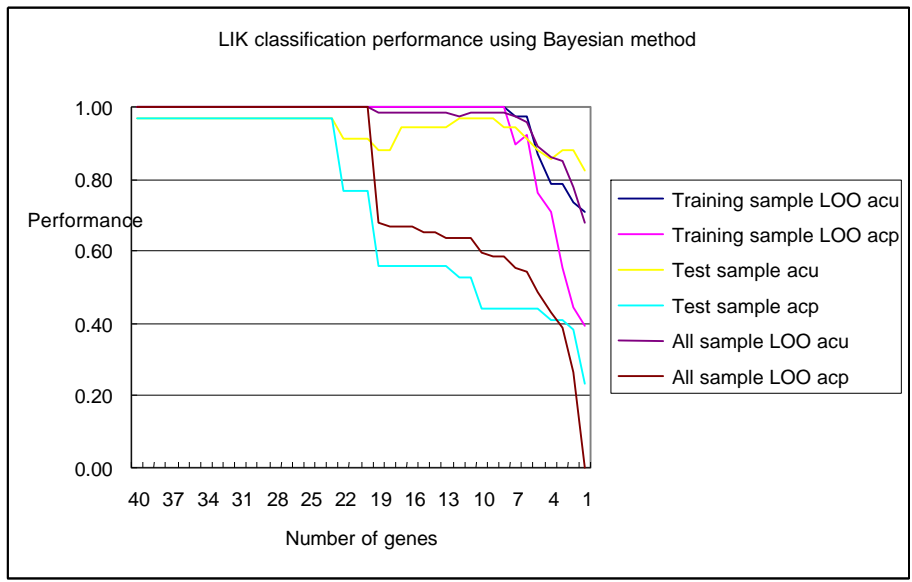

Figure 4.6: Performance of SVM and Naïve Bayesian classifiers built using genes selected according to LIK scores.

The performance of SVM and Naïve Bayesian classifiers built using the genes selected by RFE is depicted in Figure 4.7. We started with all 7129 genes in the feature set. We built an



SVM using the training samples with expression values of all the genes and measured its performance on the test samples. We also built a Naïve Bayesian classifier and measured its performance as well. The gene that had the smallest absolute weight in the SVM-constructed hyperplane was removed, and the process of training and testing was repeated with one fewer gene. This process was continued until there were no more genes to be removed.

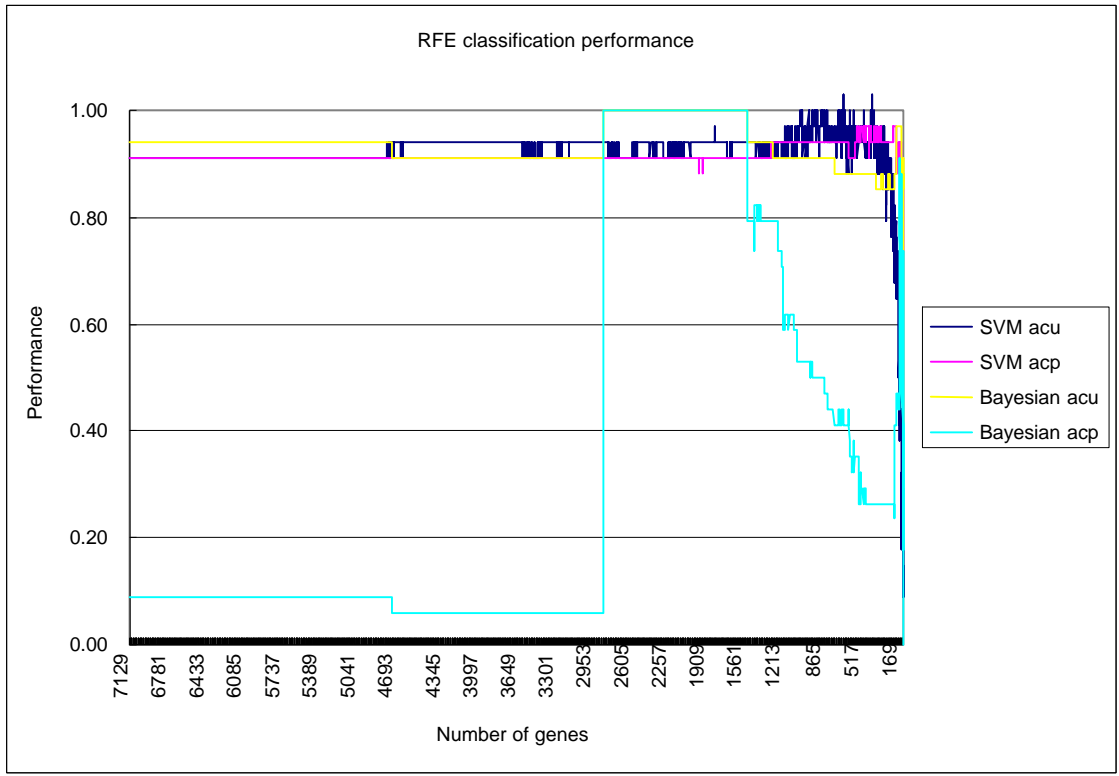

Figure 4.7: Classification performance of purely using RFE. Classification performance of SVM and Naïve Bayesian classifiers using genes selected by RFE starting from 7129 genes down to only one gene. The experimental setting was training test split and the performance measures were shown on the 34 test samples.



An interesting point to note from the results depicted in Figure 4.7 is the sharp improvement in the acceptance rate of the Bayesian classifiers when the number of genes was reduced from 2773 to 2772. The gene that was eliminated at this stage was M26602_at. The acceptance rates stayed at 100 percent when there were 2772 to 1437 genes. Further removal of genes caused the rate to deteriorate gradually. On the other hand, the performance of SVM was more stable. With more than 519 genes, both the accuracy and the acceptance rate were at least 90 percent.

We also experimented with choosing the top genes according to their LIK scores. We selected genes with LIK scores that were higher than a certain threshold. The threshold values tested were 1500, 2000, and 2500. Note that there were always more genes selected because of their high $LIK_{ALL \rightarrow AML}$ than genes selected because of their $LIK_{AML \rightarrow ALL}$ values. The best performance was obtained when the threshold was set to 1500. A total of 62 genes met this threshold value and were used to form the initial gene set for RFE. After applying RFE, we obtained a set of four genes that achieves perfect accuracy and acceptance rate on the training and test samples under all three experimental settings. The set of four selected genes is shown in Table 4.13. Two out of the four genes were the same ones as those selected using the (2 x 20) top initial genes listed in Table 4.12. These genes were U0529_rnal_at and M27891_at.  The genes M16336_s_at was also found by Keller at al. (2000) to be an important gene for classification.

The performance of the SVM classifiers with genes selected using just the RFE approach was slightly different from that reported by Guyon et al. (2002). The reason for this could be the variation in the implementation of the quadratic programming solvers. The Matlab toolbox uses Sequential Minimal Optimisation algorithm (Platt, 1999), while Guyon et al. used a



variant of the soft-margin algorithm for SVM training (Cortes, 1995). Our hybrid LIK+RFE method achieved better performance than other methods reported in the literature. To achieve perfect performance, the RFE implementation of Guyon et al. needed eight genes. When the number of genes was reduced to four, the leave-one-out results on the training samples using SVM achieved only 97 percent accuracy and 97 percent acceptance rate. SVMs trained on 38 training samples with the four selected genes achieved only 91 percent accuracy and 82 percent acceptance rate on the test samples.

| Gene assection number | Description |
|---|---|
| U05259_rna1_at | MB-1 gene |
| M16336_s_at | CD2 CD2 antigen (p50), sheep red blood cell receptor |
| M27891_at | CST3 Cystatin C (amyloid angiopathy and cerebral hemorrhage) |
| X58072_at | GATA3 GATA-binding protein 3 |

Table 4.13: The genes selected by the hybrid LIK+RFE method. The genes that have LIK scores of at least 1500 were selected initially. RFE was then applied to select these four genes that achieved perfect performance.

Using the genes selected according to their LIK scores and applying the Bayesian method, Keller et al. (2000) achieved 100 percent prediction with more than 150 genes. Hellem and Jonassen (2002) required 20 to 30 genes to obtain accurate prediction by ranking pair-wise contribution of genes to the classification. The classification of the samples was obtained by applying k-nearest neighbours, diagonal linear discriminant and Fisher's linear discriminant methods. Guyon et al. also mentioned the performance of other works on this dataset (Mukherjee et al., 2000; Chapelle et al., 2000; Weston et al., 2001). None of these works reported performance results that are as good as ours.



Besides LIK, we also tried to combine other univariate feature ranking methods with RFE. They are baseline criterion proposed by Golub et al. (1999), Fisher's criterion, and Extremal margin (Guyon et al., 2002). In the way that is similar to that in LIK+RFE, we choose the top 40 genes ranked by these three methods, then let RFE do further feature set reduction. Figure 4.8 shows the accuracy on test samples when running RFE starting from the initial feature set selected by the three univariate methods.

It can be seen from the Figure 4.8, the prediction accuracy of baseline and Fisher's criteria on test samples are similar. In the elimination process, the accuracy were kept above 80% for SVM prediction and 90% for Bayesian prediction, as long as there were more than 5 genes remaining in the set. The prediction accuracy dropped drastically when there were less than 5 genes remaining in the set. The RFE starting from genes selected by Extremal margin ranking performed better than the other two ranking methods especially in Bayesian prediction accuracy, which remained no less than 97% until there was one gene left in the gene set. Particularly, the perfect prediction was achieved when there are 5 genes left in the dataset.



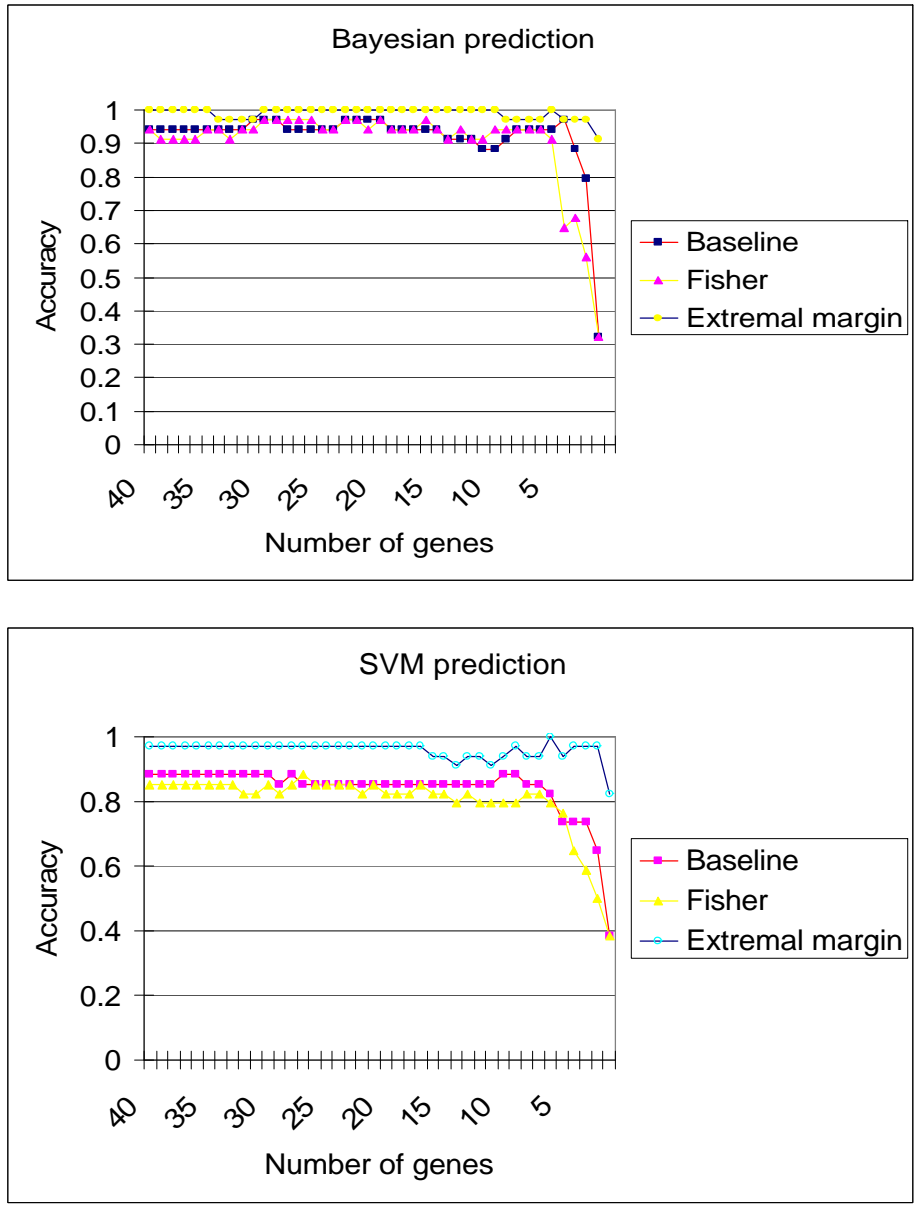

Figure 4.8: SVM and Bayesian prediction accuracy on test samples using features selected by RFE combined with baseline criterion, Fisher's criterion and Extremal margin method.



## 4.1.6.2 Small, round blue cell tumors dataset

The second dataset to test LIK+RFE is from small, round blue cell tumors (SRBCTs) (Khan et al., 2001). There are 88 samples altogether with 2308 genes, divided into 4 classes: *neuroblastoma* (NB), *rhabdomyosarcoma* (RMS), *Burkitt lymphomas* (BL) and the *Ewing family of tumors* (EWS). Because our focus is on binary classification, we decomposed the problem into 4 one-against-rest binary second type classification problems. In (Khan et al., 2001) there are 63 training samples, which consist of 23 EWS, 8 BL, 12 NB, and 20 RMS samples. There are 25 test samples, which consist of 6 EWS, 3 BL, 6 NB, 5 RMS samples, and non-SRBCTs samples. After testing the classification performance of individual problems, we then performed test on the prediction combining all the classifiers together to predict the samples that is not belong to these four classes.

We obtained the expression ratio data of Khan et al. (2001). Before we conducted our experiments, the expression values were transformed by computing their logarithmic values. Base 2 log transformation was used, as this is the usual practice employed by researchers analyzing microarray data. In Figure 4.9, the plot of the gene ranking according to their LIK scores is shown. The LIK scores were computed for differentiating EWS samples from non-EWS samples. The set of top 20 genes according to their $LIK_{EWS \rightarrow Non-EWS}$ ranking contained eight genes that were also in the set of top 20 genes according to their $LIK_{Non-EWS \rightarrow EWS}$ ranking. Hence, when RFE was applied to further eliminate genes from the feature set, it started with 32 unique genes.



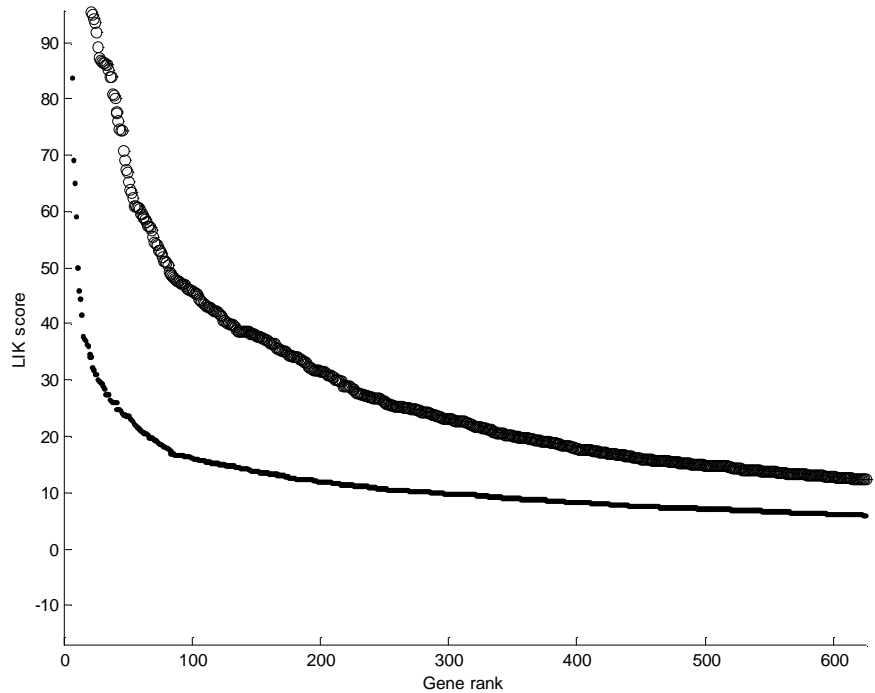

Figure 4.9: Sorted LIK scores of genes in the SRBCT dataset. Dots indicate $LIK_{EWS \to Non-EWS}$ scores and circles indicate $LIK_{Non-EWS \to EWS}$ scores.

For the other three classification problems, the plots would look very similar to Figure 4.9 and are not shown in this paper. For each of the four problems, LIK selected the top (2 x 20) genes. The number of unique genes selected by LIK and the results of the experiments from solving four binary classification problems are summarized in Table 4.14. The numbers of unique genes selected by LIK and the smallest numbers of genes required to achieve near perfect performance during the gene elimination process by RFE are shown in the second column of the table. For the three classification problems to identify EWS, BL and NB, the accuracy and the acceptance rates were at least 98 percent for all experimental settings. Those perfect performance results are highlighted in the table. For the fourth classification problem



to differentiate between RMS and non-RMS samples, the accuracy rates were at least 92 percent. However, the acceptance rate on the test samples dropped to eight percent for SVM classifier and 16 percent for Naïve Bayesian classifier, respectively.

| Classification problem | Initial/final number of genes | SVM | | | | | | Bayesian | | | | | |
|---|---|---|---|---|---|---|---|---|---|---|---|---|---|
| | | Leave-one-out on training samples | | Prediction on test samples | | Leave-one-out on all samples | | Leave-one-out on training samples | | Prediction on test samples | | Leave-one-out on all samples | |
| | | Acu | Acp | Acu | Acp | Acu | Acp | Acu | Acp | Acu | Acp | Acu | Acp |
| EWS vs non-EWS | 32/5 | 1.00 | 1.00 | 1.00 | 1.00 | 0.99 | 0.99 | **1.00** | **1.00** | **1.00** | **1.00** | **1.00** | **1.00** |
| BL vs non-BL | 37/3 | **1.00** | **1.00** | **1.00** | **1.00** | **1.00** | **1.00** | **1.00** | **1.00** | **1.00** | **1.00** | **1.00** | **1.00** |
| NB vs non-NB | 34/3 | **1.00** | **1.00** | **1.00** | **1.00** | **1.00** | **1.00** | 0.98 | 0.98 | 1.00 | 1.00 | 1.00 | 1.00 |
| RMS vs non-RMS | 34/4 | 1.00 | 1.00 | **0.92** | **0.08** | 0.99 | 0.88 | 1.00 | 1.00 | **0.92** | **0.16** | 0.97 | 0.35 |

Table 4.14: Experimental results for the SRBCT dataset using the hybrid LIK+RFE.

The poor acceptance rate obtained when predicting RMS test samples suggests that the differences in the output of the classifiers and the actual target values were high for the incorrectly predicted samples. In order to verify the predictions, we plotted the distribution of the samples according to the expression values of three genes, ImageID784224 (fibroblast growth factor receptor 4), ImageID796258 (sarcoglycan, alpha), and ImageID1409509 (troponin T1). The three were selected because their corresponding SVM weights were the largest. The plot is shown in Figure 4.10. We can clearly see that the two incorrectly classified non-RMS samples were outliers with large values for ImageID1409509 (troponin T1). These two outliers were Sk. Muscle samples TEST-9 and TEST-13, which were misclassified as RMS samples. It should be noted that there were no Sk. Muscle samples in the training dataset.



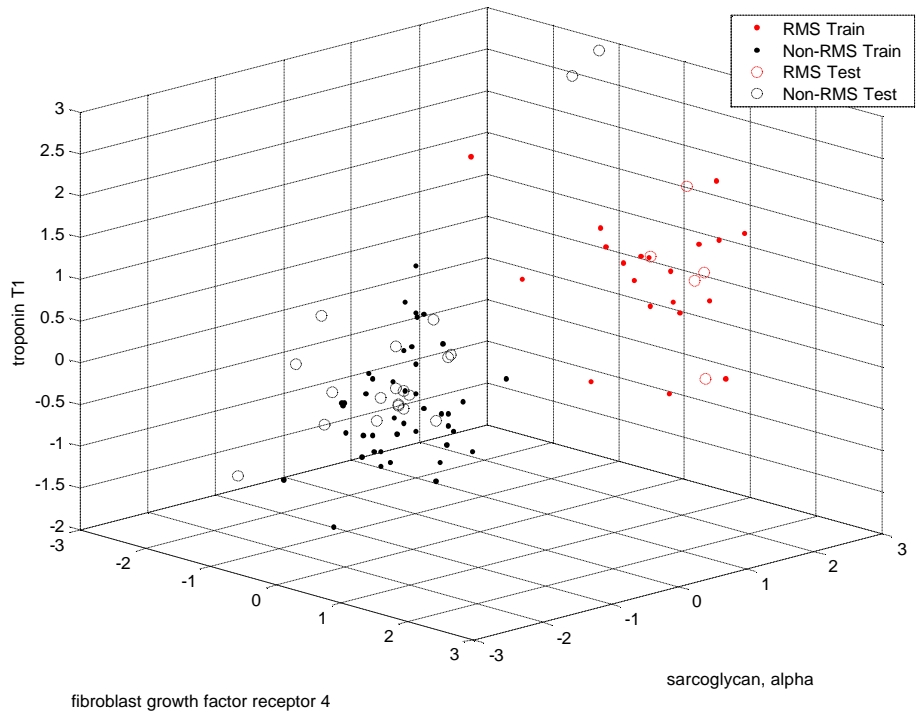

Figure 4.10: Plot of RMS and non-RMS samples. Plot of all 88 RMS and non-RMS samples according to the expression values of three of the four selected genes.

The genes selected by the hybrid LIK+RFE for each of the four classification problems are listed in Table 4.15. For the problem of differentiating EWS from non-EWS samples, our method selected five genes, all of which were also selected by Khan et al. (2001). On the other hand, to differentiate between NB and non-NB samples, only three genes were needed and none was selected by Khan et al. All together, the hybrid LIK+RFE identified 15 important genes. This number compares favorably with the total of 96 genes selected by the PCA (Principle Component Analysis) approach of Khan et al.



| Classification problem | Reported by Khan et al. (2001) | Image ID | Description |
|---|---|---|---|
| EWS vs non-EWS | Y | 377461 | caveolin 1, caveolae protein, 22kD |
| | Y | 295985 | ESTs |
| | Y | 80338 | selenium binding protein 1 |
| | Y | 52076 | olfactomedinrelated ER localized protein |
| | Y | 814260 | follicular lymphoma variant translocation 1 |
| BL vs non-BL | Y | 204545 | ESTs |
| | | 897164 | catenin (cadherin-associated protein), alpha 1 (102kD) |
| | Y | 241412 | E74-like factor 1 (ets domain transcription factor) |
| NB vs non-NB | | 45632 | glycogen synthase 1 (muscle) |
| | | 768246 | glucose-6-phosphate dehydrogenase |
| | | 810057 | cold shock domain protein A |
| RMS vs non_RMS | | 897177 | phosphoglycerate mutase 1 (brain) |
| | Y | 784224 | fibroblast growth factor receptor 4 |
| | Y | 796258 | sarcoglycan, alpha (50kD dystrophin-associated glycoprotein) |
| | Y | 1409509 | troponin T1, skeletal, slow |

Table 4.15: The genes selected by the hybrid LIK+RFE for the four binary classification problems.

We also tested the classification performance of SVM and Naïve Bayesian classifiers on genes selected based purely on their LIK scores. For comparison purpose, for each of the four problems, the number of genes was set to be the same as the corresponding final number selected by the hybrid LIK+RFE shown in Table 4.14. Table 4.16 summarizes the results. For three of the four classification problems, the performance of the classifiers was not as good as the results reported in Table 4.14. The accuracy and acceptance rates dropped to as low as 52 percent. The most unexpected results came from the fourth problem to differentiate between RMS and non-RMS samples. The SVM classifier achieved perfect accuracy and acceptance rate using four genes, while the Naïve Bayesian classifier managed to obtain at least 92 percent accuracy and acceptance rate. The four genes were ImageID461425 (MLY4), ImageID784224 (fibroblast growth factor receptor 4), ImageID296448 (insulin-like growth



factor 2), and ImageID207274 (Human DNA for insulin-like growth factor II). All these genes were among the 96 genes identified by Khan et al. (2001). Of these four, only one was selected by LIK+RFE, that is, ImageID784224.

| | | SVM | | | | | | Bayesian | | | | | |
|---|---|---|---|---|---|---|---|---|---|---|---|---|---|
| | | Leave-one-out on training samples | | Prediction on test samples | | Leave-one-out on all samples | | Leave-one-out on training samples | | Prediction on test samples | | Leave-one-out on all samples | |
| Classification problem | Number of genes | Acu | Acp | Acu | Acp | Acu | Acp | Acu | Acp | Acu | Acp | Acu | Acp |
| EWS vs non-EWS | 5 | 1.00 | 1.00 | 0.92 | 0.88 | 0.95 | 0.88 | 0.98 | 0.97 | 0.84 | 0.84 | 0.95 | 0.86 |
| BL vs non-BL | 3 | 0.95 | 0.92 | 0.88 | 0.88 | 0.97 | 0.83 | 0.98 | 0.98 | 0.88 | 0.76 | 0.93 | 0.88 |
| NB vs non-NB | 3 | 0.95 | 0.92 | 0.84 | 0.76 | 0.97 | 0.86 | 0.97 | 0.97 | 0.80 | 0.52 | 0.95 | 0.92 |
| RMS vs non-RMS | 4 | **1.00** | **1.00** | **1.00** | **1.00** | **1.00** | **1.00** | 0.97 | 0.95 | 0.92 | 0.92 | 0.97 | 0.95 |

Table 4.16: The performance of SVM and Naïve Bayesian classifiers built using the top genes selected according to their LIK scores.

In comparison, Khan et al. (2001) used neural networks for multiple classifications to achieve 93 percent EWS, 96 percent RMS, 100 percent BL and 100 percent NB diagnostic classification performance on the 88 training and test samples. Since there were four classes of training data samples, each neural network had four output units. The target outputs were binary encoded, for example, for an EWS sample the target was (EWS=1, RMS=NB=BL=0). A total of 3750 neural networks calibrated with 96 genes were required. The highest average output value from all neural networks determined the predicted class of a new sample. The Euclidean distance between the average values and the target values was computed for all samples in order to derive the probability distribution of the distances. A test sample would be diagnosed as a member of one of the four classes based on the highest average value given by the neural networks. This was provided that the distance value falls within the 95[th]



percentile of the corresponding distance probability of the predicted class. Otherwise, the diagnosis would be rejected and the sample would be classified as a non-SRBCT sample. Of the 88 samples in the training and test datasets, eight were rejected. Five of these were non-SRBCT samples in the test set, while the other three actually belonged to the correct class but their distances lay outside the threshold of the 95[th] percentile.

In order to visualize the distribution of the samples based on the expression values of the selected genes, we performed clustering of the genes using the EPCLUST program (http://ep.ebi.ac.uk/EP/EPCLUST). The default setting of the program was adopted; the average linkage clustering and uncentered correlation distance measure were used. Figure 4.11 shows the clusters. It can be seen clearly from this figure that there existed four distinct clusters corresponding to the four classes in the data. Most of the samples of a class fell into their own corresponding clusters. The five non-SRBCT samples lay between clusters. We conjecture that samples between clusters might not belong to any classes found in the training dataset. Two between-cluster samples, RMS-T7 and TEST-20 were exceptions. RMS-T7, which was nearer to the two Sk. Muscle samples TEST-9 and TEST-13 was actually an RMS sample. TEST-20, which was nearer to Prostate sample TEST-11 than to EWS cluster was actually an EWS sample. These exceptions were consistent with the neural network prediction results of Khan et al. (2001) as the neural networks predicted TEST-9 and TEST-13 to be RMS class, and they predicted TEST-20 and TEST-11 to be EWS class. Both predictions, however, did not meet the 95[th] percentile distance criterion and were therefore rejected. This indicated that these samples were also difficult to differentiate by the neural networks. Different results from our clustering and the neural network classification can be seen for test sample TEST-3, a non-SRBCT sample. The clustering placed TEST-3 between



BL and NB clusters. But the neural networks predicted this sample as an RMS sample without meeting the 95[th] percentile distance criterion.

Most of the genes selected from Leukemia and SRBCT datasets by our hybrid LIK + RFE method have some relevance to cancer according to literature search in PubMed, a document retrieval service of the National Library of Medicine of United States. However, biological experiments need to be done for further validation of the role of these genes. The performance of the method is also data dependent, as demonstrated in the significant difference in the acceptance rate of the classifiers for the first three binary classification problems and the fourth problem in the SRBCT dataset. Overall, we observe that the classification performance on the test set generally does not change much with the consecutive elimination of a few genes. The removal of one gene would not normally cause a drastic change in the performance of the classifier. Significant drops in accuracy and/or the acceptance rate are observed most frequently when a gene is removed from the optimal set.



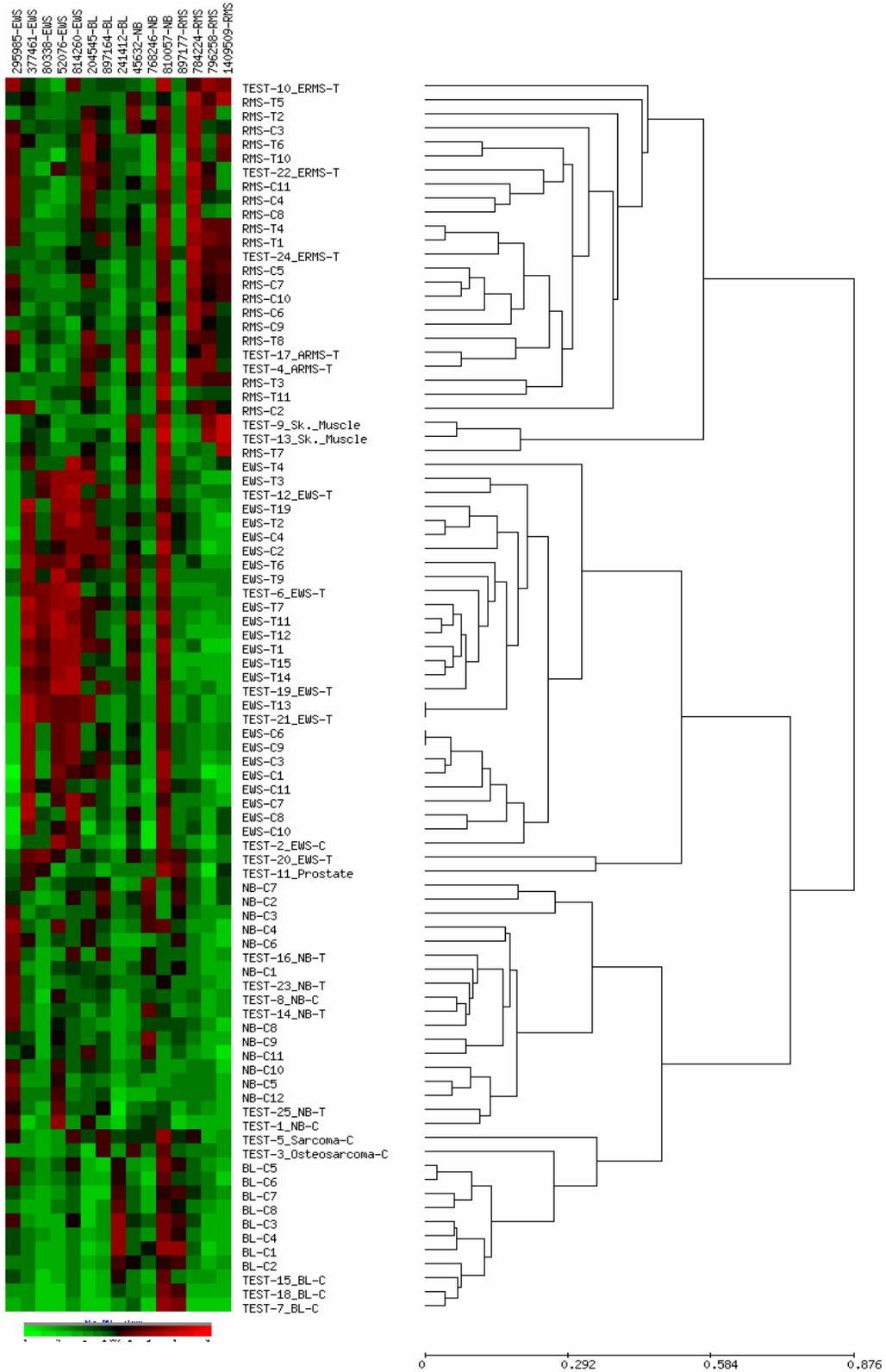

Figure 4.11: Hierarchical clustering of SRBCT samples with selected 15 genes.



### 4.1.6.3 Artificial datasets

In order to study how expression values of irrelevant genes affect the selection performance on RFE, we generated three types of artificial datasets. Each of the datasets consists of 40 training samples (20 positive class + 20 negative class) and 40 test samples (also 20 positive class + 20 negative class). All datasets consist of features that are relevant and irrelevant to the classification. The values of the irrelevant features are sampled from a normal distribution with standard deviation 1 and mean 0 regardless of the class labels of the samples. The three types of datasets are different in the way the relevant features are constructed. For the first type, the relevant features contribute independently to classification. Their values are normally distributed with standard deviation 1, and mean $x$ or $-x$ according to the class labels. For different relevant features, $x$ is randomly chosen from a uniform distribution on the interval $(0,1)$. If $x$ is big enough, univariate feature selection method is likely to be able to select the relevant features. The relevant features of the second type dataset is constructed by considering the joint effect of the features: suppose there are $k$ relevant features $X_1, \ldots, X_k$, where $X_1, \ldots X_{k-1}$ are sampled from normal distribution in the same way as that from irrelevant features. We set the remaining relevant feature $X_k = \sum_{i=1}^{k-1} X_i \pm \boldsymbol{a}$, where $\boldsymbol{a}$ is randomly sampled from a normal distribution of standard deviation 0 and mean 1, and $\boldsymbol{a}$ is added or subtracted from the sum depending on the class labels of samples. It is expected to be harder for a univariate method to select the first $k-1$ relevant features. The third type of datasets is constructed by setting the first $k-1$ relevant features the same way as that in first type datasets. But the $k$ th feature is set in the same way as that in the second type datasets. This type of datasets models the expression of the genes that related to cancer better, where the genes have certain degrees of individual contributions to cancer from observation, but they interact with other genes together to cause the disease.



We tried LIK, RFE and LIK+RFE on the datasets of all the three types above, each contained a total of 100 features, 4 of which are relevant features. In each experiment, we let the selection method select the features based on the training samples, then test the classification performance on the testing samples. For testing RFE, one feature is eliminated each time. For testing LIK, we include the same number of top ranking features of both $LIK_{a \rightarrow b}^{g}$ and $LIK_{b \rightarrow a}^{g}$ ranks. For testing the combination of LIK+RFE, we choose the top 5+5 and the top 10+10 LIK ranked features to input into RFE. We use SVM to test the classification performance. The setting of SVM for RFE and classification is same as those in Section 4.1.6.1. Before applying the methods, the datasets were normalized in the same way the Leukemia dataset. For each setting, 100 experiments were conducted, each on a different dataset generated, and the mean and standard deviation of the test result are shown in Table 4.17. The table shows the type of the datasets; the number of LIK ranked features to be used as input for RFE (lik_num), within which the number of distinct ones (num_lik_rfe); the accuracy of SVM prediction on datasets using all and relevant features (acc_all, acc_rev respectively), where acc_rev, is used to find out what is the ideal prediction performance; the corresponding number of support vectors of trained SVMs (num_sv_all and num_sv_rev respectively); for running LIK, RFE, and LIK+RFE, the highest accuracy obtained (max_rfe_acc, max_lik_acc, and max_lik_rfe_acc respectively), the smallest number of features to achieve highest accuracy (num_rfe_acc, num_lik_acc, and num_lik_rfe_acc respectively); the number of relevant features existing in selected features (num_rfe_acc_rev, num_lik_acc_rev, and num_lik_rfe_acc_rev respectively); and the percentage of times of the last relevant feature existing in these features (contain_rfe_acc_last_rev, contain_lik_acc_last_rev, and contain_lik_rfe_acc_last_rev respectively).



| lik_num | 10 | | | 20 | | |
|---|---|---|---|---|---|---|
| dataset_type | 1 | 2 | 3 | 1 | 2 | 3 |
| | | | | | | |
| num_lik_rfe | 8.57 ± 0.97 | 9.01 ± 1.00 | 8.40 ± 0.83 | 14.66 ± 1.34 | 14.97 ± 1.47 | 14.86 ± 1.20 |
| num_lik_sel | 2.71 ± 0.87 | 1.05 ± 0.59 | 2.91 ± 0.71 | 2.79 ± 0.95 | 1.24 ± 0.62 | 3.22 ± 0.66 |
| | | | | | | |
| acc_all | 0.66 ± 0.09 | 0.55 ± 0.08 | 0.71 ± 0.09 | 0.65 ± 0.11 | 0.55 ± 0.08 | 0.71 ± 0.08 |
| num_sv_all | 35.25 ± 1.56 | 36.07 ± 1.61 | 34.39 ± 1.61 | 35.14 ± 1.92 | 36.16 ± 1.64 | 34.13 ± 1.94 |
| acc_rev | 0.83 ± 0.08 | 0.81 ± 0.07 | 0.88 ± 0.06 | 0.82 ± 0.09 | 0.81 ± 0.06 | 0.89 ± 0.06 |
| num_sv_rev | 13.15 ± 6.45 | 14.87 ± 4.47 | 8.76 ± 4.21 | 12.79 ± 6.66 | 15.08 ± 4.68 | 8.29 ± 3.75 |
| | | | | | | |
| max_rfe_acc | 0.82 ± 0.08 | 0.69 ± 0.07 | 0.89 ± 0.07 | 0.82 ± 0.09 | 0.69 ± 0.07 | 0.89 ± 0.07 |
| num_rfe_acc | 9.79 ± 17.15 | 13.38 ± 17.62 | 3.51 ± 7.06 | 7.20 ± 11.85 | 11.75 ± 17.55 | 3.98 ± 8.70 |
| num_rfe_acc_rev | 2.19 ± 1.01 | 1.39 ± 0.75 | 1.29 ± 0.62 | 1.90 ± 0.89 | 1.39 ± 0.82 | 1.29 ± 0.61 |
| contain_rfe_acc_last_rev | 0.53 ± 0.50 | 0.90 ± 0.30 | 0.97 ± 0.17 | 0.50 ± 0.50 | 0.90 ± 0.30 | 0.94 ± 0.24 |
| | | | | | | |
| max_lik_acc | 0.84 ± 0.08 | 0.70 ± 0.07 | 0.91 ± 0.06 | 0.84 ± 0.08 | 0.70 ± 0.07 | 0.90 ± 0.06 |
| num_lik_acc | 10.20 ± 17.07 | 24.26 ± 27.35 | 5.20 ± 11.70 | 6.97 ± 11.14 | 18.02 ± 20.83 | 3.76 ± 5.94 |
| num_lik_acc_rev | 2.45 ± 1.00 | 1.66 ± 1.06 | 1.81 ± 0.95 | 2.08 ± 0.97 | 1.40 ± 0.90 | 1.81 ± 0.97 |
| contain_lik_acc_last_rev | 0.63 ± 0.49 | 0.94 ± 0.24 | 1.00 ± 0.00 | 0.52 ± 0.50 | 0.92 ± 0.27 | 1.00 ± 0.00 |
| | | | | | | |
| max_lik_rfe_acc | 0.81 ± 0.09 | 0.66 ± 0.09 | 0.89 ± 0.07 | 0.82 ± 0.09 | 0.66 ± 0.09 | 0.88 ± 0.07 |
| num_lik_rfe_acc | 3.17 ± 2.06 | 3.14 ± 2.56 | 1.83 ± 1.48 | 3.92 ± 3.20 | 4.24 ± 3.94 | 2.41 ± 3.15 |
| num_lik_rfe_acc_rev | 1.96 ± 0.97 | 0.84 ± 0.55 | 1.25 ± 0.58 | 1.91 ± 0.91 | 0.85 ± 0.61 | 1.38 ± 0.76 |
| contain_lik_rfe_acc_last_rev | 0.51 ± 0.50 | 0.72 ± 0.45 | 0.96 ± 0.20 | 0.41 ± 0.49 | 0.70 ± 0.46 | 0.92 ± 0.27 |

Table 4.17: Test result of LIK, RFE and LIK+RFE on artificial datasets

It can be seen from Table 4.17, as expected, in terms of the number of feature selected and the accuracy (rows max_*_acc and num_*_acc), for datasets of type 1 and 3, RFE is similar to LIK; for datasets of type 2, RFE is significantly better than LIK. But the testing results from all three types of datasets indicate that LIK+RFE selected significantly more accurate feature sets, in terms of the number of relevant features over the number of selected features, without significant loss of prediction performance (rows max_*_acc, num_*_acc and num_*_acc_rev).

It can also be seen from the table that when the number of irrelevant features is large compared with the number of relevant features, the classification of SVM becomes bad;



almost all training samples became support vectors (row num_sv_all compared with row num_sv_rev). This phenomenon also occurs when SVM is applied on the second type datasets like Leukemia and SRBCT dataset. From our experience, the number of support vectors reduces significantly only when the number of features for training is near or less than the number of training samples. We suspect the phenomenon is related to the learning capacity of SVM but we have not found theoretical basis for this.

The assumption of single or double normal distribution on irrelevant and relevant features are simple and may not accurately reflect the real situation in microarray datasets, especially, the combinatorial effect of features are not modeled. However, as can be seen from the test result from these simple datasets, it is quite likely that the weights of trained SVM on a mixture of many irrelevant features and few relevant features are unable to truly measure the contribution of the features to the classification. Some relevant features are incorrectly eliminated by RFE starting from all features. By comparison, LIK ranking is able to keep most of the relevant features for type 1 and 3 datasets, which enables RFE to perform further selection more effectively.

### 4.1.7 The combination of Likelihood method and Fisher's method

We also compared LIK+RFE with the combination of univariate and multivariate versions of Likelihood method and Fisher's method. In our experiment, a set of features was first selected by a univariate method, in the same fashion as in the experiment for LIK+RFE. A multivariate method was then used to eliminate the features recursively from that feature set. We tested the method on the Leukemia dataset and set the size of the initial set to 30 as Fisher's linear discriminant encounters matrix inversion problems if the initial gene set size is



bigger than the number of training samples. The algorithm was implemented and run on Matlab 6.1.

Figure 4.12 shows that when using the combination of F_F, L_L and F_L, the accuracy of Bayesian classification on test samples was generally better than that of SVM prediction, based on same set of genes. But the SVM prediction of L_F is better that that of Bayesian method. In both SVM and Bayesian test results, L_L outperformed the other three combinations when there are more than 15 genes remaining in the gene set. However, its performance dropped drastically when there are less than four genes remaining in the gene set. Under this situation, the combination of F_F is the best. However, none of these combinations of these four methods outperformed the combination of LIK+RFE on our test on Leukemia dataset in terms of classification accuracy based on the same number of genes.



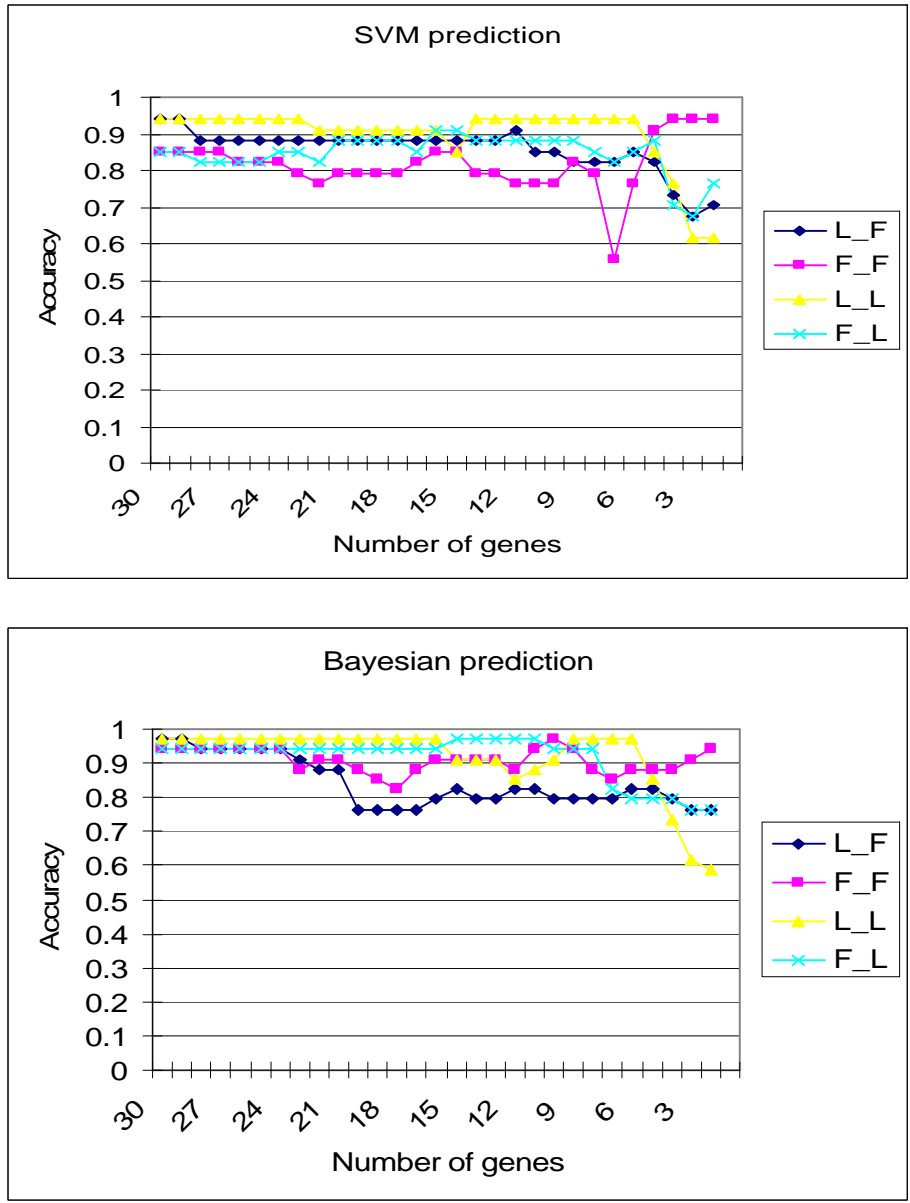

Figure 4.12: SVM and Bayesian prediction accuracy when running combination of univariate and multivariate feature selection methods. L_F: Likelihood + Fisher's linear discriminator; F_F: Fisher's criterion + Fisher's linear discriminator; L_L: Likelihood + Multivariate Likelihood Method; F_L: Fisher's criterion + Multivariate Likelihood Method.



## 4.2  First type - low dimension problems

The dataset we used that consists of the low dimension analysis problem is a Zebra fish developmental microarray dataset obtained from Lab of Functional Genomics of Institute of Molecular and Cell Biology, Singapore (Lo et al., 2003). In recent years, Zebra fish has been adopted as a model system for the studies of vertebrate development owing to some of its unique characteristics favorable for genetic studies compared to other vertebrate systems. These characteristics include reasonably short lifetime, large number of progenies, external fertilization and embryonic development, and translucent embryos (Talbot and Hopkins, 2000). In the microarray experiment, there were altogether 11,552 Expression Sequence Tag (EST) clones representing 3100 genes printed onto the microarray glass slides. According to BLAST (Basic Local Alignment Search Tool) search, 4519 of the 11,552 clones have matches to 728 distinct publicly deposited protein sequences. That is, the functions of these 4519 clones are known, and the functions of the remaining clones are unknown. The relative expression of the 11,552 clones in Zebra fishes' six developmental stages, including cleavage (E2), gastrula (E3), blastula (E4), segmentation (E5), pharyngula (E6) and hatching (E7), based on their developmental morphology, was monitored using microarray experiments, in comparison to the expression of these clones in stage unfertilized eggs (E0). A first type classification problem was constructed, which included 11,449 samples from the 11,552 clones with 6 features. A total 3887 of the 11,449 samples corresponding to the known clones were labeled according to whether they are muscle genes or not. Within these 3887 clones, 248 were clones from 17 muscle genes. We did the classification by employing SVM. The labeled clones were randomly split into two sets, 2500 for training and remaining 1387 for testing. There were 157 and 91 positive samples in the training and testing sets respectively. The remaining 7562 unlabelled samples were then used to perform prediction.



| $\log_{10} d$ | $C$ | Number of Support Vectors | True positive | False positive | True negative | False negative | Predicted positive |
|---|---|---|---|---|---|---|---|
| -3.0 | 10 | 312 | 34 | 9 | 1287 | 57 | 88 |
| -2.5 | 10 | 291 | 42 | 9 | 1287 | 49 | 100 |
| **-2.0** | **10** | **271** | **44** | **10** | **1286** | **47** | **110** |
| -1.5 | 10 | 263 | 41 | 10 | 1286 | 50 | 109 |
| -1.0 | 10 | 271 | 42 | 10 | 1286 | 49 | 104 |
| -0.5 | 10 | 379 | 42 | 14 | 1282 | 49 | 113 |
| 0.0 | 10 | 631 | 31 | 19 | 1277 | 60 | 131 |
| -3.0 | 20 | 300 | 41 | 9 | 1287 | 50 | 101 |
| -2.5 | 20 | 282 | 42 | 9 | 1287 | 49 | 107 |
| -2.0 | 20 | 267 | 42 | 10 | 1286 | 49 | 112 |
| -1.5 | 20 | 251 | 40 | 10 | 1286 | 51 | 111 |
| -1.0 | 20 | 270 | 43 | 12 | 1284 | 48 | 101 |
| -0.5 | 20 | 362 | 39 | 14 | 1282 | 52 | 128 |
| 0.0 | 20 | 611 | 33 | 26 | 1270 | 58 | 176 |
| -3.0 | 50 | 289 | 42 | 10 | 1286 | 49 | 102 |
| -2.5 | 50 | 275 | 43 | 10 | 1286 | 48 | 108 |
| -2.0 | 50 | 260 | 43 | 10 | 1286 | 48 | 112 |
| -1.5 | 50 | 248 | 41 | 9 | 1287 | 50 | 103 |
| -1.0 | 50 | 257 | 42 | 12 | 1284 | 49 | 110 |
| -0.5 | 50 | 322 | 36 | 18 | 1278 | 55 | 159 |
| 0.0 | 50 | 567 | 32 | 29 | 1267 | 59 | 238 |

Table 4.18: Test of SVM with RBF kernel using different parameters.

Table 4.18 shows the test result of the SVM using different parameters with radial basis function kernel. It can be seen that best test performance was obtained when setting $C = 10$ and $s = 0.01$. Using this parameter, the number of correctly predicted positive test samples (true positive) reached 44, which was the highest among all configurations we tried; the number of incorrectly predicted negative (false negative) samples was as low as 10, compared favorably with the lowest false negative we obtained, which is 9. The trained model under this setting is also not complex, which can be seen from the number of support vectors was as low as 271. We provided a list of the 110 positively predicted unknown clones to the biological researchers in the Lab of Functional Genomics of Institute of Molecular and Cell Biology. Ten clones were selected from these 110 clones to do further biological



validation (re-sequencing of these clones) was done. Eight of these ten clones are proven to be from muscle genes. The remaining two were found to be known genes after repeated sequencing. Although these two are not muscle genes, they are functionally related to muscle genes, therefore showed similar expression patterns to that of the other eight clones. Besides these ten clones, another positively predicted clone was re-sequenced, which is a putative novel gene. *In situ* hybridization on this clone showed that the corresponding gene truly had muscle function (Lo et al., 2003).



# 5 Conclusion and future work

## 5.1 Biological knowledge discovery process

Biological research can be treated as a knowledge discovery process, which has been greatly facilitated by the emergence of the field of Bioinformatics. Take microarray data as an example, in the discovery process, biological information is extracted out by biological studies, stored in DNA sequence trace files, scanned microarray images, descriptions of samples and descriptions of experimental conditions. The information can then be quantified or symbolized for biological data with certain structure. The biological data include sequential representations of nucleotides / proteins, gene expression matrices. Computational and statistical methods are then applied to extract biological knowledge from the data. The knowledge is in essence relationships, which could be relationship between genes from their sequential and expression similarity; relationship between gene expression and sample property, experimental condition, intermediate product, or cellular process. Machine learning plays an important role in discovering these relationships. However, the amount of knowledge that can be discovered depends on two factors: the amount and quality of the biological data available, and the suitability of the machine learning methods for various biological problems.

There are more and more researchers who work to accelerate the discovery process. There are currently two main journals and several main conferences in the bioinformatics area: Journal of Bioinformatics, Journal of Computational Biology, Pacific Symposium on Biocomputing (PSB), International Conference on Intelligent Systems for Molecular Biology (ISMB), and Annual International Conference on Research in Computational Molecular Biology



(RECOMB). There is also a conference specifically focus on microarray data analysis, which is Critical Assessment of Microarray Data Analysis (CAMDA).

Although much effort have been made for multidisciplinary collaboration in the discovery process, researchers from non-biology discipline still need to gain more insight of the nature of the biological problems, together with biologists. The real good design of machine learning methods lie in full incorporation of biological knowledge rather than simply abstract the biological problem to fit to well developed models. This criterion dictates the future direction of applying machine learning methods for biological problems.

## 5.2 Contribution, limitation and future work

This thesis focuses on the classification and feature selection problems for gene expression analysis. In the research work, we have reviewed current work in the literature and identified the classification problems. We then applied seven feature selection methods, feature extraction methods and some of their combinations for gene selection, and employed five classification methods for prediction of cancer tissue type and gene function. We improved neural network feature selector to make it more suitable to the gene selection problem for the datasets having high dimension but few samples. We also developed a multivariate version of likelihood feature selection method. We found the hybridization of Likelihood and Recursive Feature Elimination achieved significantly better gene selection performance on the benchmark Leukemia dataset than other methods. The hybridization of LIK+RFE on SRBCT dataset also significantly outperformed a neural network method proposed by other researchers.



The thesis shows a process of understanding of the nature of the problem and choosing suitable methods. We first tested whether the most informative components may contribute to the classification by applying principle component analysis and neural networks on Leukemia dataset. Result showed that those components had little discrimination ability. We then applied decision tree to find sets of rules that were able to classify all the samples. The simplicity of the rules implied the possibility to find small set of genes that have high classification ability. Due to the discrete nature of C4.5 algorithm, the small decision tree generated from training samples, with continuous expression values, did not have good generalization ability; subsequently the prediction accuracy of the tree on test samples was low.

The possibly simple underlying classification model and the deficiency of decision tree method inspired us to use information gain as a gene ranking method, but use neural network as classifier. The classification was improved. After studying of the distribution of neural network prediction errors, we found that, by employing AdaBoost technique to summarize ensemble of neural network classifiers, the classification performance could be improved. We then moved on into looking for methods that could reduce the number of genes used for classification without significant loss of prediction performance. By properly tuning the parameters, the combination of information gain and neural network feature selection achieved that goal.

We then tested other combinations of univariate selection method and multivariate selection method, including the methods that are based on extremal margin, likelihood, and Fisher's criterion. Within these combinations the hybrid of Likelihood method and Recursive Feature Elimination method (LIK+RFE) selected the most compact gene set that had prefect



prediction performance. We did systematic test on this hybrid method using other datasets of the high dimension classification problem. Test results were very promising. Applying the classification methods on the Zebra fish dataset with large number of samples was straightforward. But instead of purely validating the classification performance on known genes, the real prediction of the function of some unknown genes were confirmed by biological experiments.

Our experiments of the hybrid of LIK+RFE on SRBCT dataset showed that the feature selection and classification methods for gene expression analysis are data dependent. Our experiments also showed that, for microarray datasets of high dimension classification problem, the choice of feature selection methods are more important than the choice of classification methods. It is possible to design better selection methods or better combinations of selection methods for Leukemia dataset; and although some method used in this thesis did not achieve high selection performance on Leukemia dataset, they may do well on other datasets.

The study on linear separability (Cover, 1965) suggests that when the number of samples is small compared with the number of features, it is possible to find a number of subsets of features that can perfectly distinguish all samples. Our experiments on the leukemia dataset also support this hypothesis: we found two different gene sets consisting of just three or four genes, which can achieve perfect classification performance. Biological study shows that although many genes do not have direct relevance to the cancer under study, their expression may have subtle and systematic difference in different classes of tissues (Alon et al., 1999). Hence, a new challenge for cancer classification arises: to find as many as possible small subsets of genes that can achieve high classification performance. Using only microarray data



with these subsets of genes, we can build different classifiers and look for those that have desirable properties such as extremal margin, i.e. wide difference between the smallest output of the positive class samples and the largest output of the negative class samples. Another property could be median margin, which is the difference between the median output of the positive class samples and the median output of the negative class samples. Exhaustively enumerating and evaluating all the gene combinations is computationally NP-hard (non-deterministic Polynomial-time hard) and is feasible only when the number of relevant genes is relatively very small.

Due to its cost, microarray experiments conducted for identifying the genes that are crucial for cancer diagnosis are still scarce. The measurements obtained from the experiments are noisy. These facts make the selection of different sets of relevant genes vital. Moreover, cancer is a complex disease. It is not caused by only a few genes, but also by many other factors (Kiberstis and Roberts, 2002). So even the best selected subsets may not actually be the most crucial ones to the cancer under study. They can, however, be important candidates for a further focused study on the gene interactions within individual subsets, and the relationship between these interactions and the disease. There has been work done on the second order selection. For example, Goyun et al. (2002) found a gene pair that could have zero leave-one-out error on the training samples, but achieved poor performance on test samples. Hellem and Jonassen (2002) also evaluated the contribution of pairs of genes to the classification for the ranking of genes, but they still have to combine multiple pairs of genes to perform classification. We plan to work on finding better ways to develop methods for high order feature selection that would allow the classifiers to achieve high performance with different small sets of genes.



# Reference


- Akutsu, T., Miyano, S., and Kuhara, S., (1999). Identification of genetic networks from a small number of gene expression patterns under the Boolean network model. *Pacific Symposium on Biocomputing*, 4, 17-28.

- Aleksander, I., and Morton, H., (1990). *An Introduction to Neural Computing*. Chapman and Hall, London.

- Alon, U., Barkai, N., Notterman, D. A., Gish, K., Ybarra, S., Mack, D. and Levine, A. J. (1999). Broad patterns of gene expression revealed by clustering analysis of tumor and normal colon tissues probed by oligonucleotide arrays. *Proceedings of the National Academy of Sciences*, 96, 6745-6750.

- Alter, O., Brown, P. O., and Botstein, D., (2000). Singular value decomposition for genome-wide expression data processing and modeling. *Proceedings of the National Academy of Sciences*, 97, 10101-10106.

- Becker, S., (1991). Unsupervised learning procedures for neural networks. *International Journal of Neural Systems*, 2, 17-33.

- Bersekas, D. P., (1995). *Nonlinear Programming*. Athenas Scientific.

- Bishop, C.M., (1995). *Neural Networks for Pattern Recognition*. Clarendon Press, Oxford.

- Boser, B., Guyon, I., and Vapnik, V. N., 1992. A training algorithm for optimal margin classifiers. *Fifth Annual Workshop on Computational Learning Theory*. 144-152.

- Brazma, A., and Vilo, J., (2000). Gene expression data analysis. *FEBS Letters*, 480,17-24.





- Brown, M. P. S., Grundy, W. N., Lin, D., Sugnet, C., Ares, M., and Haussler, D., (2000). Knowledge-based analysis of microarray gene expression data by using support vector machines. *Proceedings of the National Academy of Sciences*, 97, 262-267.

- Butte, A. J., and Kohane, I. S., (2000). Mutual information relevance networks: Functional genomic clustering using pairwise entropy measurements. *Pacific Symposium on Biocomputing*, 5, 415-426.

- Cai, J., Dayanik, A., Yu, H., Hasan, N., Terauchi, T., and Grundy, W. N., (2000). Classification of Cancer Tissue Types by Support Vector Machines Using Microarray Gene Expression Data. *International Conference on Intelligent Systems for Molecular Biology*.

- Chapelle, O., Vapnik, V., Bousquet, O. and Mukherjee, S., (2000). Choosing kernel parameters for support vector machines. *AT&T Labs Technical Report*.

- Chen, T., Filkov, V., and Skiena, S. S., (1999). Identifying gene regulatory networks from experimental data. *Annual International Conference on Computational Biology*.

- Cortes, C., and Vapnik, V., (1995). Support vector networks. *Machine Learning*, 20, 273-297.

- Cover, T., (1965). Geometrical and Statistical Properties of Systems of Linear Inequalities with Applications in Pattern Recognition. *IEEE Transaction on Electronic Computer*, 14, 326-334.

- Datta, S., (2001). Exploring relationships in gene expressions: A partial least squares approach. *Gene Expression*, 9, 257-264.

- Dewey, T. G., and Bhan, A., (2001). A linear systems analysis of expression time series. *Methods of Microarray Data Analysis*, Kluwer Academic.





- D'haeseleer, P., (2000). Reconstructing Gene Networks from Large Scale Gene Expression Data. Ph.D. thesis, University of New Meixco.

- D'haeseleer, P., Wen, X., Fuhrman, S., and Somogyi, R., (1997). Mining the gene expression matrix: inferring gene relationships from large scale gene expression data. *Information processing in cells and tissues*, Paton, R. C., and Holcombe, M., Eds., Plenum Press, 203-212.

- Ewing, R. M., Kahla, A. B., Poirot, O., Lopez, F., Audic, S., and Claverie, J. M., (1999). Large-scale statistical analyses of rice ESTs reveal correlated patterns of gene expression. *Genome Research*, 9, 950-959.

- Filkov, V, Skiena, S, and Zhi, J., (2001). Analysis techniques for microarray time-series data. *Annual International Conference on Computational Biology*. 124-131.

- Fitch, W. M., and Margoliash, E., (1967). Construction of phylogenetic trees. *Science*, 155, 279-284.

- Fletcher, R., (1987). *Practical Methods of Optimization*, 2nd edition, Wiley, New York.

- Freund, Y., and Schapire, R. E., (1996). Experiments with a new boosting algorithm. *Machine learning: Proceedings of the Thirteenth International Conference*, 148-156.

- Friedman, N., Linial, M., Nachman, I., and Pe'er, D., (2000). Using Bayesian networks to analyze expression data. *Journal of Computational Biology*, 7, 601-620.

- Fuhrman, S., Cunningham, M. J., Wen, X., Zweiger, G., Seihamer, J. J., and Somogyi, R., (2000). The application of Shannon entropy in the identification of putative drug targets. *Biosystems*, 55, 5-14.

- Furey, T. S., Cristiannini, N., Duffy, N., Bednarski, D. W., Schummer, M., and Haussler, D., (2000). Support vector machine classification and validation of cancer tissue samples using microarray expression data, *Bioinformatics*, 16, 906-914.





- Golub, T. R., Slonim, D. K., Tamayo, P., Huard, C., Gassenbeek, M., Mesirov, J. P., Coller, H., Loh, M. L., Downing, J. R., Caligiuri, M. A., Bloomfield, C. D., and Lander, E. S., (1999). Molecular Classification of Cancer: Class Discovery and Class Prediction by Gene Expression Monitoring. *Science*, 286, 531-537.

- Guyon, I., Weston, J., Barnhill, S., and Vapnik, V., (2002). Gene selection for cancer classification using support vector machines. *Machine Learning*, 46, 389-422.

- Haykin, S., (1999). *Neural networks: A comprehensive foundation*. Prentice Hall.

- Hellem, T. and Jonassen, I., (2002). New feature subset selection procedures for classification of expression profiles. *Genome Biology*, 3(4), research0017.1-0017.11

- Hertz, J., Krogh, A., and Pla,er, R. G., (1991). *Introduction to the Theory of Neural Computation*. Addison-Wesley.

- Hieter, P., and Boguski, M., (1997). Functional genomics: it's all how you read it. *Science*, 278, 601-602.

- Holter, N. S., Maritan, A., Cieplak, M., Fedoroff, N. V., and Banavar, J. R., (2001). Dynamic modeling of gene expression data. Proceedings of the National Academy of *Sciences*, 98, 1693-1698.

- Huang, S., (1999). Gene expression profiling, genetic networks, and cellular states: an integrating concept for tumorigenesis and drug discovery. *Journal of Molecular Medicine*, 77, 469-480.

- Hwang, K. B., Cho, D. Y., Park, S. W., Kim, S. D., and Zhang, B. T., (2001). Applying machine learning techniques to analysis of gene expression data: cancer diagnosis. *Methods of Microarray Data Analysis*. Kluwer Academic, 167-182.

- Ideker, T. E., Thorsson, V. and Karp, R. M., (2000). Discovery of regulatory interactions through pertubation: inference and experimental design. *Pacific Symposium on Biocomputing*, 5, 302-313.





- Kauffman, S. A., (1969). Metabolic stability and epigenesis in randomly connected nets. *Journal of Theoretical Biology*, 22, 437-467.

- Keller, A. D., Schummer, M., Ruzzo, W. L., and Hood, L., (2000). Bayesian classification of DNA array expression data. Technical Report, University of Washington - Computer Science Engineering - 2000-08-01.

- Khan, J., Wei, J. S., Ringner, M., Saal, L. H., Ladanyi, M, Westermann, F, Berthold, F., Schwab, M., Antonescu, C. R., Peterson, C., and Meltzer, P. S., (2001). Classification and diagnostic prediction of cancers using gene expression profiling and artificial neural networks. *Nature Medicine*, 7, 673-679.

- Kiberstis, P., and Roberts, L., (2002). It's Not Just the Genes. *Science*, 296, 685.

- Kitano, H., (2002). Systems biology: a brief overview. *Science*, 295, 1662-1664.

- Klevecz, R. R., (2000). Dynamic architecture of the yeast cell cycle uncovered by wavelet decomposition of expression microarray data. *Functional and Integrative Genomics*, 1, 186-192.

- Li, W., (2002). Zipf's law in importance of genes for cancer classification using microarray Data. *Journal of Theoretical Biology*, 219, 539-51.

- Liang, S., Fuhrman, S., and Somogyi, R., (1998). REVEAL, A general reverse engineering algorithm for inference of genetic network architectures, *Pacific Symposium on Biocomputing*.

- Liu, H., and Motoda, H., (1998). *Feature selection for knowledge discovery and data mining.* Kluwer Academic Publishers, 1998.

- Lo, J., Lee, S., Xu, M., Liu, F., Ruan, H., Eun, A., He, Y., Ma, W., Wang, W., Wen, Z., and Peng, J., (2003). 15,000 Unique Zebrafish EST Clusters and Their Use in Microarray for Profiling Gene Expression Patterns During Embryogenesis. *Genome Research*, 13, 455-466.





- Maki, Y., Tominaga, D., Okamoto, M., Watanabe, S., and Eguchi, Y., (2001). Development of a System for the Inference of Large Scale Genetic Networks. *Pacific Symposium on Biocomputing*, 6, 446-458.

- Michaels, G. S., Carr, D. B., Askenazi, M., Fuhrman, S., Wen, X., and Somogyi, R., (1998). Cluster analysis and data visualization of large-scale gene expression data. *Pacific Symposium on Biocomputing*, 3, 42-53.

- Michigan, (1999). Challenges and Opportunities in Understanding the Complexity of Living Systems. University of Michigan, report into initiatives of Life Sciences Commission within context of biocomplexity.

- Muirhead, R. J., (1982). *Aspects of Multivariate Statistical Theory*. Wiley series in probability and mathematical statistics. Wiley, New York.

- Mukherjee, S., Tamayo, P., Slonim, D., Verri, A., Golub, T., Messirov, J. P., and Poggio, T., (2000). Support vector machine classification of microarray data. *AI memo, CBCL paper* 182. MIT.

- Pe'er, D., Regev, A., Elidan, G., and Friedman, N., (2001). Inferring subnetworks from perturbed expression profiles. *International Conference on Intelligent Systems for Molecular Biology*.

- Pineda, F. J., (1987). Generalization of back-propagation to recurrent neural networks. *Physical Review Letters*, 59, 2229-2232.

- Platt, J., (1999). Fast training of SVMs using sequential minimal optimisation. *Advances in Kernel Methods: Support Vector Learning*. MIT press, Cambridge, MA, 185-208.

- Quinlan, J. R., (1993). *C4.5: Programming for Machine Learning*. Morgan Kaufmann Publishers.





- Raychaudhuri, S., Stuart, J. M., and Altman, R. B., (2000). Principal components analysis to summarize microarray experiments: application to sporulation time series. *Pacific Symposium on Biocomputing*, 5, 452-463.

- Samsonova, M. G., and Serov, V. N., (1999). NetWork: An interactive interface to the tools for analysis of genetic network structure and dynamics. *Pacific Symposium on Biocomputing*.

- Savageau, M. A., (1976). *Biochemical Systems analysis: a study of function and design in molecular biology*. Addison-Wesley, Reading.

- Setiono, R., and Liu, H., (1997). Neural-network feature selector. *IEEE Transactions on Neural Networks*, 8, 654-662.

- Slonim, D., Tamayo, P., Mesirov, J., Golub, T. R., and Lander, E., (2000). Class prediction and discovery using gene expression data. *Annual International Conference on Computational Biology*.

- Someren, E. P. V., Wessels, L. F. A. and Reinders, M. J. T., (2001). Genetic Network Models: A Comparative Study. *Proceedings of SPIE, Micro-arrays: Optical Technologies and Informatics*.

- Someren, E. V., Wessels, L. F. A., and Reinders, M. J. T., (2000). Linear modeling of genetic networks from experimental data. *International Conference on Intelligent Systems for Molecular Biology*.

- Somogyi, R., Fuhrman, S., Askenazi, M. and Wuensche, A., (1996). The gene expression matrix: towards the extraction of genetic network architectures. *Proceedings of Second World Congress of Nonlinear Analysis*, 30, 1815-1824.

- Spellman, P. T., Sherlock, G., Zhang, M. Q., Iyer, V. R., Anders, K., Eisen, M. B., Brown, P. O., Botstein, D., and Futcher, B., (1998). Comprehensive identification of





cell cycle-regulated genes of the yeast sacccharomyces cerevisiae by microarray hybridization. *Molecular Biology of the Cell*, 9, 3272-3297.

- Stone, M., and Brooks, R. J., (1992). Continuum regression: cross-validated sequentially constructed prediction embracing ordinary least squares, partial least squares and principal component regression. *Journal of the Royal Statistical Society B*, 52, 237-269: 1990. Corrigendum 54, 906-907: 1992.

- Szallasi, Z., (1999). Genetic network analysis in light of massively parallel biological data acquisition. *Pacific Symposium on Biocomputing*.

- Talbot, W.S., and Hopkins, N., (2000). Zebra fish mutations and functional analysis of the vertebrate genome. *Genes and Development*, 14, 755-762.

- Tibshirani, R., Hastie, T., Eisen, M., Ross, D., Botstein, D., Brown, P., (1999). Clustering methods for the analysis of DNA microarray data. Technical Report, Stanford University.

- Vapnik, V., (1998). *Statistical Learning Theory*. Wiley, New York.

- Vohradsky, J., (2001). Neural network model of gene expression. *FASEB Journal*, 15, 846-854.

- Weigend, A. S., Rumelhart, D. E., and Huberman, B. A., (1991). Generalization by weight-elimination with application to forecasting. *Advances in Neural Information Processing Systems*, Lippmann, R. P., Moody, J., and Touretzky, D. S., Eds., Morgan Kaufmann, 3, 875-882.

- Wen, X., Fuhrman, S., Michaels, G. S., Carr, D. B., Smith, S., Barker, J. L., and Somogyi, R., (1998). Large-scale temporal gene expression mapping of CNS development. *Proceedings of the National Academy of Sciences*, 95, 334-339.

- Werbos, P. J., (1990). Backpropagation through time: what it does and how to do it. *Proceedings of the IEEE*, 78, 1550-1560.





- Wessels, L. F. A., Someren, E. P. V., and Reinders, M. J. T., (2001). A Comparison of Genetic Network Models. *Pacific Symposium on Biocomputing*, 508-519, Hawai.

- Weston, J., Mukherjee, S., Chapelle, O., Pontil, M., Poggio, T. and Vapnik, V., (2001). Feature Selection for SVMs. *Advances in Neural Information Processing Systems*, 13, 668-674.

- Yang, Y. H., Dudoit, S., Luu, P., and Speed, T. P., (2001). Normalization for cDNA Microarray Data. Microarrays: *Optical Technologies and Informatics, Proceedings of SPIE*.

- Yeung, K. Y., Haynor, D. R., Ruzzo, W. L., (2001). Validating clustering for gene expression data. *Bioinformatics*, 17, 309-318.

- Zhang, B. T., Ohm, P., and Muhlenbein, H., (1997). Evolutionary Induction of Sparse Neural Trees. *Evolutionary Computation*, 5, 213-236.

- Zipf, G. F., (1965) [1935]. *Psycho-Biology of Languages*. Mass. MIT Press.